\documentclass[11pt,a4paper]{article} 
\usepackage{jheppub}
\usepackage{multirow}
\usepackage{fix-cm}
\usepackage{epsfig}
\usepackage{amsmath}
\usepackage{color}
\usepackage{graphicx}
\usepackage{dcolumn}
\usepackage{bm}

\def\beq{\begin{equation}}
\def\eeq{\end{equation}}
\def\beqa{\begin{eqnarray}}
\def\eeqa{\end{eqnarray}}

\def \as {\relax\ifmmode\alpha_s\else{$\alpha_s${ }}\fi}

\newcommand{\lsim}{
\mathrel{\hbox{\rlap{\hbox{\lower4pt\hbox{$\sim$}}}\hbox{$<$}}}}
\newcommand{\gsim}{
\mathrel{\hbox{\rlap{\hbox{\lower4pt\hbox{$\sim$}}}\hbox{$>$}}}}

\keywords{renormalization, perturbative QCD, resummation, exponentiation, Wilson lines, eikonal approximation, soft singularities}

\title{On the renormalization of multiparton webs}

\author[a]{Einan Gardi,} 
\author[a]{Jennifer M. Smillie}
\author[b]{and Chris D. White}

\affiliation[a]{The Tait Institute, School of Physics and Astronomy, 
The University of Edinburgh, \\ 
Edinburgh EH9 3JZ, Scotland, UK}
\affiliation[b]{School of Physics and Astronomy, 
Scottish Universities Physics Alliance, University of Glasgow, \\
Glasgow G12 8QQ, Scotland, UK}

\emailAdd{Einan.Gardi@ed.ac.uk}
\,\emailAdd{j.m.smillie@ed.ac.uk}
\emailAdd{Christopher.White@glasgow.ac.uk}

\abstract{We consider the recently developed diagrammatic approach to soft-gluon exponentiation in multiparton scattering amplitudes, where the exponent is written as a sum of \emph{webs} -- closed sets of diagrams whose colour and kinematic parts are entangled via mixing matrices.
A complementary approach to exponentiation is based on the multiplicative renormalizability of intersecting Wilson lines, and their subsequent finite anomalous dimension.
Relating this framework to that of webs, we derive renormalization constraints expressing all multiple poles of any given web in terms of lower-order webs. 
We examine these constraints explicitly up to four loops, and find that they are realised through the action of the web mixing matrices in conjunction with the fact that multiple pole terms in each diagram reduce to sums of products of lower-loop integrals. 
Relevant singularities of multi-eikonal amplitudes up to three loops are calculated in dimensional regularization using an exponential infrared regulator.
 Finally, we formulate a new conjecture for web mixing matrices, involving a weighted sum over column entries.
Our results form an important step in understanding non-Abelian exponentiation in multiparton amplitudes, and pave the way for higher-loop computations of the soft anomalous dimension. 
}
\begin{document}
\begin{flushright}
Edinburgh 2011/20
\vspace*{-25pt}
\end{flushright}
\maketitle
\allowdisplaybreaks 

\section{Introduction\label{sec:introduction}}

The infrared singularities of scattering amplitudes \cite{Korchemsky:1985xj,Ivanov:1985np,Korchemsky:1987wg,Korchemsky:1988hd,Korchemsky:1988si,Korchemsky:1992xv,Gardi:2005yi,Becher:2006qw,Becher:2005pd,Korchemskaya:1994qp,Botts:1989kf,Contopanagos:1996nh,Kidonakis:1998nf,Kidonakis:1997gm,Kidonakis:2009zc,Sterman:2002qn,Aybat:2006mz,Aybat:2006wq,Laenen:2008ux,Kyrieleis:2005dt,Sjodahl:2008fz,Seymour:2008xr,Kidonakis:2009ev,Mitov:2009sv,Becher:2009kw,Beneke:2009rj,Czakon:2009zw,Ferroglia:2009ep,Ferroglia:2009ii,Chiu:2009mg,Mitov:2010xw,Ferroglia:2010mi,Becher:2009cu,Gardi:2009qi,Becher:2009qa,Dixon:2008gr,Dixon:2009gx,Dixon:2009ur,Bierenbaum:2011gg,Laenen:2008gt} are important both for pragmatic collider phenomenology applications and for theoretical reasons. 
The singularities govern the structure of large logarithmic corrections to cross sections
to all orders in perturbation theory, allowing the possibility to resum such
terms, see e.g.~\cite{Sterman:1986aj,Catani:1992ua,Contopanagos:1993yq,Catani:1989ne,Oderda:1999im,Kidonakis:1998bk,Laenen:1998qw,Laenen:2000ij,Bozzi:2007pn,Catani:1996yz,Gardi:2007ma,Andersen:2005mj,Gardi:2002bg,Gardi:2001ny,Cacciari:2002xb,Gardi:2002xm,Becher:2006mr,Becher:2007ty,Becher:2008cf,Ahrens:2008nc,Bozzi:2008bb,Kang:2008zzd,Idilbi:2009cc,Almeida:2009jt,Moch:2009mu,Moch:2009my,Mantry:2009qz,Beenakker:2010fw,Beneke:2010gm,Papaefstathiou:2010bw,Chien:2010kc}. 
They have also been explored in a variety of more formal contexts, such as supersymmetric gauge theories and quantum 
gravity~\cite{Weinberg:1965nx,Giddings:2010pp,Bork:2009nc,Naculich:2008ew,Donoghue:1999qh,Dunbar:1995ed,Naculich:2009cv,Naculich:2008ys,Naculich:2011ry,White:2011yy}, 
in order to examine the higher loop structure of those theories and to 
elucidate the relationships between them. 

It is well known that infrared (IR) singularities can be traced to the emission of soft and collinear gluons, where the latter contribute only in the case of massless partons. Consequently an amplitude may be factorised into hard, soft and jet functions, in such a way that the hard function, describing high-virtuality exchanges at any loop order, is infrared finite, the soft function collects all non-collinear long-distance singularities, while a jet is assigned to each massless parton, collecting the corresponding collinear singularities.
This separation is useful because the soft and jet functions which carry the singularities have a simple and universal structure, which does not depend on the details of the hard process. Most importantly, these functions \emph{exponentiate}. It is this property that provides the basis for resummation, allowing multiple emissions to be derived from a single emission.

In non-Abelian gauge theories there is a crucial distinction between scattering
amplitudes with two coloured partons, and those involving several coloured partons. 
In the former case, only one possible colour flow is present, such that the two partons must necessarily be coupled by a colour
singlet hard interaction. In the latter case, more than one possible colour flow may be
present. If there are $L$ partons, one may decompose the amplitude as
\begin{equation}
{\cal M}_{i_1\cdots i_L}=\sum_I{\cal M}_I\,(c^I)_{i_1\cdots i_L}
\label{Mdef}
\end{equation}
where the  $\{c^I\}$ are a basis of colour tensors, and $i_n$ the colour index
associated with the $n^{\rm th}$ parton. In this colour flow basis the amplitude is represented by a vector with components ${\cal M}_I$. Consequently, the factorization between hard and soft modes turns into a matrix problem~\cite{Korchemskaya:1994qp,Kidonakis:1997gm,Kidonakis:1998nf,Sterman:2002qn,Aybat:2006wq,Aybat:2006mz,Becher:2009cu,Gardi:2009qi,Becher:2009qa,Kidonakis:2009ev,Mitov:2009sv,Becher:2009kw,Beneke:2009rj,Czakon:2009zw,Ferroglia:2009ep,Ferroglia:2009ii,Kidonakis:2009zc,Chiu:2009mg,Mitov:2010xw,Ferroglia:2010mi,Kyrieleis:2005dt,Sjodahl:2008fz,Seymour:2008xr}: the soft function becomes a matrix in colour-flow space, which acts on the hard interaction, itself a vector in this space. This adds a degree of combinatoric
complexity which is absent in the two parton case, and as a consequence the
IR singularity structure of multiparton amplitudes has been explored 
only recently relative to that of the colour-singlet case. 

In this paper we study soft-gluon exponentiation in multiparton amplitudes. The basic tool is the soft (or eikonal) approximation, in which each external hard parton $i$ is replaced by a corresponding semi-infinite Wilson line, extending from the origin -- where the hard interaction takes place -- to infinity, in a direction $\beta_i$, the direction of motion of parton $i$.  
The eikonalized amplitude so obtained is much simpler than the original one, and yet it has the exact same soft singularities. 
The simplicity of the eikonal amplitude is due to the fact that it only depends on the (Minkowskian) angles between the hard partons and on their colours -- but neither on their energies, nor their spins.  
Collinear singularities do depend on these, and they are not fully captured by the eikonal approximation.  Importantly, however, collinear singularities -- in contrast to soft ones -- are associated with individual massless partons: they do not depend on the colour flow in the hard process and can be computed using two-parton amplitudes (the Sudakov form factor).   
Thus, for the purpose of studying the exponentiation of soft gluons and addressing the complication posed by colour in the multiparton case, we may indeed use the eikonal approximation. In our analysis it will be useful to avoid collinear singularities altogether, and we do this by tilting the Wilson lines off the lightcone, such that $\beta_i^2<0$ for any $i$. At the end one may return to the massless case by considering the $\beta_i^2\to 0$ limit.

The conventional approach to soft-gluon exponentiation is based on the fact that products of Wilson lines renormalize multiplicatively~\cite{Polyakov:1980ca,Arefeva:1980zd,Dotsenko:1979wb,Brandt:1981kf}. Exponentiation is obtained as a solution of a renormalization-group equation involving a finite anomalous dimension function, the so-called soft anomalous dimension. 
This approach underlies much of our present understanding of the infrared singularity of amplitudes ~\cite{Korchemsky:1985xj,Ivanov:1985np,Korchemsky:1987wg,Korchemsky:1988hd,Korchemsky:1988si,Korchemsky:1992xv,Gardi:2005yi,Becher:2006qw,Becher:2005pd,Korchemskaya:1994qp,Botts:1989kf,Contopanagos:1996nh,Kidonakis:1997gm,Kidonakis:1998nf,Sterman:2002qn,Aybat:2006mz,Laenen:2008ux,Kyrieleis:2005dt,Sjodahl:2008fz,Seymour:2008xr,Kidonakis:2009ev,Mitov:2009sv,Becher:2009kw,Beneke:2009rj,Czakon:2009zw,Ferroglia:2009ep,Ferroglia:2009ii,Chiu:2009mg,Mitov:2010xw,Ferroglia:2010mi,Becher:2009cu,Gardi:2009qi,Becher:2009qa,Dixon:2008gr,Dixon:2009gx,Dixon:2009ur}.
Significant progress was made in recent years in determining the soft anomalous dimension to 
two-loop order and furthermore, gaining some insight into its all order properties in the massless case~\cite{Becher:2009cu,Gardi:2009qi,Becher:2009qa,Dixon:2009gx,Dixon:2009ur}. It is clear, however, that new techniques are needed to push this programme further.

A complementary approach to exponentiation is the diagrammatic one. This approach aims at a direct computation of the exponent of the eikonal amplitude.
In the two parton case, it has long been established that the exponent may be written as a sum of a restricted set of Feynman diagrams -- so-called {\it webs} -- which are irreducible with respect to cutting the external parton lines~\cite{Gatheral:1983cz,Frenkel:1984pz,Sterman:1981jc}\footnote{See~\cite{Laenen:2008gt} for a rederivation of these results, which provides a more elegant solution for the exponentiated colour factors.}. 
A direct consequence of irreducibility is that, aside from running coupling corrections, each web has only a single pole in dimensional regularization.
This goes hand in hand with the fact that the exponent must be expressible 
as an integral over a finite anomalous dimension.
Recently, the diagrammatic picture has been extended to the multiparton 
case by two independent groups of authors~\cite{Gardi:2010rn,Mitov:2010rp,
Gardi:2011wa}. Both groups find that multiparton webs are no longer
irreducible, and as a consequence may contain higher-order poles, corresponding to subdivergences of the multi-eikonal vertex.
It is not a priori clear how the singularity structure of multiparton webs can be consistent with
multiplicative renormalizability and a finite anomalous dimension. 

The main goal of the present paper is to combine the renormalization picture with that of webs, in order to gain further insight into the singularity structure of the soft gluon 
exponent in multiparton scattering. 
We will see in particular that renormalization of the eikonal amplitude 
implies a set of constraint equations for multiparton webs, which determine all multiple poles in any given web in terms of lower-order webs.
Single poles remain the only genuinely new content of webs which is needed to determine the soft anomalous dimension at a given order. We will explicitly examine these constraints up
to four-loop order, and then study how they are realised in a number of non-trivial examples. In order to do this, we will introduce a new exponential regulator for disentangling
UV and IR poles in eikonal calculations at arbitrary orders in perturbation theory. 
We will also use what we learn from this analysis to formulate a new conjecture on the \emph{web mixing matrices} which govern the mixing of colour and kinematic information in multiparton 
webs~\cite{Gardi:2010rn,Gardi:2011wa}.

The structure of the paper is as follows. In the rest of this introduction,
we review known results relating to renormalization of products of Wilson lines (section \ref{sec:Wilson_lines}) and multiparton webs (section~\ref{sec:multiparton}), which
will be useful in what follows. 
In section~\ref{sec:renormalization} we discuss the renormalization of multiparton webs in detail, and formulate constraint equations relating the singularities of webs across different orders in perturbation theory, alongside expressions for the soft anomalous dimension coefficients in terms of webs. 
In section~\ref{sec:3_loop_examples} we introduce the infrared regulator we use in order to compute ultraviolet singularities in eikonal amplitudes, and then examine how the renormalization constraints of the previous section are realised using specific three-loop examples, where certain technical details are collected in appendices. 
We show that for the leading singularity, ${\cal O}(\epsilon^{-3})$, the kinematic dependence of each diagram factorises into three one-loop kinematic factors, while the next-to-leading singularity, ${\cal O}(\epsilon^{-2})$, can be written as a sum of products of lower-order diagrams. These properties, in conjunction with the action of the web mixing matrix, provide the key to understanding how the renormalization constraints are satisfied.
Next, in section~\ref{sec:factorization} we show that the factorization of the leading singularity can be generalised to all orders for the class of diagrams made exclusively of individual gluon exchanges.
Leading singularities in different diagrams $D$ in such webs differ only by a combinatorial factor $s(D)$ counting the number of orders in which gluon subdiagrams can be sequentially shrunk to the origin.
In section~\ref{sec:conjecture}, we formulate a new conjecture regarding web mixing matrices, involving a sum rule over column entries, weighted by the above $s(D)$. For the class of webs made of single-gluon exchanges this sum-rule directly explains the cancellation of the leading singularities required by renormalization. It appears, however, that the sum rule itself is completely general, and holds for any web, similarly to other properties of the web mixing matrices~\cite{Gardi:2010rn} which already have general combinatorial proofs~\cite{Gardi:2011wa}.
In section~\ref{sec:conclusions} we summarize our conclusions.

\subsection{Wilson lines in multiparton scattering\label{sec:Wilson_lines}}

As stated in the previous section, the conventional approach to soft-gluon exponentiation in scattering amplitudes is based on multiplicative renormalizability of the corresponding eikonal amplitude. In this section, we review this subject, preparing for the analysis in section~\ref{sec:renormalization}.
Throughout this paper we consider fixed-angle scattering amplitudes with hard momenta $p_i$,  such that $|p_i\cdot p_j| \gg \Lambda_{\rm QCD}^2$, where all invariants are taken simultaneously large. 
The emission of soft gluons corresponds to the eikonal approximation, in which the momenta
of the emitted gauge bosons may be neglected with respect to the hard 
momenta~$p_i$. In this limit, each external particle may be replaced, as far as the interaction with soft gluons is concerned, by the semi-infinite Wilson line
\begin{equation}
\Phi_{\beta_i}\equiv {\bf P}\,\exp\left\{{\rm i}g_s\int_0^{\infty}d\lambda \beta\cdot A(\lambda\beta)\right\},
\label{Wilsondef}
\end{equation}
where the direction $\beta_i$ is set by the respective hard parton momenta $p_i$. Similarly, the colour representation in which the gauge field is defined is determined by that of parton~$i$. The eikonal approximation is useful because it simplifies the calculation while capturing all soft singularities. Throughout this paper we consider Wilson lines which are off the lightcone, $\beta_i^2\neq 0$ (corresponding to massive external particles), thus avoiding any collinear singularities. The soft singularities are then captured by the eikonal amplitude
\begin{align}
\label{S_ren_def}
{\cal S}_{\rm ren.}\left(\gamma_{ij},\epsilon_{\rm IR}\right)
&\equiv\left<0\left|\Phi_{\beta_1}\otimes\Phi_{\beta_2}\otimes\ldots\otimes\Phi_{\beta_L} \right|0\right>_{\rm ren.}
\end{align}
where the colour indices of different Wilson lines remain open. Upon projecting on the colour tensors introduced in (\ref{Mdef}), eq.~(\ref{S_ren_def}) becomes a matrix in colour flow space. 
This matrix acts on
the hard interaction (finite as $\epsilon\rightarrow 0$) which is a vector in this space:
\begin{align}
{\cal M}^{I} (p_i,\epsilon_{\rm IR})= {\cal S}_{\rm ren.}^{IJ}\left(\gamma_{ij},\epsilon_{\rm IR}\right){\cal H}_J(p_i)\,\,.
\end{align}
In contrast to the full amplitude or the hard function, which depend on the momenta $p_i$, the kinematic dependence of the soft function (\ref{S_ren_def}) is entirely through the cusp parameters\footnote{The dependence on $\gamma_{ij}$ at leading order is often expressed via the corresponding cusp angle, see eq.~(\ref{eq:F1m1}).}, 
\begin{equation}
\label{gamma_ij}
\gamma_{ij}={2\beta_i\cdot\beta_j}/{\sqrt{\beta_i^2\beta_j^2}}\,,
\end{equation} 
which, similarly to the eikonal Feynman rules, are invariant under rescaling of the velocities.
Furthermore the soft function is independent of the spins of the emitting particles. 

Note that the operator in (\ref{S_ren_def}) has been renormalized: all its ultraviolet singularities, including in particular those related to the multi-eikonal vertex, have been removed. Following Refs.~\cite{Polyakov:1980ca,Arefeva:1980zd,Dotsenko:1979wb,Brandt:1981kf} we know that it renormalizes multiplicatively, namely there exists a factor $Z$ which collects ultraviolet 
counterterms at each order, and such that 
${\cal S}_{\rm ren.}={\cal S}\,Z$ is ultraviolet finite. Similarly to ${\cal
  S}$, the $Z$ factor is a matrix in colour flow space, so their product should be understood as a matrix product (we suppress the colour space indices throughout). 
Clearly, upon renormalization the eikonal amplitude must acquire some renormalization-scale dependence. Furthermore, it will be convenient to distinguish between the renormalization scale of the coupling $\mu_R$ and the renormalization scale $\mu$ introduced through the $Z$ factor in the process of renormalizing the multi-eikonal vertex.

Considered in pure dimensional regularization, all loop corrections in ${\cal S}$ involve just scaleless integrals, and therefore vanish identically. Therefore \emph{before renormalization, the result is trivial}, and one has
\begin{align}
\begin{split}
\label{S_ren_UV_and_IR}
{\cal S}_{\rm ren.}\left(\gamma_{ij},\alpha_s(\mu_R^2),\epsilon_{\rm IR},\mu\right)
&
={\cal S}_{{\rm UV}+{\rm IR}}\,
Z\left(\gamma_{ij},\alpha_s(\mu_R^2),\epsilon_{\rm UV},\mu\right)
\\&
=Z\left(\gamma_{ij},\alpha_s(\mu_R^2),\epsilon_{\rm UV},\mu\right)\,.
\end{split}
\end{align}
Thus, in dimensional regularization the soft (infrared) singularities we seek to know are given by the $Z$ factor. As recognized by different authors~\cite{Korchemsky:1985xj,Korchemsky:1987wg,Kidonakis:1998nf,Kidonakis:1997gm,Becher:2009cu},
this allows one to explore soft singularities to all loops by studying the \emph{ultraviolet} renormalization factor of the multi-eikonal vertex in (\ref{S_ren_def}).

The analysis of the renormalization of cusp or cross singularities of Wilson lines forms the basis of much theoretical work in recent years leading to substantial progress in understanding the structure of infrared singularities in multi-leg amplitudes.  As we review below, these singularities may be encoded in an integral over a so-called `soft anomalous dimension' $\Gamma$.
The soft anomalous dimension has been fully determined to two-loop order, with both massless~\cite{Aybat:2006wq,Aybat:2006mz} and massive partons~\cite{Kidonakis:2009ev,Mitov:2009sv,Becher:2009kw,Beneke:2009rj,Czakon:2009zw,Ferroglia:2009ep,Ferroglia:2009ii,Chiu:2009mg,Mitov:2010xw,Ferroglia:2010mi,Bierenbaum:2011gg}.
Moreover, in the massless case, stringent all-order constraints were 
derived~\cite{Becher:2009cu,Gardi:2009qi,Becher:2009qa} based on factorization and rescaling symmetry, leading to a remarkable possibility, 
namely that all soft singularities in any multi-leg amplitude take the form of a sum over colour dipoles formed by any pair of hard coloured partons.
Despite recent progress~\cite{Dixon:2008gr,Gardi:2009qi,Dixon:2009gx,Becher:2009cu,Becher:2009qa,Dixon:2009ur,Gardi:2009zv,Gehrmann:2010ue}, the basic questions of whether the sum-over-dipoles formula receives corrections, and at what loop order, remain so far unanswered. Further progress in understanding the singularity structure of multiparton scattering amplitudes in both the massless and massive cases requires new techniques to facilitate higher-loop computations, or a new viewpoint to give further insight into the all-loop structure of the anomalous dimension. 

Our discussion leading to (\ref{S_ren_UV_and_IR}) has been quite formal. In order to explicitly compute the counterterms entering $Z$, and as suggested by eq.~(\ref{S_ren_UV_and_IR}), we need to introduce an infrared regulator which disentangles UV and IR singularities, and which does not break the symmetries of the problem. With the regulator in place, all poles in $\epsilon\equiv \epsilon_{\rm UV}$ are strictly of ultraviolet origin, and using $D=4-2\epsilon$ with $\epsilon>0$ will render the integrals well defined. The infrared regulator we use in this paper will be described in section~\ref{sec:regulator} below. For our general discussion here it is sufficient to assume that such a regularization exists. Denoting the energy scale associated with the infrared regulator by $m$, and the scale associated with the renormalization of the multi-eikonal vertex by $\mu$, the regulated version of eq.~(\ref{S_ren_UV_and_IR}) takes the form:
\begin{equation}
\label{S_reg_ren_UV_and_IR}
{\cal S}_{\rm ren.}\left(\gamma_{ij},\alpha_s(\mu_R^2),\mu,m\right)
={\cal S}  \left(\gamma_{ij},\alpha_s(\mu_R^2),\epsilon, m\right)\,
Z\left(\gamma_{ij},\alpha_s(\mu_R^2),\epsilon,\mu\right)\,.
\end{equation}
We emphasize that while the $m$-dependent ${\cal S}$ and ${\cal S}_{\rm ren.}$ here are different from their nonregularized counterparts in eq.~(\ref{S_ren_UV_and_IR}), the $Z$ factor, which by definition depends on the ultraviolet only, is identically the same.
By acting on ${\cal S}$ the $Z$ factor removes its $\epsilon\to 0$ singularities, while making it $\mu$ dependent. Although not indicated in (\ref{S_reg_ren_UV_and_IR}), ${\cal S}_{\rm ren.}$, while finite, usually retains some dependence of $\epsilon$ through positive powers. This is precisely the situation we will encounter here using a minimal subtraction scheme, where counterterms are pure poles.

To proceed one introduces an anomalous dimension $\Gamma$, defined by
\begin{equation}
\frac{dZ}{d\ln \mu}=- Z \Gamma\,.
\label{Gamma_def}
\end{equation}
Similarly to $Z$ itself, $\Gamma$ is a function of the kinematic variables, $\gamma_{ij}$, as well as of $\alpha_s(\mu_R^2)$ and $\mu$.
Importantly, however, $\Gamma$ \emph{must be finite}, as we now show. 
In section~\ref{sec:renormalization} we will further show that in a minimal subtraction scheme, $\Gamma$ (unlike ${\cal S}_{\rm ren.}$) is independent of~$\epsilon$, aside from its dependence through the running coupling; thus the perturbative coefficients $\Gamma^{(n)}$ at any order $\alpha_s^n$ are truly $\epsilon$-independent.
 
To show that $\Gamma$ must be finite let us first invert (\ref{S_reg_ren_UV_and_IR}) getting
\begin{equation}
\label{S_reg_ren_UV_and_IR2}
{\cal S}  \left(\epsilon, m\right)=
{\cal S}_{\rm ren.}\left(\mu,m\right)
Z^{-1}\left(\epsilon,\mu\right)
\,,
\end{equation}
where we suppressed the dependence on $\gamma_{ij}$ and $\alpha_s(\mu_R^2)$. We note that while both ${\cal S}_{\rm ren.}$ and $Z^{-1}$ on the r.h.s. depend on $\mu$, the l.h.s. does not: $\mu$ is only introduced through the renormalization procedure of the multi-eikonal vertex.
Therefore, differentiating with respect to $\mu$ we have:
\begin{equation}
\label{scale_invariance_of_S}
0\,=\,\frac{d{\cal S}_{\rm ren}(\mu,m)}{d\ln \mu}\,Z^{-1} (\epsilon,\mu)
+ {\cal S}_{\rm ren}(\mu,m)\,\frac{dZ^{-1} (\epsilon,\mu)}{d\ln\mu}\,.
\end{equation}
From (\ref{Gamma_def}) we have:
\[
\Gamma\equiv -Z^{-1}\frac{dZ}{d\ln \mu}=\frac{dZ^{-1}}{d\ln \mu}Z\qquad\Longrightarrow\qquad 
\frac{dZ^{-1}}{d\ln \mu}=\Gamma\,Z^{-1}\,.
\]
Substituting this into (\ref{scale_invariance_of_S}) and multiplying by $Z$ from the right we obtain:
\begin{equation}
\label{scale_invariance_of_S_}
\frac{d{\cal S}_{\rm ren}(\mu,m)}{d\ln \mu}\,
=- {\cal S}_{\rm ren}(\mu,m)\,\Gamma
 \,.
\end{equation}
This implies that $\Gamma$ also describes the scale dependence of the \emph{renormalized} correlator, which is by construction free of any ultraviolet singularities. It follows that 
\[
\Gamma=
-{\cal S}_{\rm ren}^{-1}(\mu,m)\,\frac{d{\cal S}_{\rm ren}(\mu,m)}{d\ln \mu}\,\] 
must itself be ultraviolet finite (finite in 4 dimensions). Since $Z$ is composed of ultraviolet poles, the fact that $\Gamma$ in (\ref{Gamma_def}) is ultraviolet finite
is a strong constraint on the singularity structure of $Z$. 

The next observation is that the only scales on which $\Gamma$ can depend are renormalization scales:\footnote{Our argument is nothing but the standard `separation of variables' argument,
where the anomalous dimension admitting both (\ref{Gamma_def}) and (\ref{scale_invariance_of_S_}) can only depend on the common arguments of the ultraviolet object $Z$ and the infrared one ${\cal S}_{\rm ren}$.} the renormalization scale of the coupling, $\mu_R$, and $\mu$ (in particular, given the definition (\ref{Gamma_def}) $\Gamma$ cannot depend on the infrared scale $m$). Therefore, if we choose $\mu_R=\mu$, the only scale dependence that survives in $\Gamma$ is through the argument of the coupling. Thus we have:
\begin{equation}
\Gamma=\Gamma\left(\gamma_{ij},\alpha_s(\mu^2,\epsilon)\right)\,,
\end{equation}
where we have emphasised that working in $D=4-2\epsilon$ dimensions, epsilon dependence occurs through the running coupling $\alpha_s=\alpha_s (\mu^2,\epsilon)$, which obeys the following renormalization-group equation:
\begin{equation}
\label{beta_function}
\frac{d\alpha_s}{d\ln\mu^2}=-\alpha_s\left[\epsilon+b_0\alpha_s+b_1\alpha_s^2+b_2\alpha_s^3+\cdots\right]\,,
\qquad\qquad
b_0=\frac{1}{4\pi}\left(\frac{11}{3}C_A-\frac{4}{3}T_R N_f\right)\,.
\end{equation}

Having established the finiteness of the anomalous dimension, the exponentiation of soft-gluon singularities is a direct consequence of the defining evolution equation (\ref{Gamma_def}). However, the situation is complicated in multiparton amplitudes by the fact that $\Gamma$ is a matrix in colour space. This issue was addressed in section 3 of \cite{Mitov:2010rp} (see also section 6 in \cite{Gardi:2010rn}) and will be developed further here, in section~\ref{sec:renormalization}.
Before addressing the general multiparton case it is useful to first recall the structure of the exponent in the colour singlet case. 
Equation~(\ref{Gamma_def}) can then be readily integrated to yield an exponential form:
\begin{equation}
\label{Z_exp_colou_singlet}
Z_{\rm colour\,\, singlet}=\,\exp\left\{\frac12\int_{\lambda^2}^{\infty}\frac{d\lambda^2}{\lambda^2}\, \Gamma\left(\gamma_{ij},\alpha_s(\lambda^2,\epsilon)\right)\right\}\,,
\end{equation}
where we used the initial condition at $\mu^2\to \infty$, where the coupling vanishes due to asymptotic freedom.

Similarly to~\cite{magnea:1990zb} the singularities in the exponent are generated here through integration over the $D$-dimensional coupling (note that in our case these are ultraviolet singularities). 
Substituting the expansion of the anomalous dimension
\begin{align}
\label{Gamma_expansion}
\Gamma=\sum_{n=1}^{\infty}\Gamma^{(n)}\alpha_s^n
\end{align}
it is easy to see that in the conformal case, where $b_i=0$, namely where $d\alpha_s/{d\ln\mu^2}=-\alpha_s\,\epsilon$, one obtains:
\begin{equation}
\label{Z_exp_expanded_1l}
\left.Z_{\rm colour\,\, singlet}\right\vert_{b_i=0}=\,\exp\left\{\frac{1}{2\epsilon}\,\sum_{n=1}^{\infty}\frac{1}{n}\, \alpha_s^n\,\Gamma^{(n)} \right\}\,.
\end{equation}
Thus, put simply, in the colour singlet case, the exponent (in a conformal theory) has \emph{only simple poles} in $\epsilon$ to all orders in perturbation theory. 
This result, which is straightforward to derive, is central to our understanding of the infrared singularity structure of gauge theory amplitudes.
 
As we show in detail in section~\ref{sec:renormalization}, upon relaxing the two simplifying assumptions we took above, namely including running-coupling corrections, and going to the multiparton case, this structure gets modified, giving rise to higher-order poles in the exponent as follows:
\begin{align}
\label{Z_exp_expanded_1l_multiparton}
\begin{split}
Z=&\,\exp\left\{
\,\frac{1}{2\epsilon}\,\Gamma^{(1)}\,\alpha_s
+\,\left(\frac{1}{4\epsilon}\,\Gamma^{(2)}-\frac{b_0}{4\epsilon^2}\,\Gamma^{(1)}\right)\,\alpha_s^2
\right.\\&
\left.
+\,\left(\frac{1}{6\epsilon}\,\Gamma^{(3)}
+\frac{1}{48\epsilon^2}\left[\Gamma^{(1)},\Gamma^{(2)}\right]-\frac{1}{6\epsilon^2}
\left(b_0\Gamma^{(2)}+b_1\Gamma^{(1)}\right)+\frac{b_0^2}{6\epsilon^3}\Gamma^{(1)}\right)\,\alpha_s^3
\,+\,{\cal O}(\alpha_s^4)\,
 \right\}\,.
\end{split}
\end{align}
Indeed, running-coupling corrections generate multiple poles starting from two
loops, and in addition, due to the non-commuting nature of $\Gamma^{(n)}$ in the
multiparton case, one finds multiple pole terms involving commutators starting at
three loops.  Importantly, however, \emph{all multiple poles in $\epsilon$, at
  any order, are determined by lower-order coefficients}. This in turn implies
non-trivial relations between the webs of different orders, relations we shall
determine in section~\ref{sec:renormalization} and explore through examples in section~\ref{sec:3_loop_examples}.
 
\subsection{Multiparton webs\label{sec:multiparton}}

Having reviewed the use of Wilson lines in multiparton scattering in the previous section, in this section we discuss the recently developed diagrammatic approach to soft gluon exponentiation, briefly describing the results of \cite{Gardi:2010rn,Mitov:2010rp,Gardi:2011wa}.
In the case of Wilson loops -- or alternatively amplitudes with a colour-singlet hard interaction -- diagrammatic exponentiation 
was first studied in the 1980s by Gatheral \cite{Gatheral:1983cz} and by Frenkel and Taylor~\cite{Frenkel:1984pz}. In this case, as already mentioned briefly above, only irreducible\footnote{We distinguish between reducible and irreducible (or colour-connected) diagrams. Reducible diagrams are ones whose colour factor can be written as a product of the colour factors of its subdiagrams; see e.g. \cite{Berger:2003zh,Gardi:2010rn,Gardi:2011wa}.} 
diagrams contribute to the exponent.
By contrast, in the multiparton case, the exponent receives contributions from reducible diagrams as well. Refs.~\cite{Gardi:2010rn,Mitov:2010rp,Gardi:2011wa} have shown that these contributions are formed through mixing between kinematic factors ${\cal F}(D)$ and colour factors $C(D')$ of closed sets of diagrams, thus generalizing the concept of webs to the multiparton case.
A web $W_{(n_1,n_2,\ldots,n_L)}$ represents the contribution to the exponent from a set of diagrams which are related to each other by permutations of the gluon attachments to each Wilson line, where $n_l$ is the number of attachments to line $l$, with $l=1,\ldots,L$.
A multiparton web is then given by  
\begin{equation}
W_{(n_1,n_2,\ldots,n_L)}\,\equiv\, \sum_{D}{\cal F}(D)\,\widetilde{C}(D)=\sum_{D,D'}{\cal F}(D) \,R_{DD'}\,C(D')\,,
\label{setmix}
\end{equation}
where $\widetilde{C}(D)$ is the so-called Exponentiated Colour Factor (ECF) corresponding to diagram $D$ and the sums are over all the diagrams in the set.
A general algorithm for computing the relevant \emph{web mixing matrices} $R$ was developed in~\cite{Gardi:2010rn}, and a corresponding closed-form combinatoric expression was established in terms of a sum over decompositions of a diagram into complete sets of its subdiagrams (eq.~(10) in~\cite{Gardi:2011wa}). 

In \cite{Gardi:2010rn} the web mixing matrices were computed for a range of non-trivial 3 and 4-loop examples, revealing interesting mathematical properties: these matrices are idempotent, $R^2=R$, and have zero-sum rows, $\sum_{D'} R_{DD'}=0$, $\forall D$. These properties, which were later proven to be completely general~\cite{Gardi:2011wa}, are crucial for understanding the colour structure of the exponent, the interplay between colour and kinematic dependence, and the singularity structure. 

The generalization of the familiar properties of colour-singlet webs, namely their ``maximal non-Abelian'' colour structure and the absence of subdivergences, to the multiparton case is non-trivial. Recall that colour-singlet webs are individually irreducible (or colour-connected) diagrams. Each diagram that enters the exponent does so with a ``maximally non-Abelian" ECF.
These diagrams have just one overall ultraviolet pole which is associated with the cusp, and any other subdivergence may only contribute to the renormalization of the strong coupling.
Indeed there is a one-to-one correspondence between the irreducibility of the colour structure and the absence of subdivergences: as one can easily verify graphically, any diagram which has a reducible colour structure can be shown to have subdivergences, since its innermost subdiagrams can be sequentially shrunk towards the cusp (the hard interaction) without affecting the attachments of the outer gluons. Each such shrinking step gives rise to an ultraviolet singularity, a $1/\epsilon$ pole in dimensional regularization. The converse is also true, namely any irreducible diagram would have just a single overall divergence associated with the cusp. 

On the face of it this picture is entirely broken in the multiparton case.
As shown in \cite{Gardi:2010rn,Mitov:2010rp,Gardi:2011wa} only some multiparton webs are individual irreducible diagrams, while most are instead formed as in (\ref{setmix}) through a sum over a set of diagrams, many of which have a reducible colour structure, and therefore have subdivergences.
One of the most interesting observations in \cite{Gardi:2010rn}, however, was that
web properties do generalise to the multiparton case upon considering the set of diagrams, rather than individual ones. These properties are realised through the action of the web mixing matrix. Being idempotent, $R$ acts as a projection operator on the space of colour and kinematic factors of the diagrams in the set, thus has eigenvalues 0 and 1 (each with nontrivial multiplicity in general).
 Equation~(\ref{setmix}) then states that contributions to the exponents are formed through particular linear combinations of kinematic factors multiplying corresponding linear combinations of colour factors. The number of such combinations is determined by the rank of $R$, corresponding to the multiplicity of its unit eigenvalue, while the remaining combinations, corresponding to its zero eigenvalue, are projected out. 
The non-Abelian nature of these colour factors is realised by forming anti-symmetric combinations; in particular, the fully symmetric part of the colour factor in each web is projected out owing to the zero-sum rows property. This property is expected to hold separately for subsets of diagrams with different planarity~\cite{Gardi:2011wa}.

\begin{figure}[tb]
\begin{center}
\scalebox{1}{\includegraphics{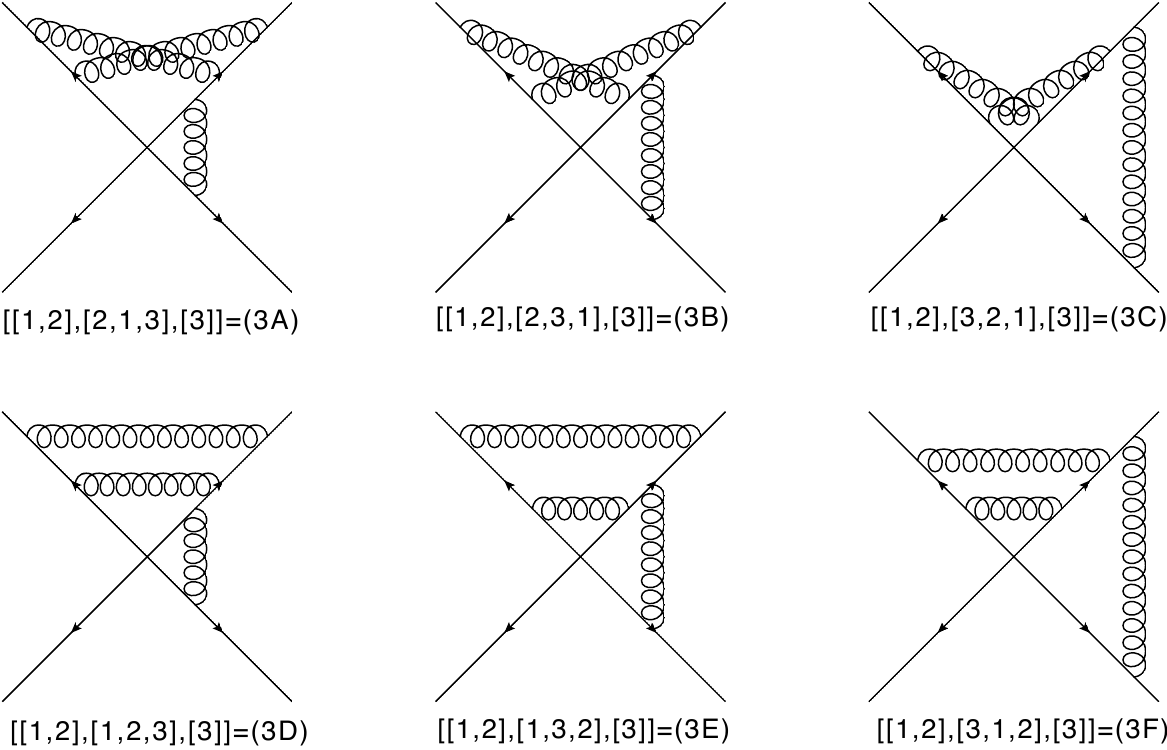}}
\caption{The six 3-loop diagrams forming the 2-3-1 web in which three eikonal lines are linked by three gluon exchanges.}
\label{3lsix}
\end{center}
\end{figure}

The pole structure of multiparton webs is rather intricate, but crucially, it is strongly constrained by renormalization. The simplest case to analyse is that of the leading singularities, namely ${\cal O}(1/\epsilon^n)$ poles at ${\cal O}(\alpha_s^n)$, corresponding to the maximal subdivergence of each diagram. As already noted in~\cite{Gardi:2010rn} the only ${\cal O}(1/\epsilon^n)$ singularities are the ones purely associated with the running coupling (specifically, they are given by eq.~(\ref{wnn}) below). Therefore, the maximal subdivergences in webs with two or more connected pieces must cancel between different diagrams through the operation of the web mixing matrix. For example, in $W_{(2,3,1)}$ of figure \ref{3lsix} one expects~\cite{Gardi:2010rn} cancellation of ${\cal O}(\epsilon^{-3})$ poles among the lower three diagrams (diagrams ($3A$)--($3C$) have only subleading divergences), and
we shall verify these cancellations explicitly in section \ref{sec:3_loop_examples} below.
We shall revisit this in section~\ref{sec:conjecture}, where we provide a general explanation of the cancellation mechanism noting a general property of the web mixing matrices in the form of a weighted sum over column entries.

A more complicated structure arises for subleading poles. Starting at ${\cal O}(1/\epsilon^{n-1})$ at ${\cal O}(\alpha_s^n)$, for $n\geq 3$, renormalization of multiparton webs involves commutators of lower-orders terms, as in eq.~(\ref{Z_exp_expanded_1l_multiparton}). This implies that very particular multiple poles should survive in multiparton webs. As we shall see below, to be consistent with renormalization, these must correspond to commutators between lower-order webs, which are formed from subdiagrams of the diagrams in the set. We shall see how this structure arises in section~\ref{sec:3_loop_examples}.

In summary, multiparton webs are crucially different to the two parton case in
that they are closed sets of diagrams rather than single diagrams. Each
individual diagram may be reducible, and thus may contain subdivergences
associated with the renormalization of the multi-eikonal vertex.  While in the
two parton case all multiple poles in $\epsilon$ arise from the running of the
coupling, in the multiparton case additional multiple poles appear due to the
non-commutativity of the multi-eikonal vertex counterterms.  Nevertheless,
similarly to the two parton case, all multiple poles at a given order are fully
determined by lower orders. This highly constrained singularity structure must
be matched by diagrammatic exponentiation, and therefore by the web mixing
matrices.  The main aim of the present paper is to investigate how this
structure is realised. We begin by deriving the singularity structure of webs
from the renormalization of the multi-eikonal vertex, which is the subject of
the following section.

\section{Renormalization of multiparton webs\label{sec:renormalization}}

In the previous section, we reviewed the present state of knowledge regarding
soft gluon exponentiation in multiparton amplitudes, including the recently 
developed diagrammatic approach. In this section, we will further 
develop the analysis of the renormalization of the multi-eikonal vertex that
originated in~\cite{Mitov:2010rp}, and was also discussed 
in~\cite{Gardi:2010rn}. We will see in particular that the requirement of 
multiplicative renormalizability of the eikonal scattering amplitude places severe constraints
on the singularity structure of multiparton webs. These lead to a series of consistency relations expressing the multiple poles of webs at
a given order in terms of commutators of lower-order web poles. We will develop
such equations explicitly, and check these in the following section.

Let us now examine the renormalization of the eikonal vertex along the lines of section~3 in \cite{Mitov:2010rp} and 6.1 in \cite{Gardi:2010rn}. As already
remarked above, eikonal diagrams are themselves scaleless, and thus give zero
in dimensional regularization due to the cancellation of UV and IR poles. At 
one-loop order it is relatively straightforward to disentangle these 
singularities. At higher loop orders, the identification of UV and
IR contributions becomes ambiguous, and in order to compute the UV singularities which comprise the $Z$ factor in dimensional regularization, it becomes necessary to implement an
additional regulator which eliminates all IR divergences. 
We will indeed introduce such a regulator in section~\ref{sec:3_loop_examples}, where we explicitly set about calculating specific
diagrams. With this regulator in place all integrals are mathematically well-defined for $D<4$ namely for $\epsilon=(4-D)/2>0$. In this section, we assume that such a regulator has already been
introduced, such that any remaining poles are of UV origin.

As described in the introduction, multiplicative renormalizability of the
multi-eikonal vertex implies that one may introduce a renormalization factor $Z$ according
to eq.~\eqref{S_reg_ren_UV_and_IR} such that $Z$ absorbs all the ultraviolet
singularities of ${\cal S}(\epsilon)$, making the renormalized eikonal amplitude
${\cal S}_{\rm ren}(\mu)$ \emph{finite}, and at the same time
$\mu$-dependent.
Note that both
${\cal S}$ and $Z$ are matrix-valued in colour flow space, consistently with the
left-hand side. The ordering of the factors on the right-hand side of
eq.~(\ref{S_reg_ren_UV_and_IR}) is important because of their non-commuting
nature. This equation constrains the singularity structure that ${\cal
  S}(\epsilon)$ may have, thus effectively constraining the corresponding webs.

Let us see how these constraints arise. Both the eikonal amplitude ${\cal
  S}(\epsilon)$ and the $Z$ factor exponentiate. That is, one may write~\cite{Mitov:2010rp}
\begin{equation}
{\cal S}(\epsilon)={\rm e}^{w},
\qquad 
Z (\epsilon,\mu) ={\rm e}^{\zeta},
\label{SZexp}
\end{equation}
where the first exponent can be decomposed in perturbation theory as 
\begin{equation}
w=\sum_{n=1}^{\infty}w^{(n)} \,\alpha_s^n\,,
\label{wzdef}
\end{equation}
and $\alpha_s$ (here and below) denotes the renormalized coupling in $D=4-2\epsilon$ dimensions $\alpha_s=\alpha_s(\mu^2,\epsilon)$.
The coefficients on the right-hand side collect all non-renormalized webs at a 
given order in $\alpha_s$, i.e.
\begin{equation}
w^{(n)}\,\alpha_s^n\,=\,\sum_{n_1,\ldots,n_L} W_{(n_1,\ldots,n_L)}^{(n)}\,,
\label{wndef}
\end{equation}
where we have used the notation of
section~\ref{sec:multiparton}, adding a superscript to indicate the loop order,
$W_{(n_1,\ldots,n_L)}^{(n)}={\cal O}(\alpha_s^n)$.  At any given order ${\cal
  O}(\alpha_s^n)$ one must have $n_1+\ldots n_L\leq 2n$ due to the fact that
each gluon emission from an external (eikonal) line carries a factor of $g_s$.
The inequality arises from the fact that gluons may also be connected (by
multiple gluon vertices or fermion bubbles) away from the external lines, which
introduces extra powers of the coupling.

The second exponent in eq.~(\ref{SZexp}) may also be decomposed in perturbation
theory, as
\begin{equation}
\quad \zeta=\sum_{n=1}^{\infty}\zeta^{(n)}\,\alpha_s^n\,,
\label{zetadef}
\end{equation}
where $\zeta^{(n)}$ is the counterterm at ${\cal O}(\alpha_s^n)$. 
In general the finite part of the counterterms depends on the renormalization scheme. Here we use the $\overline{\rm MS}$ scheme throughout, meaning in particular that $\zeta^{(n)}$ does not have any finite parts: it is comprised of \emph{only} negative powers of $\epsilon$.  

As first observed in~\cite{Mitov:2010rp}, combining the exponents in eq.~(\ref{SZexp}) according
to eq.~\eqref{S_reg_ren_UV_and_IR2} leads to an exponential form for ${\cal
  S}_{\rm ren}$, where the exponent contains an infinite number of commutator terms arising
from the Baker-Campbell-Hausdorff formula:
\begin{align}
\label{S_ren_BCH}
\begin{split}
{\cal S}_{\rm ren}(\mu)&=  \exp \left\{ w (\epsilon) \right\} \, \exp \left\{ 
\zeta(\epsilon,\mu)  \right\} \\&= 
\exp\left\{
w+\zeta 
+\frac12 [w,\zeta] 
+\frac{1}{12}\Big([w,[w,\zeta]] - [\zeta,[w,\zeta]]\Big)
-\frac{1}{24}[\zeta,[w,[w,\zeta]]] +\cdots
\right\}\,.
\end{split}
\end{align}
This complication is completely absent in the case of two partons connected
by a colour singlet interaction, as in that case both ${\cal S}(\epsilon)$
and $Z$ are not matrix-valued. All commutators in eq.~(\ref{S_ren_BCH}) then
vanish, consistent with the fact that, aside from running-coupling terms, webs in two-parton scattering have only single poles (a consequence of their irreducibility). In the 
multiparton case, the commutators are non-zero in general, and lead to a
rich structure of multiple epsilon poles even in the conformal case.

Let us now examine the exponent of eq.~(\ref{S_ren_BCH}) at the first few 
orders in perturbation theory. In terms of the quantities introduced in
eqs.~(\ref{wzdef}, \ref{zetadef}) one finds, up to four loop order,
\begin{subequations}
\label{w_n_ren}
\begin{align}
w_{\rm ren}^{(1)}&=w^{(1)}+\zeta^{(1)};\label{webren1}\\
w_{\rm ren}^{(2)}&=w^{(2)}+\zeta^{(2)}+\frac{1}{2}\left[w^{(1)},\zeta^{(1)}\right];\label{webren2}\\
w_{\rm ren}^{(3)}&=w^{(3)}+\zeta^{(3)}+\frac{1}{2}\left(\left[w^{(1)},\zeta^{(2)}\right]
+\left[w^{(2)},\zeta^{(1)}\right]\right)+\frac{1}{12}\left[w^{(1)}-\zeta^{(1)},
\left[w^{(1)},\zeta^{(1)}\right]\right] \label{webren3}
\\
\begin{split}\label{webren4}
w_{\rm ren}^{(4)}&=w^{(4)}+\zeta^{(4)}
+\frac12\left(
\left[w^{(1)},\zeta^{(3)}\right]+
\left[w^{(2)},\zeta^{(2)}\right]
+
\left[w^{(3)},\zeta^{(1)}\right]
\right)
\\&+\frac{1}{12}\left(
\left[w^{(1)}-\zeta^{(1)}, \left[w^{(1)},\zeta^{(2)}\right]+\left[w^{(2)},\zeta^{(1)}\right]\right]+\left[w^{(2)}-\zeta^{(2)}, \left[w^{(1)},\zeta^{(1)}\right]\right]
\right)
\\&-\frac{1}{24}\left[\zeta^{(1)},\left[w^{(1)}, \left[w^{(1)},\zeta^{(1)}\right]\right]\right]\,
\,.
\end{split}
\end{align}
\end{subequations}
These equations express the sum of renormalized webs at each order in terms of
lower-order counterterms and non-renormalized webs. We can already see how this
constrains multiparton webs: the right-hand side of each equation must be finite
as $\epsilon\rightarrow 0$, implying the cancellation of all $\epsilon$ poles in
the particular combinations which occur. 
The real power of these relations, however, is the fact the counterterm $\zeta$ is associated with a 
\emph{finite anomalous dimension}. This implies that $\zeta$ must have a very particular pole structure, which in turn constrains the singularities of the webs. Our next task is therefore to inject this information into (\ref{w_n_ren}). To do this, we begin by defining the anomalous dimension matrix according to eq.~(\ref{Gamma_def})
\begin{subequations}
\begin{align}
\label{Gamma_}
\Gamma\equiv -Z^{-1}\frac{dZ}{d\ln \mu}&
=-\int_0^1d\tau\, {\rm e}^{-\tau\zeta}\,\frac{d\zeta}{d\ln\mu}\,{\rm e}^{\tau \zeta}
\\&
\label{Gamma_commut}
=-\frac{d\zeta}{d\ln\mu}
+\frac{1}{2!}\left[\zeta,\frac{d\zeta}{d\ln\mu}\right]
-\frac{1}{3!}\left[\zeta,\left[\zeta,\frac{d\zeta}{d\ln\mu}\right]\right]
+\cdots
\,. 
\end{align}
\end{subequations}
We have used a known operator identity to relate the derivative of $Z={\rm
  e}^{\zeta}$ to that of $\zeta$ and then the Hadamard lemma to express the
result in terms of commutators, see e.g.~\cite{Hadamard,Fuchs:1997jv}. As for
the webs and counterterms, one may expand the anomalous dimension matrix in an
analogous way, defining
\begin{equation}
\Gamma=\sum_{n=1}^{\infty}\Gamma^{(n)}\alpha_s^n=\Gamma^{(1)}\alpha_s
+\Gamma^{(2)}\alpha_s^2
+\Gamma^{(3)}\alpha_s^3
+\Gamma^{(4)}\alpha_s^4+\cdots.
\end{equation}
One may calculate the first few orders explicitly in terms of the expansion coefficients of $\zeta$ and the QCD beta function (\ref{beta_function}). To this end we first write 
\begin{eqnarray}
\frac{d\zeta}{d\ln{\mu}}&=&2\frac{d\alpha_s}{d\ln\mu^2}\,\frac{d\zeta}  {d\alpha_s}
\nonumber
\\&=&-2\alpha_s\left[\epsilon+b_0\alpha_s+b_1\alpha_s^2+b_2\alpha_s^3+\cdots\right]
\left(\zeta^{(1)}+2\alpha_s\zeta^{(2)}+3\alpha_s^2\zeta^{(3)}
+4\alpha_s^3\zeta^{(4)}+\cdots\right)
\nonumber
\\&=& \nonumber
-2\Bigg[
\epsilon\zeta^{(1)}\alpha_s +\left(2\epsilon\zeta^{(2)}+\zeta^{(1)}b_0\right)\alpha_s^2
+\left(3\epsilon\zeta^{(3)}+2\zeta^{(2)}b_0+\zeta^{(1)}b_1\right)\alpha_s^3
\\& &\qquad 
+\left(4\epsilon\zeta^{(4)}+3\zeta^{(3)}b_0+2\zeta^{(2)}b_1+\zeta^{(1)}b_2\right)
\alpha_s^4
+\cdots
\Bigg]\,.
\end{eqnarray}
Using (\ref{Gamma_commut}) we now obtain:
\begin{subequations}
\label{Gamma_n_zeta_n}
\begin{align}
\frac12\Gamma^{(1)}&=\epsilon\zeta^{(1)}
\\
\frac12\Gamma^{(2)}&=2\epsilon\zeta^{(2)}+b_0\zeta^{(1)}
\\
\frac12\Gamma^{(3)}&=\epsilon\left(3\zeta^{(3)}-\frac12\left[\zeta^{(1)},\zeta^{(2)}\right]\right)+2\zeta^{(2)}b_0+\zeta^{(1)}b_1
\\
\begin{split}
\frac12\Gamma^{(4)}&=\epsilon\left(4\zeta^{(4)}-\left[\zeta^{(1)},\zeta^{(3)}\right]+\frac16\left[\zeta^{(1)},\left[\zeta^{(1)},\zeta^{(2)}\right]\right]\right)
\\&\qquad
 +\left(3\zeta^{(3)}-\frac12\left[\zeta^{(1)},\zeta^{(2)}\right]\right)b_0+2\zeta^{(2)}b_1+\zeta^{(1)}b_2\,
.
\end{split}
\end{align}
\end{subequations}
According to the standard separation of variables argument, hard-soft factorization
implies that $\Gamma$, as defined in (\ref{Gamma_}), must be independent of both an infrared cutoff and an ultraviolet cutoff, thus it must be finite in four dimensions\footnote{Recall that we consider Wilson lines which are off the light cone, so that no collinear singularities are present.} ($\epsilon\to 0$). Knowing that $\Gamma^{(n)}$ are finite, and working in a minimal subtraction scheme where the counterterms $\zeta^{(n)}$ include, by definition, only negative powers of $\epsilon$ (no ${\cal O}(\epsilon^i)$ for $i\geq0$), we see that
$\Gamma^{(n)}$ are independent of $\epsilon$: neither negative nor positive powers are allowed. It is therefore very convenient to express the singularity structure of $\zeta^{(n)}$ (and also $w^{(n)}$ below) in terms of $\Gamma^{(n)}$. 

Inverting the relations in (\ref{Gamma_n_zeta_n}) we get:
\begin{subequations}
\label{zeta_n_Gamma_n}
\begin{align}
\zeta^{(1)}&=\frac{1}{2\epsilon}\,\Gamma^{(1)}
\label{zeta_n_Gamma_n_1}
\\
\zeta^{(2)}&=\frac{1}{4\epsilon}\,\Gamma^{(2)}-\frac{b_0}{4\epsilon^2}\,\Gamma^{(1)} \label{zeta_n_Gamma_n_2}
\\
\zeta^{(3)}&=\frac{1}{6\epsilon}\,\Gamma^{(3)}
+\frac{1}{48\epsilon^2}\left[\Gamma^{(1)},\Gamma^{(2)}\right]-\frac{1}{6\epsilon^2}
\left(b_0\Gamma^{(2)}+b_1\Gamma^{(1)}\right)+\frac{b_0^2}{6\epsilon^3}\Gamma^{(1)}
\label{zeta_n_Gamma_n_3}
\\
\begin{split}
\label{zeta_n_Gamma_n_4}
\zeta^{(4)}&=\frac{1}{8\epsilon}\,\Gamma^{(4)}
+\frac{1}{48\epsilon^2}\left[\Gamma^{(1)},\Gamma^{(3)}\right]
-\frac{b_0}{8\epsilon^2}\Gamma^{(3)}
+\frac{1}{8\epsilon^2}\left(\frac{b_0^2}{\epsilon}-b_1\right)\Gamma^{(2)}
\\
&
+\frac{1}{8\epsilon^2}\left(-\frac{b_0^3}{\epsilon^2}+\frac{2b_0b_1}{\epsilon}-b_2\right)\Gamma^{(1)}
-\frac{b_0}{48\epsilon^3}\left[\Gamma^{(1)},\Gamma^{(2)}\right]
,
\end{split}
\end{align}
\end{subequations}
where we note that owing to a cancellation there are no nested commutator terms in $\zeta^{(4)}$.
Finally substituting the expressions for $\zeta^{(n)}$ from (\ref{zeta_n_Gamma_n}) into eqs.~(\ref{w_n_ren}) we extract the following results for the singularity structure of the non-renormalized webs $w^{(n)}$:
\begin{subequations}
\label{w_n_Gamma_n}
\begin{align}
w^{(1)}&=w_{\rm ren}^{(1)}-\frac{1}{2\epsilon}\,\Gamma^{(1)}
\label{w_1_Gamma_1}
\\
w^{(2)}&=w_{\rm ren}^{(2)} -\frac{1}{4\epsilon}\Gamma^{(2)}
-\frac{1}{4\epsilon}\left[w_{\rm ren}^{(1)},\Gamma^{(1)}\right]
+\frac{b_0}{4\epsilon^2}\Gamma^{(1)}
\label{w_2_Gamma_2}
\\
\begin{split}
\label{w_3_Gamma_3}
w^{(3)}&=w_{\rm ren}^{(3)}-\frac{1}{6\epsilon}\Gamma^{(3)}
-\frac{1}{8\epsilon}\left[w_{\rm ren}^{(1)},\Gamma^{(2)}\right]
-\frac{1}{4\epsilon}\left[w_{\rm ren}^{(2)},\Gamma^{(1)}\right]
-\frac{1}{24\epsilon}
\left[w_{\rm ren}^{(1)},\left[w_{\rm ren}^{(1)},\Gamma^{(1)}\right]\right]
\\
&-\frac{1}{48\epsilon^2}\left[\Gamma^{(1)},\Gamma^{(2)}\right]
-\frac{1}{48\epsilon^2}
\left[\Gamma^{(1)},\left[w_{\rm ren}^{(1)},\Gamma^{(1)}\right]\right]
+\frac{b_0}{8\epsilon^2}\left[w_{\rm ren}^{(1)},\Gamma^{(1)}\right]
\\
&
+\frac{1}{6\epsilon^2}\left(b_0\Gamma^{(2)}+b_1\Gamma^{(1)}\right)
-\frac{b_0^2}{6\epsilon^3}\Gamma^{(1)}
\end{split}
\\
\begin{split}
\label{w_4_Gamma_4}
w^{(4)}&=w_{\rm ren}^{(4)}-\frac{1}{8\epsilon}\Gamma^{(4)}
-\frac{1}{12\epsilon} \left[w_{\rm ren}^{(1)},\Gamma^{(3)}\right]
-\frac{1}{4\epsilon} \left[w_{\rm ren}^{(3)},\Gamma^{(1)}\right]
-\frac{1}{8\epsilon} \left[w_{\rm ren}^{(2)},\Gamma^{(2)}\right]
\\
&
-\frac{1}{24\,\epsilon} \left[w_{\rm ren}^{(2)},\left[w_{\rm ren}^{(1)},\Gamma^{(1)}\right]\right]
-\frac{1}{48\,\epsilon} \left[w_{\rm ren}^{(1)},\left[w_{\rm ren}^{(1)},\Gamma^{(2)}\right]\right]
-\frac{1}{24\,\epsilon} \left[w_{\rm ren}^{(1)},\left[w_{\rm ren}^{(2)},\Gamma^{(1)}\right]\right]
\\&
-\frac{1}{48\,\epsilon^2}\left[\Gamma^{(1)},\Gamma^{(3)}\right]
+\frac{b_0}{8\epsilon^2}\Gamma^{(3)}
-\frac{1}{8\epsilon^2}\left(\frac{b_0^2}{\epsilon}-b_1\right)\Gamma^{(2)}
+\frac{1}{8\epsilon^2}\left(\frac{b_0^3}{\epsilon^2}-\frac{2b_1b_0}{\epsilon} +b_2\right)\,\Gamma^{(1)}
\\
&
-\frac{1}{96\,\epsilon^2}\left[w_{\rm ren}^{(1)},\left[\Gamma^{(1)},\left[w_{\rm ren}^{(1)},\Gamma^{(1)}\right]\right]\right]
+\frac{b_0}{12\epsilon^2}\left[w_{\rm ren}^{(1)},\Gamma^{(2)}\right]
+\frac{b_1}{12\epsilon^2}\left[w_{\rm ren}^{(1)},\Gamma^{(1)}\right]
\\&
+\frac{b_0}{8\epsilon^2}\left[w_{\rm ren}^{(2)},\Gamma^{(1)}\right]
+\frac{b_0}{48\epsilon^2}\left[w_{\rm ren}^{(1)},\left[w_{\rm ren}^{(1)},\Gamma^{(1)}\right]\right]
-\frac{1}{48\,\epsilon^2} \left[\Gamma^{(1)},\left[w_{\rm ren}^{(1)},\Gamma^{(2)}\right]\right]
\\&
-\frac{1}{48\,\epsilon^2}\left[\Gamma^{(1)},\left[w_{\rm ren}^{(2)},\Gamma^{(1)}\right]\right]
+\frac{b_0}{48\epsilon^3}\left[\Gamma^{(1)},\Gamma^{(2)}\right]
-\frac{b_0^2}{12\epsilon^3}\left[w_{\rm ren}^{(1)},\Gamma^{(1)}\right]
\\&
+\frac{b_0}{48\epsilon^3}\left[\Gamma^{(1)},\left[w_{\rm ren}^{(1)},\Gamma^{(1)}\right]\right]
\,
\,.
\end{split}
\end{align}
\end{subequations}
In eqs. (\ref{w_n_Gamma_n}) one can see that higher-order powers in $\epsilon$ (i.e. not just single poles) are indeed present in the non-renormalized webs, as we described in the previous section. However, these are fully determined by lower orders. 
Indeed, as we see below, one can explicitly express all multiple poles at any given order in terms of lower-order webs.
Finally, the new anomalous dimension coefficient $\Gamma^{(n)}$ will be determined at each order from the coefficient of the single ${\cal O}(1/\epsilon)$ pole in the non-renormalized webs $w^{(n)}$ of that order. 
To explicitly determine this structure we introduce a further notation for the $\epsilon$ expansion of (non-renormalized) webs:
\begin{align}
\label{wnk-2}
\begin{split}
w^{(n)}&=\sum_{k=1}^{n} w^{(n,-k)}\,\epsilon^{-k}+\sum_{k=0}^{\infty} w^{(n,k)}\,\epsilon^{k},
\end{split}
\end{align}
such that $w^{(n,-k)}$ and $w^{(n,k)}$ are the coefficients of the $k^{\rm th}$ negative and positive powers of $\epsilon$ at ${\cal O}(\alpha_s^n)$.  
By inserting (\ref{wnk-2}) into (\ref{w_n_Gamma_n}) and extracting the coefficients of $\epsilon^{-k}$ for $k\geq 2$  we obtain the relations
\begin{subequations}
\label{w_nk_relations}
\begin{align}
w^{(2,-2)}&=-\frac12 b_0 w^{(1,-1)}
\label{w_22}
\\
w^{(3,-3)}&=\frac13 b_0^2 w^{(1,-1)}
\label{w_33}
\\
w^{(3,-2)}&=\left(-\frac23w^{(2,-1)}-\frac{1}{12}\left[w^{(1,-1)},w^{(1,0)}\right]\right)\,b_0 -\frac13b_1w^{(1,-1)}
+\frac16\left[w^{(2,-1)},w^{(1,-1)}\right]
\label{w_32}
\\
w^{(4,-4)}&=-\frac14 b_0^3 w^{(1,-1)}
\label{w_44}
\\
w^{(4,-3)}&=\left(\frac12 w^{(2,-1)}-\frac{1}{12}\left[w^{(1,0)},w^{(1,-1)}\right]
\right) b_0^2
+\left(-\frac16\left[w^{(2,-1)},w^{(1,-1)}\right]
+\frac12 b_1w^{(1,-1)}\right)b_0
\label{w_43}
\\
\begin{split}
w^{(4,-2)}&=-\frac{1}{48}\left[w^{(1,1)},w^{(1,-1)}\right]\,b_0^2
+\left(\frac{1}{24}\left[w^{(1,0)},w^{(2,-1)}\right]
+\frac{1}{8}\left[w^{(2,0)},w^{(1,-1)}\right]
-\frac34w^{(3,-1)}
\right.
\\&
\left.
+\frac{1}{48}\left[w^{(1,-1)},\left[w^{(1,1)},w^{(1,-1)}\right]\right]
\right)\,b_0
+\left(-\frac12w^{(2,-1)}-\frac{1}{12}\left[w^{(1,-1)},w^{(1,0)}\right]\right)\,b_1
\label{w_42}
\\&
-\frac{1}{4}b_2w^{(1,-1)}
+\frac{1}{4}\left[w^{(3,-1)},w^{(1,-1)}\right]
+\frac{1}{24}\left[w^{(1,-1)},\left[w^{(2,0)},w^{(1,-1)}\right]\right]
\\&
-\frac{1}{24}\left[w^{(1,-1)},\left[w^{(1,0)},w^{(2,-1)}\right]\right].
\end{split}
\end{align}
\end{subequations}
These achieve the above objective of encapsulating the constraints derived from renormalization in a set of consistency conditions involving only coefficients of non-renormalized webs. Equations~(\ref{w_nk_relations}) can be checked explicitly in specific examples, which forms the subject of the following section.

Before doing this, however, we give explicit expressions for the first few orders of the anomalous dimension matrix in terms of the web coefficients of eq.~(\ref{wnk-2}). These can be calculated after extracting the coefficient of $\epsilon^{-1}$ in eqs.~(\ref{w_n_Gamma_n}), and one finds
{\allowdisplaybreaks 
\begin{subequations}
\label{Gamma_n}
\begin{align}
\Gamma^{(1)}&=-2w^{(1,-1)}\\
\Gamma^{(2)}&=-4w^{(2,-1)}-2\left[w^{(1,-1)},w^{(1,0)}\right]\\
\begin{split}
\label{Gamma_3}
\Gamma^{(3)}&=-6w^{(3,-1)}+\frac32b_0\left[w^{(1,-1)},w^{(1,1)}\right]
+3\left[w^{(1,0)},w^{(2,-1)}\right]
+3\left[w^{(2,0)},w^{(1,-1)}\right]
\\&
+\left[w^{(1,0)},\left[w^{(1,-1)},w^{(1,0)}\right]\right]
-\left[w^{(1,-1)},\left[w^{(1,-1)},w^{(1,1)}\right]\right]
\end{split}
\\
\begin{split}
\Gamma^{(4)}&=-8w^{(4,-1)}
+\frac43\left[w^{(1,2)},w^{(1,-1)}\right]b_0^2
+\left(-2\left[w^{(2,1)},w^{(1,-1)}\right]-
\frac83\left[w^{(1,1)},w^{(2,-1)}\right]
\right.
\\&
\left.
+\left[w^{(1,1)},\left[w^{(1,0)},w^{(1,-1)}\right]\right]
-\frac23\left[w^{(1,0)},\left[w^{(1,-1)},w^{(1,1)}\right]\right]
+\frac43\left[w^{(1,-1)},\left[w^{(1,-1)},w^{(1,2)}\right]\right]
\right)b_0
\\&
+\frac43 \left[w^{(1,-1)},w^{(1,1)}\right]\,b_1
+4\left[w^{(1,0)},w^{(3,-1)}\right]
+4\left[w^{(3,0)},w^{(1,-1)}\right]+4\left[w^{(2,0)},w^{(2,-1)}\right]
\\&
+2\left[w^{(1,1)},\left[w^{(2,-1)},w^{(1,-1)}\right]\right]
+\frac83\left[w^{(1,-1)},\left[w^{(1,1)},w^{(2,-1)}\right]\right]
-\frac43\left[w^{(2,0)},\left[w^{(1,0)},w^{(1,-1)}\right]\right]
\\&
-\frac43\left[w^{(1,0)},\left[w^{(2,0)},w^{(1,-1)}\right]\right]
+\frac43\left[w^{(1,-1)},\left[w^{(2,1)},w^{(1,-1)}\right]\right]
-\frac43\left[w^{(1,0)},\left[w^{(1,0)},w^{(2,-1)}\right]\right]
\\&
-\frac13\left[w^{(1,-1)},\left[w^{(1,-1)},\left[w^{(1,0)},w^{(1,1)}\right]\right]\right]
-\frac13\left[w^{(1,-1)},\left[w^{(1,0)},\left[w^{(1,-1)},w^{(1,1)}\right]\right]\right]
\\&
+\left[w^{(1,0)},\left[w^{(1,-1)},\left[w^{(1,-1)},w^{(1,1)}\right]\right]\right]
+\frac13\left[w^{(1,0)},\left[w^{(1,0)},\left[w^{(1,0)},w^{(1,-1)}\right]\right]\right]
\\&
-\frac13\left[w^{(1,-1)},\left[w^{(1,-1)},\left[w^{(1,-1)},w^{(1,2)}\right]\right]\right]\,.
\end{split}
\end{align}
\end{subequations}
}
As mentioned above, the coefficients $\Gamma^{(n)}$ themselves contain neither
positive nor negative powers of $\epsilon$. However, an important observation is
that higher-order coefficients in the webs, suppressed by positive powers of $\epsilon$, do in fact enter the expressions for the anomalous dimension coefficients. For example, both $w^{(1,1)}$ and $w^{(1,2)}$ appear in eq.~(\ref{Gamma_n}).

Let us now discuss the emerging picture for the pole structure of webs ($w$) at the multi-loop level. According to (\ref{w_22}), at two loops only running coupling corrections can generate a double pole. 
In fact, it is true to all-loop orders that the maximal divergence 
$w^{(n,-n)}$ is simply given by the one-loop web (single gluon exchange) with running coupling insertions:
\begin{equation}
\label{wnn}
w^{(n,-n)}=\frac{1}{n} \,\left(-b_0\right)^{n-1}\, w^{(1,-1)}\,,
\end{equation}
where $b_0$ is the leading coefficient of the beta function defined in eq.~(\ref{beta_function}).
In particular, as explained in \cite{Gardi:2010rn}, commutator terms cannot contribute to the leading singularity, and in a conformal theory we have, at any order,
\begin{equation}
\label{wnn_no_beta}
 \left.w^{(n,-n)}\right\vert_{b_n= 0}\,=\,0\,\,.
\end{equation}

Equations~(\ref{wnn}) and (\ref{wnn_no_beta}) can be proven by induction. While we do not present a detailed proof, we briefly explain why it holds, repeating   
the argument already given in section 6.1 of \cite{Gardi:2010rn}. 
According to eq.~(\ref{webren1}) above, the one-loop counterterm is just the negative of the simple pole of the one-loop web, 
\begin{equation}
\zeta^{(1)}=-\frac{1}{\epsilon}\,w^{(1,-1)}=\frac{1}{2\epsilon}\Gamma^{(1)}.
\label{zeta1pole}
\end{equation}
Consequently at two loops the commutator term in eq.~(\ref{webren2}) $\left[w^{(1)},\zeta^{(1)}\right]$ does not contribute at ${\cal O}(\epsilon^{-2})$. Indeed $\zeta^{(2)}$ in (\ref{zeta_n_Gamma_n_2}) receives no commutator contribution, and the ${\cal O}(\epsilon^{-2})$ on the r.h.s of~(\ref{webren2}) must be a pure running coupling term, proportional to $b_0\Gamma^{(1)}$, which cancels between $w^{(2)}$ and $\zeta^{(2)}$. 
Proceeding to three loops, one can easily see using eq.~(\ref{webren3}) how this iterates: none of the commutator terms can contribute to the maximal singularity, ${\cal O}(\epsilon^{-3})$, and indeed the only ${\cal O}(\epsilon^{-3})$ term that appears in $\zeta^{(3)}$ in (\ref{zeta_n_Gamma_n_3}) is proportional to $\Gamma^{(1)}$, with two insertions of $b_0$. 

With eq.~(\ref{wnn_no_beta}) at hand it is useful to return to the diagrammatic picture of the webs via (\ref{wndef}), and recall that the contribution of each set of diagrams to $w^{(n)}$ takes the form of a linear combination of kinematic factors times the corresponding linear combination of colour factors, as determined by the mixing matrix $R_{(n_1,\ldots,n_L)}^{(n)}$:
\begin{equation}
w^{(n)}\,\alpha_s^n\,=\,\sum_{n_1,\ldots,n_L} W_{(n_1,\ldots,n_L)}^{(n)}
\,=\,\sum_{n_1,\ldots,n_L}\,\,\sum_{D,D'}{\cal F}(D) \,R_{(n_1,\ldots,n_L)}^{(n)}(D,D')\,C(D')\,.
\label{setmix_}
\end{equation}
Thus, the vanishing of the leading singularity in the conformal limit,  eq.~(\ref{wnn_no_beta}), clearly invites the conjecture that 
\begin{equation}
\label{conjecture}
\sum_{D}
\,\left.{\cal F}(D)\right\vert_{{\cal O}(\epsilon^{-n})} \,R_{(n_1,\ldots,n_L)}^{(n)}(D,D')\,=\,0\,,\qquad \quad \forall D'\,,
\end{equation}
namely that the entries of the mixing matrix are precisely such that the leading poles cancel out.
This deduction relies upon the fact that the colour factors of individual diagrams in the web are linearly independent, and on the assumption that there are no further cancellations between different webs contributing at a given order $(n)$. Although this has not been proven in general, one does not expect cancellation of poles between webs. In particular, connected multiparton webs readily admit eq.~(\ref{wnn_no_beta}) on their own, and, as will be verified in non-trivial three-loop examples in the next section, so do reducible webs composed of single-gluon exchange subdiagrams. 

As we shall see in section~\ref{sec:conjecture}, owing to the factorization properties of the leading singularities in a special class of diagrams consisting of individual gluon exchanges, the conjecture of (\ref{conjecture}) directly translates into a sum-rule on the columns of the mixing matrix itself, which we also verify in a variety of examples.

Let us now proceed to consider subleading singularities. According to the above analysis, at each order $n$, commutator terms start contributing at ${\cal O}(\epsilon^{n-1})$. The first such contribution is at three loops
where a double pole is present in the non-renormalized web even when $b_n=0$; in (\ref{w_32}) we have:
\begin{align}
\begin{split}
\label{eq:prize}
  \left.w^{(3,-2)}\right\vert_{b_n= 0}&=
  \,\,
  -\frac{1}{6}\left[w^{(1,-1)},w^{(2,-1)}
  \right]\,.
\end{split}
\end{align}
This simple relation applies to webs that do not include running coupling 
corrections. We will verify it explicitly for two three-loop examples in the following section.

Considering four-loop webs, one sees directly from eq.~(\ref{w_4_Gamma_4}) that only running coupling corrections generate poles at ${\cal O}(\epsilon^{-4})$ and ${\cal O}(\epsilon^{-3})$, namely
\begin{subequations}
\begin{align}
\label{eq:prize_4a}\left.w^{(4,-4)}\right\vert_{b_n=0}=0\,, \\
\label{eq:prize_4b}
  \left.w^{(4,-3)}\right\vert_{b_n= 0}=0\,.
\end{align}
\end{subequations}
While the absence of the leading singularity (\ref{eq:prize_4a}) in the conformal case is expected according to (\ref{wnn_no_beta}), the absence of the next-to-leading one in (\ref{eq:prize_4b}) looks surprising.  Indeed, a term such as $[w_{1, -1},[w_{1, -1},w_{2, -1}]]$ could have a priori appeared in $w^{(4,-3)}$, but the calculation shows that it is absent. The cancellation of this term appears to be coincidental, and does not generalise to higher loops.  We have checked explicitly that for five-loop webs\footnote{The expression for the five-loop anomalous dimension coefficient is rather lengthy, however, and we do not report this here.} one has
\begin{equation}
\left.w^{(5,-4)}\right\vert_{b_0=0} = 
\frac{1}{360}\, [w_{1, -1},[w_{1, -1},[w_{1, -1},w_{2, -1}]]],
\label{5loopweb}
\end{equation}
thus showing that ${\cal O}(\epsilon^{-4})$ poles at five loops are present even in the conformal case.
 
Finally, the ${\cal O}(\epsilon^{-2})$ terms at four loops, in (\ref{w_42}), without running coupling corrections read:
\begin{align}
\begin{split}
\label{eq:w4_double_pole_no_beta_terms}
\left.w^{(4,-2)}\right\vert_{b_n= 0}&=
\frac{1}{4}\left[w^{(3,-1)},w^{(1,-1)}\right]
+\frac{1}{24}\left[w^{(1,-1)},\left[w^{(2,0)},w^{(1,-1)}\right]\right]
\\&
-\frac{1}{24}\left[w^{(1,-1)},\left[w^{(1,0)},w^{(2,-1)}\right]\right]\,.
\end{split}
\end{align}

In this section, we have examined in detail the consequences of the multiplicative renormalizability of multi-eikonal vertices, and have derived a number of important results. The expressions~(\ref{Gamma_n}) show explicitly how the soft anomalous dimension matrix may be constructed, up to four loops, from the coefficients of multiparton webs, thus clarifying how the diagrammatic approach to soft gluon exponentiation can be used in practice to compute the soft anomalous dimension governing soft singularities in multiparton amplitudes. Furthermore, we have gained detailed understanding of the singularity structure of multipartons webs.
We formulated consistency relations which determine all multiple pole terms in webs in terms of lower-order webs. 
We have seen that multiple poles are generated both by running coupling contributions and by commutators. The simplest example of the latter is given by eq.~(\ref{eq:prize}) which expresses the double pole of three-parton webs (which do not contribute to running coupling corrections) in terms of single poles of lower-order webs. 
In the next section we will examine how this relation is satisfied in the context of specific three-loop webs. The ensuing calculations will also allow us to study leading poles at arbitrary order for a special class of diagrams in section~\ref{sec:factorization}.

\section{The singularity structure of multiparton webs: examples \label{sec:3_loop_examples}}

In the previous section, we examined in detail the renormalization of multiparton webs, and found that it was possible to relate all multiple poles of multiloop webs to poles of lower-order webs. 
The main difference with respect to the colour singlet case is that multiple poles remain non-trivial in the conformal limit owing to the presence of commutators. The first non-trivial example is the double pole at three loops, where renormalization implies the relation of eq.~(\ref{eq:prize}). 

In this section we focus on two specific examples of three-loop webs, compute
their singularities, and confirm that these are consistent with the relations of
the previous section.  Beyond illustrating the relations per se, this
investigation is important to reaffirm the general picture proposed
in~\cite{Gardi:2010rn} that multiparton webs, namely the closed sets of diagrams
whose colour and kinematic factors mix, do indeed share the same properties of
connected webs. In particular, each such web has on its own, a singularity
structure that is consistent with renormalization, eq.~(\ref{w_nk_relations}) above. Equivalently, no
cancellations between webs are needed. This is not a priori guaranteed, since
webs are defined based on diagrams, and as such do not in general form
gauge-invariant objects.  Discussing the cancellation of the leading
singularities in (\ref{setmix_}) in the previous section, we have already seen
that it is crucial to verify that the properties of $w^{(n)}$, the sum of all
webs at a given order, which follow from renormalization, are indeed realised on
a web-by-web basis.  The question we address here is more general: we expect
that all the renormalization constraints will be fulfilled by individual webs,
such that it would be possible to renormalize individual webs. We must therefore
check that also subleading poles are consistent with the
constraints, and to this end we will examine in detail how eq.~(\ref{eq:prize})
is satisfied.

As we have already mentioned, any calculation of a multiloop eikonal diagram
requires the use of a regulator for disentangling infrared and ultraviolet
singularities. For the purposes of this paper, we introduce a method for
doing this, which we describe in the following subsection.  It will likely be
useful for future calculations and thus of general interest. 

\subsection{Calculating webs: an exponential regulator~\label{sec:regulator}}

As discussed in eq.~(\ref{S_ren_UV_and_IR}) the calculation of Wilson line amplitudes gives rise to scaleless integrals, which vanish in dimensional regularisation. This can be understood as
the cancellation of IR and UV singularities, where the coefficients of the
latter are relevant for determining the renormalization of the multi-eikonal vertex. 
At one loop, it is possible to
separate unambiguously the IR and UV poles. At higher loops, this separation can
become ambiguous. 
Indeed, previous two-loop calculations of the cusp anomalous dimension~\cite{Korchemsky:1987wg}, and $2\to 2$ eikonal scattering in forward kinematics~\cite{Korchemskaya:1994qp}, have employed regulators which explicitly suppress IR
divergences, such that all remaining poles in dimensional regularisation are manifestly of UV origin. 
The advantage of such an approach is clear: one may
then calculate higher loop diagrams including multiple
variable transformations without having to keep track of how such
transformations mix UV and IR behaviour. Most importantly, the obtained UV poles, leading as well as subleading, are unambiguous. To this end refs.~\cite{Korchemsky:1987wg,Korchemskaya:1994qp} have  employed a fictional gluon mass up to two-loop order. 

In this paper we introduce a regulator which is convenient for multiloop, multi-eikonal calculations. We work directly in configuration space and introduce an \emph{exponential regulator} along each of the Wilson lines by changing the Feynman rule representing the emission of a gluon from a Wilson line  as follows:
\begin{equation}
(ig_s)\,\beta_i^\mu\int_0^\infty d\lambda \,\left(\cdots\right) \qquad\,\,\ \longrightarrow\qquad\,\,
(ig_s)\,\beta_i^\mu\int_0^\infty d\lambda\,e^{-m\lambda\sqrt{-\beta_i^2}} \,\left(\cdots\right)
\label{FRmod}
\end{equation}
where the integral is defined in the region where $m\sqrt{-\beta_i^2}>0$, so that the extra exponential factor smoothly cuts off any long distance ($\lambda\rightarrow\infty$) contribution. This prescription is guaranteed to yield an infrared-finite result because the rest of the integrand in any Feynman diagram can never diverge exponentially for $\lambda\rightarrow\infty$. The exponential damping thus removes all IR singularities, such that all remaining poles are of UV ($\lambda\rightarrow0$) origin. These UV singularities and the corresponding $Z$ factor remain unaffected by the regulator.

This Feynman rule can be formally obtained from a modified definition of the Wilson line of eq.~(\ref{Wilsondef}):
\begin{equation}
\Phi_{\beta_i}^{(m)}\equiv {\bf P}\,\exp\left\{{\rm i}g_s\int_0^{\infty}d\lambda\
  \beta_i\cdot A(\lambda\beta_i)\,e^{-m\lambda\sqrt{-\beta_i^2}}\right\},
\label{Wilsondef2}
\end{equation}
where the exponential damping is now built into the Wilson line which provides the
source for the soft gluon field. Thus instead of computing the webs with
ordinary Wilson lines, we compute them starting with a product of
exponentially-damped Wilson lines, all involving the same infrared scale $m$:
\begin{align}
\label{S_def_m}
{\cal S}\left(\gamma_{ij},\alpha_s(\mu,\epsilon), \epsilon, m\right)
&\equiv\left<0\left|\Phi_{\beta_1}^{(m)}\otimes\Phi_{\beta_2}^{(m)}\otimes\ldots\otimes\Phi_{\beta_L}^{(m)} \right|0\right>\,.
\end{align}
We emphasise that (\ref{S_def_m}) and (\ref{Wilsondef2}), in contrast to (\ref{S_ren_def})  and~(\ref{Wilsondef}), are not thought of as an effective physical description of the interactions of the hard partons with soft gluons. Rather the modified Wilson line (\ref{Wilsondef2}) is merely a technical tool that allows one to compute the UV singularities that comprise the $Z$ factor one seeks to determine. Feynman diagrams and webs corresponding to (\ref{S_def_m}) will acquire the $m$ dependence through the Feynman rule (\ref{FRmod}), but, as already emphasised in (\ref{S_reg_ren_UV_and_IR}) the $Z$ factor will not.

The chief advantages of this regulator are twofold. Firstly, there are no problems applying it at any loop order. Secondly, the regulator does not break the important symmetry of Wilson loop computations, namely invariance under rescaling:
\begin{equation}
\lambda\rightarrow \kappa \lambda,\quad \beta_i\rightarrow \beta_i/\kappa\,.
\end{equation}

\begin{figure}
\begin{center}
\scalebox{0.7}{\includegraphics{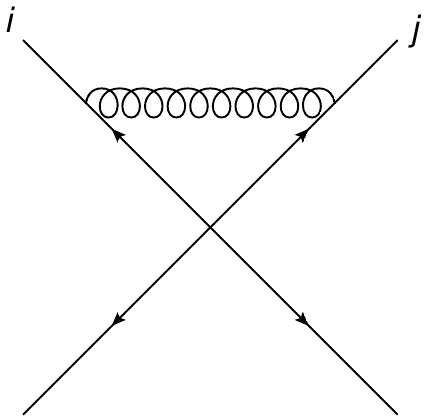}}
\caption{One loop web, where the gluon is emitted between partons $i$ and $j$,
whose kinematic part is given by eq.~(\ref{eq:Fijoneloop}).}
\label{1loopfig}
\end{center}
\end{figure}
As an illustration, and in order to introduce notation that will be used later on in multiloop diagrams, we consider here the calculation of the one-loop diagram in figure~\ref{1loopfig} with this regulator in place. 
We denote the one-loop diagram with a gluon exchanged between legs $i$ and $j$ by 
\begin{equation}
1_{ij}={\rm T}_{i}\cdot {\rm T}_{j}\,\,{\cal F}^{(1)}_{ij}(\gamma_{ij},\mu^2/m^2,\epsilon)\,,
\end{equation}
where we have explicitly written the colour factor. 
Using the eikonal Feynman rule of eq.~(\ref{FRmod}) as well as the configuration-space gluon propagator 
\begin{equation}
D_{\mu\nu}(x)=-{\cal N}\,g_{\mu\nu}\,
\left(-x^2\right)^{\epsilon-1}\,,\,\qquad \quad
{\cal N}\equiv
\frac{\Gamma(1-\epsilon)}{4\pi^{2-\epsilon}}\,,
\label{propdef}
\end{equation}
the kinematic factor takes the form: 
\begin{align}
  \label{eq:Fijoneloop}
  \begin{split}
  {\cal F}^{(1)}_{ij}(\gamma_{ij},\mu^2/m^2,\epsilon)
  &=\mu^{2\epsilon} g_s^2\,{\cal N}\,\beta_i\cdot\beta_j\,
\int_0^{\infty} ds \,\int_0^{\infty} dt\,
  \Big(-(s\beta_i-t\beta_j)^2\Big)^{\epsilon-1}
\,{\rm e}^{-m\left(s\sqrt{-\beta_i^2}+t\sqrt{-\beta_j^2}\right)}
  \\
  &=\left(\frac{\mu^2}{m^2}\right)^{\epsilon} \frac{g_s^2}{2}\,{\cal N}
  \,\Gamma(2\epsilon) \,\gamma_{ij}\int_0^1dx \, P(x,\gamma_{ij})\,\, 
  \\
  &=\left(\frac{\mu^2}{m^2}\right)^{\epsilon} \frac{g_s^2}{2}\,{\cal N}
  \,\Gamma(2\epsilon) \,\gamma_{ij}\, \,\,
  _2F_1\left([1,1-\epsilon],[3/2],\frac12+\frac{\gamma_{ij}}{4}\right)\,,
  \end{split}
\end{align}
where, for later convenience, we defined the propagator-related function,
\begin{equation}
P(x,\gamma_{ij})\equiv\Big[x^2+(1-x)^2-x(1-x)\gamma_{ij}\Big]^{\epsilon-1}\,.
\label{Pdef}
\end{equation} 
The kinematic dependence is expressed in terms of the cusp parameter $\gamma_{ij}$, defined in eq.~(\ref{gamma_ij}). A more elaborate example of a three-loop calculation is given in section~\ref{sec:F3D}.

In the following sections we shall make use of the leading term in the $\epsilon$ expansion of
eq.~\eqref{eq:Fijoneloop}, which we denote by
\begin{align}
  \label{eq:F1m1}
\begin{split}
  \mathcal{F}^{(1,-1)}(1_{ij}) &= \frac{g_s^2 \gamma_{ij}}{16 \pi^2 \epsilon}
  \int_0^1 {\rm d}x\ P^0(x,\gamma_{ij})\\
&=
\frac{g_s^2 }{16 \pi^2 \epsilon}
\Big[-2\,\xi_{ij}\,\coth(\xi_{ij})\Big]\,,\qquad \quad\text{where}\,\qquad \quad  \xi_{ij}=\cosh^{-1}(-\gamma_{ij}/2)
\,,
\end{split}
\end{align}
and in the first line $P^0(x,\gamma_{ij})$ is the $\epsilon\to 0$ limit of $P(x,\gamma_{ij})$ in  (\ref{Pdef}). This is the familiar result in the one-loop calculation of the cusp anomalous dimension (see e.g.~\cite{Korchemsky:1987wg}) where $\xi_{ij}$ is the Minkowski space cusp angle.

It is straightforward to expand the hypergeometric function in \eqref{eq:Fijoneloop} to subleading orders in $\epsilon$. Thanks to the IR regulator this expansion is under control, and yields a unique result. 

Having described the exponential regulator, we can now set about studying the singularity structure of specific three-loop webs. This is the subject of the following sections.

\subsection{The three-leg three-loop web $W_{(2,3,1)}^{(3)}$~\label{3lw}}

In this section we consider the three-loop web composed of the six diagrams in figure~\ref{3lsix}. This web was first studied in~\cite{Gardi:2010rn}, and has a number of interesting properties which make it useful as a testing ground for studying renormalization constraints. In particular, not all diagrams have a leading divergence: whereas diagrams ($3D$)--($3F$) are ${\cal O}(\epsilon^{-3})$, diagrams ($3A$)--($3C$) are both ${\cal O}(\epsilon^{-2})$, and diagram ($3B$) is ${\cal O}(\epsilon^{-1})$ i.e. has no subdivergences at all. The contribution of this web to the exponent of the eikonal scattering amplitude, including the relevant mixing matrix, is~\cite{Gardi:2010rn}
\begin{equation}
W_{(2,3,1)}^{(3)}=\left(\begin{array}{c}{\cal F}(3A)\\{\cal F}(3B)\\{\cal F}(3C)\\{\cal F}(3D)\\{\cal F}(3E)\\{\cal F}(3F)\end{array}\right)^T\frac{1}{6}\left(\begin{array}{rrrrrr}3&0&-3&-2&-2&4\\-3&6&-3&1&-2&1\\-3&0&3&4&-2&-2\\0&0&0&1&-2&1\\0&0&0&-2&4&-2\\0&0&0&1&-2&1\end{array}\right)\left(\begin{array}{c}C(3A)\\C(3B)\\C(3C)\\C(3D)\\C(3E)\\C(3F)\end{array}\right)\,.
\label{1-6mat}
\end{equation}
We now need to calculate the kinematic parts ${\cal F}(3A)$ etc. of each
diagram. After combining these as in eq.~(\ref{1-6mat}), we should find that the
leading divergence of the web cancels, and the remaining terms are such that
eq.~(\ref{eq:prize}) is verified, where the right-hand side of that equation
contains only those lower-order webs that combine to make the diagrams of
figure~\ref{3lsix}. This will then amount to the statement that the
renormalization of the exponent is consistent on a web-by-web basis, with no mixing
between different closed sets of diagrams.

All of the kinematic integrals we need to compute in this paper can be computed using the same general recipe. We thus outline this in detail for one specific example in the following subsection
and summarise the results for the other kinematic factors in appendix~\ref{app:Fresults}.
The details of this method will be useful in section~\ref{sec:factorization}.  

\subsubsection{Calculation of ${\cal F}(3D)$}
\label{sec:F3D}
\begin{figure}
\begin{center}
\scalebox{0.7}{\includegraphics{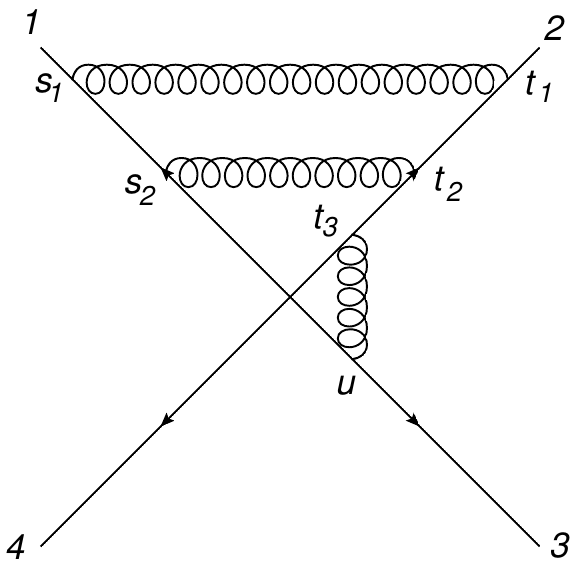}}
\caption{Diagram ($3D$) from figure~\ref{3lsix} showing the labels for gluon
  emissions used in eq.~(\ref{F3D1}).}
\label{3Dfig}
\end{center}
\end{figure}

In order to illustrate how to calculate the kinematic factors needed throughout the paper, we consider the explicit example of ($3D$) from figure~\ref{3lsix}, which has a leading (${\cal O}(\epsilon^{-3})$) divergence. Using the eikonal Feynman rule of eq.~(\ref{FRmod}) as well as the configuration space gluon propagator (\ref{propdef}) one finds
\begin{align}
{\cal F}(3D)&=g_s^6(\mu^2)^{3\epsilon}{\cal N}^3(\beta_1\cdot\beta_2)^2
(\beta_2\cdot\beta_3)
\int_0^\infty ds_1\int_0^\infty ds_2\int_0^\infty dt_1
\int_0^\infty dt_2\int_0^\infty dt_3\int_0^\infty du \notag\\
&\times e^{-m\left(\sqrt{-\beta_1^2}(s_1+s_2)+\sqrt{-\beta_2^2}(t_1+t_2+t_3)
+\sqrt{-\beta_3^2}u\right)}\notag\\
&\times
\Big[(s_1\beta_1-t_1\beta_2)^2(s_2\beta_1-t_2\beta_2)^2
(t_3\beta_2-u\beta_3)^2\Big]^{\epsilon-1}  \,
\Theta[s_1-s_2]\,\Theta[t_1-t_2]\,\Theta[t_2-t_3],
\label{F3D1}
\end{align}
where $\mu$ is the dimensional regularisation scale, and we have labelled the gluon emissions according to figure~\ref{3Dfig}. 
Here the variables $s_i$, $t_i$ and $u$ measure the distance along each Wilson
line at which gluons are emitted, and we use Heaviside functions to implement
the ordering constraints implied by the topology of the diagram. 
Equation~(\ref{F3D1}) can be made more symmetric by rescaling the distance
parameters according to $s_i\rightarrow s_i/\sqrt{-\beta_1^2}$ etc. to give
\begin{align}
{\cal F}(3D)&=\frac{g_s^6}{8}(\mu^2)^{3\epsilon}{\cal N}^3\gamma_{12}^2
\gamma_{23}\int_0^\infty ds_1\int_0^\infty ds_2\int_0^\infty dt_1
\int_0^\infty dt_2\int_0^\infty dt_3\int_0^\infty du \notag\\
&\times e^{-m(s_1+s_2+t_1+t_2+t_3)}
\Big[(s_1^2+t_1^2-s_1\,t_1\gamma_{12})(s_2^2+t_2^2-s_2\,t_2\gamma_{12})
(t_3^2+u^2-t_3\,u\gamma_{23})\Big]^{\epsilon-1}\notag\\
&\times\Theta[s_1-s_2]\,\Theta[t_1-t_2]\,\Theta[t_2-t_3],
\label{F3D2}
\end{align}
where we have used again the cusp parameters defined in eq.~(\ref{gamma_ij}). 

As a next step, one may transform according to 
\begin{equation}
\left(\begin{array}{c}s_1\\t_1\end{array}\right)=
\lambda\left(\begin{array}{c}x\\1-x\end{array}\right),\quad
\left(\begin{array}{c}s_2\\t_2\end{array}\right)=
\mu\left(\begin{array}{c}y\\1-y\end{array}\right),\quad
\left(\begin{array}{c}t_3\\u\end{array}\right)=
\alpha\left(\begin{array}{c}w\\1-w\end{array}\right),
\label{trans1}
\end{equation}
after which eq.~(\ref{F3D2}) becomes
\begin{align}
{\cal F}(3D)&=\frac{g_s^6}{8}(\mu^2)^{3\epsilon}{\cal N}^3\gamma_{12}^2
\gamma_{23}
\int_0^1 dx\int_0^1 dy\int_0^1 dw \,
P(x,\gamma_{12})P(y,\gamma_{12})P(w,\gamma_{23})
\notag\\
&\qquad\times 
\int_0^\infty d\lambda\int_0^\infty d\mu\int_0^\infty d\alpha\,\,
(\lambda\mu\alpha)^{2\epsilon-1}e^{-m(\lambda+\mu+\alpha)}
\notag\\
&\qquad\times\Theta[\lambda x-\mu y]\,\Theta[\lambda(1-x)-\mu(1-y)]\,
\Theta[\mu(1-y)-\alpha w],
\label{F3D3}
\end{align}
where $P(x,\gamma_{ij})$ is defined in eq.~(\ref{Pdef}).

One may proceed further using the additional transformations
\begin{equation}
\left(\begin{array}{c}\mu\\\lambda\end{array}\right)=
\sigma\left(\begin{array}{c}z\\1-z\end{array}\right),\quad
\left(\begin{array}{c}\alpha\\\sigma\end{array}\right)=
\beta\left(\begin{array}{c}r\\1-r\end{array}\right),
\label{trans2}
\end{equation}
so that eq.~(\ref{F3D3}) becomes
\begin{align}
\label{F3D4}
\begin{split}
{\cal F}(3D)&=\frac{g_s^6}{8}(\mu^2)^{3\epsilon}{\cal N}^3\gamma_{12}^2
\gamma_{23}\int_0^\infty d\beta\,\beta^{6\epsilon-1}\,e^{-m\beta}
\\
&\qquad\times
\int_0^1 dx\int_0^1 dy\int_0^1 dw \,P(x,\gamma_{12})P(y,\gamma_{12})P(w,\gamma_{23})
\\
&\qquad\times\int_0^1 dz [z(1-z)]^{2\epsilon-1}\int_0^1dr\,r^{2\epsilon-1}
(1-r)^{4\epsilon-1}
\\
&\qquad\times\Theta\left[\frac{x}{y}-\frac{z}{1-z}\right]
\Theta\left[\frac{(1-x)}{(1-y)}-\frac{z}{1-z}\right]
\Theta\left[\frac{z(1-y)}{w}-\frac{r}{1-r}\right].
\end{split}
\end{align}
The $\beta$ integral gives a singular gamma function $\Gamma(6\epsilon)$. This represents the
overall singularity obtained by simultaneously shrinking all gluons to the origin in 
figure~\ref{3Dfig}. The remaining integrals over $z$ and $r$ both give 
additional singularities, as expected from the fact that diagram ($3D$) has
subdivergences. One may carry out the $r$ integral by transforming to
$\rho=r/(1-r)$, and one finds
\begin{align}
\label{rint1}
\begin{split}
\int_0^1 dr\,r^{2\epsilon-1}\,(1-r)^{4\epsilon-1}
\,&\Theta\left[\frac{z(1-y)}{w}-\frac{r}{1-r}\right]
=\,\int_0^{z(1-y)/w} d\rho\,\rho^{2\epsilon-1}(1+\rho)^{-6\epsilon}
\\
&=\,\frac{1}{2\epsilon}\left(\frac{z(1-y)}{w}\right)^{2\epsilon}
\phantom{!}_2F_1\left(2\epsilon,6\epsilon;2\epsilon+1;-\frac{z(1-y)}{w}\right)
\\
&=\,\frac{1}{2\epsilon}\left(\frac{z(1-y)}{w}\right)^{2\epsilon}
\left[1+{\cal O}(\epsilon^2)\right],
\end{split}
\end{align}
where $\phantom{!}_2F_1(a,b;c;z)$ is the hypergeometric function. Note that,
for the purposes of this paper in which we wish to check the 
relation~(\ref{eq:prize}) for the first subleading pole, we may neglect higher
order $\epsilon$ terms from the hypergeometric function as shown in the last 
line of eq.~(\ref{rint1}). This leads to a drastic simplification.

The $z$ integral in eq.~(\ref{F3D4}) may be carried out in a similar fashion
after transforming to $\zeta= z/(1-z)$, and one ultimately finds
\begin{align}
      \mathcal{F}^{(3)}(3D)\ &=\ \mathcal{C}_3\
      \frac{\gamma_{12}^2\gamma_{23}}{8\epsilon^2} \int_0^1{\rm d}x{\rm d}y{\rm
        d}w\ \left( \frac{1-y}{w} \right)^{2\epsilon} \left( {\rm
        min}\left(\frac{x}{y},\frac{1-x}{1-y}\right) \right)^{4\epsilon}
\notag\\ &
    \qquad \qquad \times P(x,\gamma_{12})P(y,\gamma_{12})
    P(w,\gamma_{23})\,\,\left(1+{\cal O}(\epsilon^2)\right),
\label{F3Dres}
\end{align}
where ${\cal C}_3$ is the overall normalisation factor:
\begin{align}
  \label{eq:C3def}
  \mathcal{C}_3=\left(\frac{\mu^2}{m^2} \right)^{3\epsilon}
    \frac{g_s^6}8\ \frac{\Gamma(1-\epsilon)^3\Gamma(6\epsilon)}{(4\pi^{2-\epsilon})^3}.
\end{align}
In calculating ${\cal F}(3D)$, eq.~(\ref{F3Dres}) is already sufficient to
express the multiple poles of diagram ($3D$) in terms of poles of lower-order webs, which
is all that is needed to verify eq.~(\ref{eq:prize}). 

Before proceeding with this analysis, let us summarise the method we have used above, which can be generally applied in calculating any multi-eikonal diagram comprising of individual gluon exchanges. 
The general recipe for extracting ${\cal F}(D)$ up to and including the first subleading pole
for a general diagram $D$ belonging to this class is
\begin{enumerate}
\item Insert a factor $e^{-mt\sqrt{-\beta^2}}$ for all gluon emissions from an 
external line of 4-velocity~$\beta$, where $t$ is the appropriate distance 
parameter.
\item Represent the ordering of successive emissions along a given line by 
Heaviside functions.
\item Rescale all distance parameters $t\rightarrow t/\sqrt{-\beta^2}$. 
\item If $s_1$ and $s_2$ are parameters associated with the same gluon 
exchange, transform to the variables $s_1 = \lambda x$, 
$s_2 = \lambda(1-x)$.
Thus, for each separate subdiagram (here a single gluon exchange) we have a $\lambda$-type variable which measures the distance of this subdiagram from the hard interaction in units of the infrared regulator $1/m$.
\item Repeat a similar transformation, as in (\ref{trans2}), for each successive pair of 
$\lambda$-type variables, until only one remains. The integral over this
parameter will give an overall gamma function representing the overall UV divergence of the diagram.
At each stage in this process, one trades two dimensionful parameters representing the distance of two separate subdiagrams into one common dimensionful parameter and one dimensionless parameter $z$ measuring the relative distance. 
\item Eliminate the Heaviside functions by integrating over the $z$-type 
variables introduced in the previous step. This introduces 
hypergeometric functions, which can be expanded in $\epsilon$ in order to
extract the leading and first subleading pole of ${\cal F}(D)$. 
\end{enumerate}
Using the above prescription, we have calculated the leading and subleading 
poles of all of the diagrams which we need for the rest of the paper. These
results are collected in appendix~\ref{app:Fresults}. 
We shall further apply this technique in section \ref{sec:factorization} to derive a factorised expression for the leading pole of any diagram consisting exclusively of single gluon exchanges.

This method can also be extended to the next subleading order in epsilon
so as to compute the contribution of these webs to the soft anomalous dimension coefficient $\Gamma^{(3)}$ according to eq.~(\ref{Gamma_3}).
Note that this calculation is technically more involved owing to the presence of higher-order terms from the expansion of hypergeometric functions such as that of eq.~(\ref{rint1}).

\subsubsection{Expansion of Poles}
\label{sec:expansion-poles}

In the previous section we showed in detail how to calculate the kinematic
parts of all web diagrams encountered throughout the paper, collecting other
necessary results in appendix~\ref{app:Fresults}. In this section, we 
expand these expressions, and show that the multiple poles
of the three-loop web of figure~\ref{3lsix} can be expressed in terms of
sums of products of poles of lower-order webs. These results will then be
used in the following section to verify the relation~(\ref{eq:prize}).

We use a notation for the expansion of the $n$-loop kinematic factor
$\mathcal{F}^{(n)}$ which is analagous to eq.~(\ref{wnk-2}):
\begin{equation}
  \label{eq:Fnexp}
  {\cal F}^{(n)}(D) = \sum_{k=-n}^\infty \mathcal{F}^{(n,k)}(D)\ \epsilon^k. 
\end{equation}
Then from eq.~(\ref{eq:3loops}) one finds
\begin{equation}
{\cal F}^{(3,-3)}(3A)={\cal F}^{(3,-3)}(3B)={\cal F}^{(3,-3)}(3C)=0,
\label{F3ABC0}
\end{equation}
as expected from the fact that these diagrams are not maximally reducible. 
Furthermore one finds
\begin{align}
  \label{eq:lo3loop}
  \begin{split}
  &\mathcal{F}^{(3,-3)}(3D)=\mathcal{F}^{(3,-3)}(3E)=\mathcal{F}^{(3,-3)}(3F)\\
  &=\frac{g_s^6\gamma_{12}^2\gamma_{23}}{2^{13}\times3 \pi^6\epsilon^3}  \int_0^1
  {\rm d}x\ {\rm d}y\ {\rm d}w\ P^0(x,\gamma_{12}) P^0(y,\gamma_{12})P^0(w,\gamma_{23}),
  \end{split}
\end{align}
where, as above, $P^0(x,\gamma_{ij})$ is shorthand for $P(x,\gamma_{ij})$ evaluated at
$\epsilon=0$.

Comparing this with the leading pole for the one-loop diagram of
figure~\ref{1loopfig}, where the gluon connects lines $i$ and $j$,
eq.~(\ref{eq:F1m1}), we may rewrite eq.~(\ref{eq:lo3loop}) as
\begin{align}
  \label{eq:loresult}
  \mathcal{F}^{(3,-3)}(3D) = \mathcal{F}^{(3,-3)}(3E) = \mathcal{F}^{(3,-3)}(3F)
  = \frac16\ \left(\mathcal{F}^{(1,-1)}(1_{12})\right)^2\
  \mathcal{F}^{(1,-1)}(1_{23}).
\end{align}
This result exemplifies the fact that the leading poles of maximally-reducible diagrams can be expressed as products of poles of their irreducible subdiagrams. 
Furthermore, we observe that in this particular case each of the diagrams $(3D)$--$(3F)$ has the same numerical coefficient in front of the product of one-loop subgraphs. We will see the reason for this in section~\ref{sec:factorization}. 

\begin{figure}
\begin{center}
\scalebox{0.8}{\includegraphics{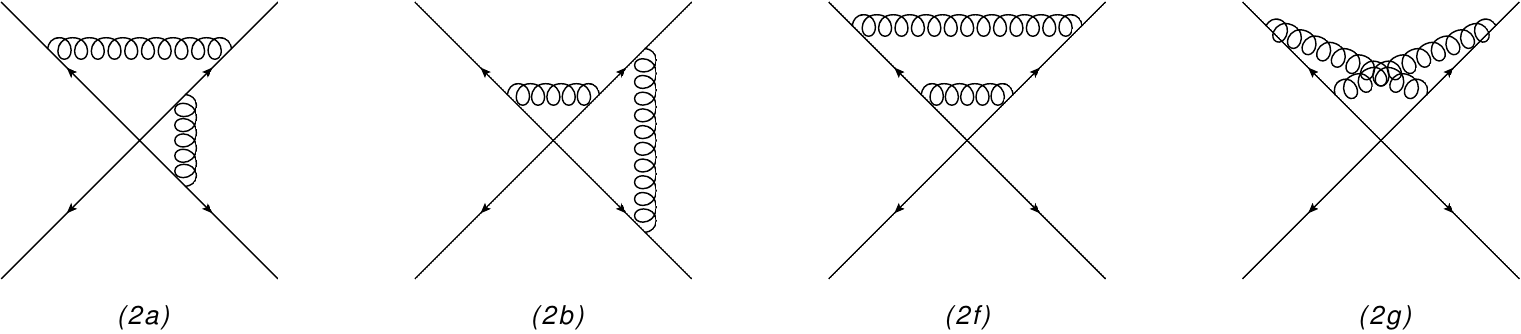}}
\caption{The two-loop webs which contribute to $W_{(2,3,1)}$, labelled as
  in~\cite{Gardi:2010rn}.}
\label{2six}
\end{center}
\end{figure}
Analogous results are obtained for the two-loop results in this web shown in
figure~\ref{2six} and labelled as in~\cite{Gardi:2010rn}.  The full results are
given in appendix~\ref{app:Fresults}.  Their leading pole is
at ${\cal O}(\epsilon^{-2})$ and can be written in terms of the poles of one-loop diagrams
as follows:
\begin{subequations}
\label{eq:LP2loop}
\begin{align}  
  \mathcal{F}^{(2,-2)}(2a) &= \mathcal{F}^{(2,-2)}(2b) = \frac12\
  \mathcal{F}^{(1,-1)}_{12} \mathcal{F}^{(1,-1)}_{23},
\\
  \mathcal{F}^{(2,-2)}(2f) &= \frac12 \left(\mathcal{F}^{(1,-1)}_{12}\right)^2\,,
\qquad\quad  \mathcal{F}^{(2,-2)}(2g) = 0\,.
\end{align}
\end{subequations}
Note that diagram ($2g$) has zero contribution at ${\cal O}(\epsilon^{-2})$ as we expect from
irreducibility. Again we see a factorisation by which higher-order poles can
be written in terms of lower-order web diagrams, where at leading pole a
unique product occurs on the right-hand side. By comparing 
eq.~(\ref{eq:LP2loop}) with eq.~(\ref{eq:loresult}), one sees that the 
coefficient of the product of lower-order web poles appears to be given by 
$1/n!$, where $n$ is the number of single-gluon subdiagrams. We return to this 
point in section~\ref{sec:factorization}.

Having examined the leading divergence, we now consider the web of 
figure~\ref{3lsix} at the first subleading pole  
$\sim{\cal O}(\epsilon^{-2})$. Looking at the kinematic
factors in eq.~(\ref{eq:3loops}), the next-to-leading pole contributions can be 
divided into three distinct categories, arising from
\begin{enumerate}
\item the expansion of $\mathcal{C}_3$;
\item the expansion of the propagator functions $P(x,\gamma_{ij})$; 
\item the expansion of the $a^{b\epsilon}$ terms in the integrand, where $a$ is a 
function of $\{x,y,z,w\}$.
\end{enumerate}
The first two of these are identical for the diagrams in the web, and only
contribute at ${\cal O}(\epsilon^{-2})$ for the diagrams which have non-zero
$\epsilon^{-3}$ poles. They will then be proportional to~$\epsilon$ times 
that leading pole contribution in each case.  The observed cancellation of the 
$\epsilon^{-3}$ poles therefore means that the contributions from (1) and (2) 
above will also cancel when the diagrams are combined and we therefore do not 
consider them further here.  

We instead focus on the contributions from (3) and write these in terms of
$P^0(x,\gamma_{ij})$ and the residue of the pole in $\mathcal{C}_3$,
\begin{align}
  \label{eq:C3minus1}
  \mathcal{C}_3^{(-1)}\ = \ \frac{g_s^6}{2^{10}\times 3\pi^6}.
\end{align}
The results are:
\begin{subequations}
\label{eq:3loops2}
\begin{align}
\begin{split}  
    \mathcal{F}^{(3,-2)}(3A)\ &=\ \mathcal{C}_3^{(-1)}\
    \frac{\gamma_{12}^2\gamma_{23}}{2\epsilon^2} \int_0^1{\rm d}x{\rm d}y{\rm
      d}w\ \left( \log\left(\frac{x}{y}\right) - \log\left( 
        \frac{1-x}{1-y}\right) \right) \\ & \qquad \qquad \times \Theta(x-y)
      \ P^0(x,\gamma_{12})P^0(y,\gamma_{12}) P^0(w,\gamma_{23});
\end{split}
 \\ 
    \mathcal{F}^{(3,-2)}(3B)\ &=\ 0; \\
\begin{split}
      \mathcal{F}^{(3,-2)}(3C)\ &=\ \mathcal{C}_3^{(-1)}\
      \frac{\gamma_{12}^2\gamma_{23}}{4\epsilon^2} \int_0^1{\rm d}x{\rm d}y{\rm
        d}w\ \left(
        \log\left(\frac{x}{y}\right) - \log\left(\frac{1-x}{1-y}\right) \right)
      \\ & \qquad \qquad \times \Theta(x-y)\
      P^0(x,\gamma_{12})P^0(y,\gamma_{12}) P^0(w,\gamma_{23}); 
\\&
\end{split}
\\
\begin{split}
      \mathcal{F}^{(3,-2)}(3D)\ &=\ \mathcal{C}_3^{(-1)}\
      \frac{\gamma_{12}^2\gamma_{23}}{4\epsilon^2} \int_0^1{\rm d}x{\rm d}y{\rm
        d}w\ \left( \log\left( \frac{1-y}{w} \right)
+2\log\left( {\rm
        min}\left(\frac{x}{y},\frac{1-x}{1-y}\right) \right) \right)\\ &
    \qquad \qquad \times P^0(x,\gamma_{12})P^0(y,\gamma_{12})
    P^0(w,\gamma_{23}); 
\end{split}
\\
\begin{split}
    \mathcal{F}^{(3,-2)}(3E)\ &=\ \mathcal{C}_3^{(-1)}\
    \frac{\gamma_{12}^2\gamma_{23}}{4\epsilon^2} \int_0^1{\rm d}x{\rm d}y{\rm
      d}w\ \left( 2\log\left( \frac{1-x}{w} \right) -\log\left( \frac{1-y}{w}
      \right)\right)\\ &
    \qquad \qquad \times P^0(x,\gamma_{12})P^0(y,\gamma_{12}) P^0(w,\gamma_{23}); 
\end{split}
 \\
\begin{split}
    \mathcal{F}^{(3,-2)}(3F)\ &=\ \mathcal{C}_3^{(-1)}\
    \frac{\gamma_{12}^2\gamma_{23}}{4\epsilon^2} \int_0^1{\rm d}x{\rm d}y{\rm
      d}w\ \left( 2\log\left( \frac{w}{1-x} \right) \right. \\ & \hspace{5.5cm}
    \left. + \log\left( {\rm
        min}\left(\frac{x}{y},\frac{1-x}{1-y}\right) \right) \right)\\ &
    \qquad \qquad \times P^0(x,\gamma_{12})P^0(y,\gamma_{12}) P^0(w,\gamma_{23}).
\end{split}
\end{align}
\end{subequations}
We expect the contribution from ($3B$) at this order to be zero given that the 
diagram is completely irreducible, thus has only a single epsilon pole.  
As was the case at leading pole, one may write the above expressions 
in terms of poles of lower-order diagrams i.e. in terms of the relevant
$\mathcal{F}^{(1,-1)}(1_{ij})$ and $\mathcal{F}^{(2,-1)}(2n)$
contributions. To this end we first consider the subleading poles of the
two-loop webs, whose leading poles have already been written as products
of one-loop leading poles in eq.~\eqref{eq:LP2loop}.  Again at two-loops, the
contributions at next-to-leading pole from the expansion of 
$\mathcal{C}_2$ and $P(x,\gamma_{ij})$
cancel among themselves, as was discussed at three-loops.  We therefore
concentrate here on the type-(3) contributions above, written in terms of
$P^0(x,\gamma_{ij})$ and the residue of the $\epsilon^{-1}$ pole in
$\mathcal{C}_2$
\begin{align}
  \label{eq:C2minus1}
  \mathcal{C}_2^{(-1)}\ = \ \frac{g_s^4}{2^8 \pi^4}.
\end{align}
The two-loop subleading poles are then
\begin{subequations}
  \label{eq:F2NLP}
\begin{align}
        \mathcal{F}^{(2,-1)}(2a)\ &=\ \mathcal{C}_2^{(-1)}\
    \frac{\gamma_{12} \gamma_{23}}{\epsilon}\int_0^1 {\rm d}x {\rm d}w\
    \log\left(\frac{1-x}w\right) P^0(x,\gamma_{12})P^0(w,\gamma_{23}) \label{eq:F2NLP_2a}\\
    \mathcal{F}^{(2,-1)}(2b)\ &=\ \mathcal{C}_2^{(-1)}\
    \frac{\gamma_{12} \gamma_{23}}{\epsilon}\int_0^1 {\rm d}x {\rm d}w\
    \log\left(\frac{w}{1-x}\right) P^0(x,\gamma_{12})P^0(w,\gamma_{23})
\label{eq:F2NLP_2b}\\
    \mathcal{F}^{(2,-1)}(2f)\ &=\ \mathcal{C}_2^{(-1)}\
    \frac{\gamma_{12}^2}{\epsilon}\int_0^1 {\rm d}x {\rm d}y\
    \ \log\left( {\rm
        min}\left(\frac{x}{y},\frac{1-x}{1-y}\right)\right)\
    P^0(x,\gamma_{12})P^0(y,\gamma_{12})
\label{eq:F2NLP_2f}
\\
\begin{split}
\label{eq:F2NLP_2g}
    \mathcal{F}^{(2,-1)}(2g)\, &=\, \mathcal{C}_2^{(-1)}\ \frac{\gamma_{12}^2}{\epsilon}
    \int_0^1 {\rm d}x {\rm d}y\, \, \left( \log\left(\frac{x}{y}\right) - \log\left(\frac{1-x}{1-y}\right) \right) 
\\ & \hspace{4cm} \times \Theta(x-y)  P^0(x,\gamma_{12})P^0(y \gamma_{12}).
\end{split}
\end{align}
\end{subequations}
By inspection, $\mathcal{F}^{(2,-1)}(2a)=-\mathcal{F}^{(2,-1)}(2b)$ and by
explicit calculation, one can also show that
$\mathcal{F}^{(2,-1)}(2f)=-\mathcal{F}^{(2,-1)}(2g)$.

\begin{figure}[tb]
\begin{center}
\scalebox{.5}{\includegraphics{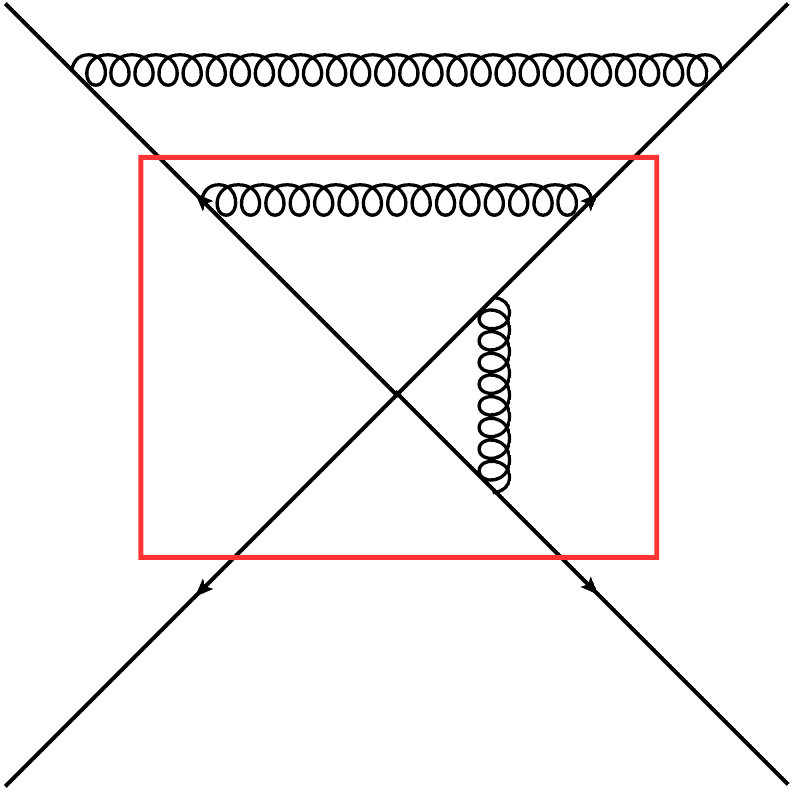}}
\hspace*{50pt}
\scalebox{.5}{\includegraphics{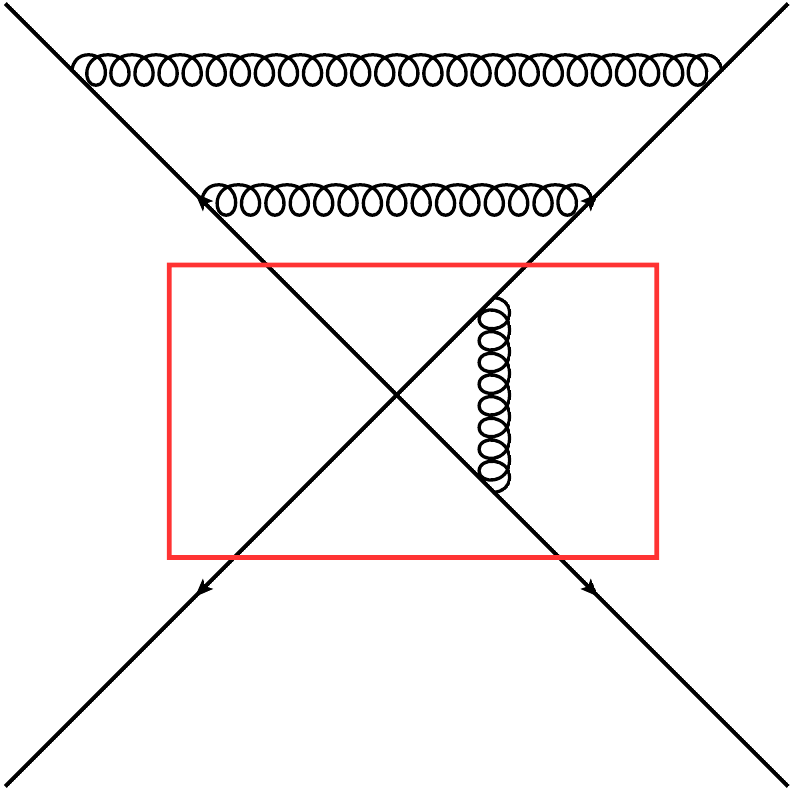}}
\caption{The two possible decompositions of diagram $3D$ into a product of a two-loop subdiagram times a one-loop one, corresponding to the two terms in (\ref{3D_decomp}). Note that the latter equation is expressed in terms of $(2g)$, while the actual subdiagram is $(2f)$. However, the two are related by $\mathcal{F}^{(2,-1)}(2f)=-\mathcal{F}^{(2,-1)}(2g)$. }
\label{3Ddecomposition}
\end{center}
\end{figure}

Using these results, we can express the three-loop results at this 
order in terms of diagrams with fewer loops as follows:
\begin{subequations}
  \label{eq:master3NLP}
\begin{align}
  \mathcal{F}^{(3,-2)}(3A)\ &=\ \frac23 \mathcal{F}^{(2,-1)}(2g)\
  \mathcal{F}^{(1,-1)}(1_{23})\\
  \mathcal{F}^{(3,-2)}(3B)\ &=\ 0 \\
  \mathcal{F}^{(3,-2)}(3C)\ &=\ \frac13 \mathcal{F}^{(2,-1)}(2g)\
  \mathcal{F}^{(1,-1)}(1_{23})\\
  \mathcal{F}^{(3,-2)}(3D)\ &=\ \frac13
  \left(\mathcal{F}^{(2,-1)}(2a)\mathcal{F}^{(1,-1)}(1_{12})\ - \
    2\mathcal{F}^{(2,-1)}(2g)\mathcal{F}^{(1,-1)}(1_{23}) \right) \label{3D_decomp}\\
  \mathcal{F}^{(3,-2)}(3E)\ &=\ \frac13 
    \mathcal{F}^{(2,-1)}(2a)
  \mathcal{F}^{(1,-1)}(1_{12})\\
  \mathcal{F}^{(3,-2)}(3F)\ &=\ \frac13
  \left(-2\mathcal{F}^{(2,-1)}(2a)\mathcal{F}^{(1,-1)}(1_{12})\ - \
    \mathcal{F}^{(2,-1)}(2g)\mathcal{F}^{(1,-1)}(1_{23}) \right).
\end{align}
\end{subequations}
This confirms the statement made above that poles of higher loop webs can
be written as sums of products of poles of lower-order webs. Unlike the 
case of the leading pole, however, there is in general more than one permissible
product on the right-hand side of each equation corresponding to the
different ways of decomposing each web diagram into subdiagrams. This decomposition is illustrated in figure~\ref{3Ddecomposition} for the case of diagram $3D$.
Furthermore, the coefficients in front of each product of lower-order
diagrams are not simple factorials, as they were in the leading pole case. 
Armed with eq.~(\ref{eq:master3NLP}), we now have everything necessary to 
verify the relation~(\ref{eq:prize}) for this particular three-loop web.

\subsubsection{Verifying the renormalization constraint}
\label{sec:verify}
In the previous section, we used the results collected in 
appendix~\ref{app:Fresults} to decompose the higher-order poles of the web
diagrams in figure~\ref{3lsix} in terms of pole coefficients of lower-order
webs. In this section, we use these results to verify the renormalization
constraint expressed in eq.~(\ref{eq:prize}). 

To proceed, one may first expand eq.~(\ref{1-6mat}) and collect coefficients
of each kinematic factor to get
\begin{align}
  \label{eq:colourcombs}
  \begin{split}
    W_{(2,3,1)}^{(3)}&=\frac16 \big[ 3C(3A)-3C(3C)-2C(3C)-2C(3E)+4C(3F) \big]
    \mathcal{F}(3A) \\
    &+ \frac16 \big[ -3C(3A)+6C(3B)-3C(3C)+C(3D)-2C(3E)+C(3F)\big]
    \mathcal{F}(3B) \\
    &+ \frac16 \big[-3C(3A)+3C(3C)+4C(3D)-2C(3E)-2C(3F) \big] \mathcal{F}(3C)
    \\
    &+ \frac16 \big[C(3D)-2C(3E)+C(3F) \big] \mathcal{F}(3D) \\
    &+ \frac13 \big[-C(3D)+2C(3E)-C(3F) \big] \mathcal{F}(3E) \\ 
    &+ \frac16 \big[ C(3D)-2C(3E)+C(3F) \big] \mathcal{F}(3F).
  \end{split}
\end{align}
It can be seen immediately after substituting the results of 
eqs.~(\ref{F3ABC0}, \ref{eq:loresult}) that the leading poles do
indeed cancel to give zero contribution at $\mathcal{O}(\epsilon^{-3})$. That
is, this cancellation is purely a consequence of the properties of web mixing
matrices, and does not depend on additional commutators of lower-order webs, which only start to contribute at next-to-leading pole.

We now consider the next-to-leading pole contributions and find by substituting
eq.~\eqref{eq:master3NLP} into eq.~\eqref{eq:colourcombs}:
\begin{align}
  \label{eq:Wnlp}
  \begin{split}
    W_{(2,3,1)}^{(3,-2)}=&-\frac16 \Big( C(3D)-2C(3E)+C(3F) \Big) \mathcal{F}^{(2,-1)}(2a)
    \mathcal{F}^{(1,-1)}(1_{12}) \\
    & \ +\frac16 \Big( C(3A)-C(3C)-C(3D)+C(3F) \Big) \mathcal{F}^{(2,-1)}(2g)
    \mathcal{F}^{(1,-1)}(1_{23}).
  \end{split}
\end{align}
Whilst many cancellations have taken place, a non-zero contribution
remains which we expect from the renormalisation arguments of
section~\ref{sec:renormalization}.

\begin{figure}
\begin{center}
\scalebox{0.7}{\includegraphics{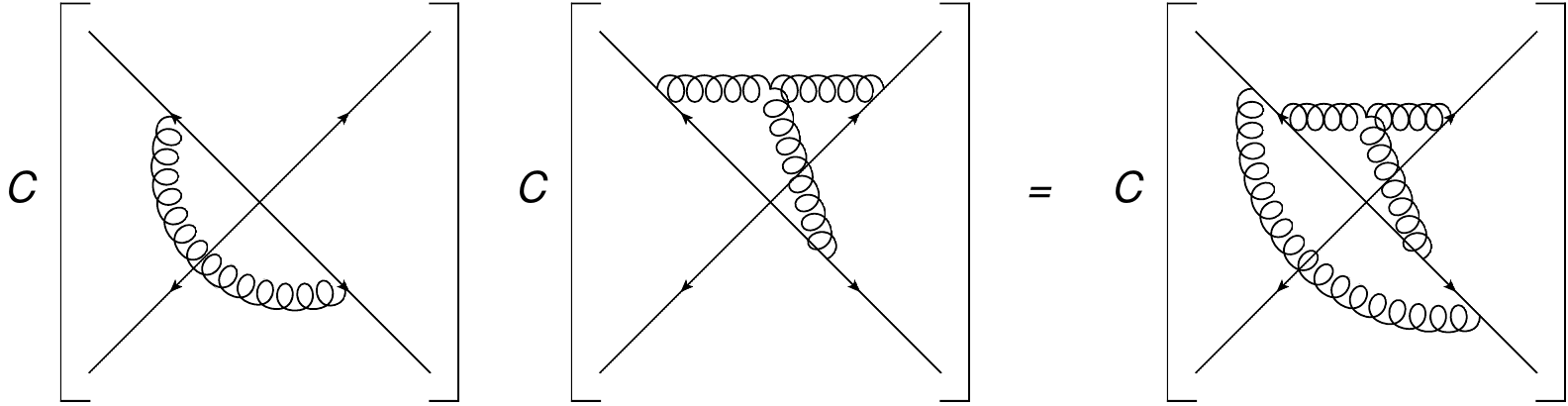}}
\caption{Illustration of the multiplication rule for colour factors of 
multiparton diagrams, as given in eq.~(\ref{colrule}).}
\label{colrulefig}
\end{center}
\end{figure}

We may simplify eq.~(\ref{eq:Wnlp}) by using the multiplication rule
\begin{equation}
C(AB)=C(A)C(B)
\label{colrule}
\end{equation}
for colour factors of diagrams, where $AB$ is the diagram obtained by ordering
subdiagram $B$, then $A$, outwards from the hard interaction (see e.g. 
figure~\ref{colrulefig}). 
This allows us to write the colour factors for the three
loop web diagrams in terms of the colour factors of lower-order diagrams.
In particular, for the combinations of colour factors appearing in (\ref{eq:Wnlp}), one has
\begin{subequations}
\begin{align}
  \label{eq:colourfacts}
  \begin{split}
    C(3D)-2C(3E)+C(3F) &= \Big(C(3D)-C(3E)\Big)\ - \ \Big(C(3E)-C(3F)\Big) \\
    &= C(1_{12}) \Big(C(2a)-C(2b)\Big) - \Big(C(2a)-C(2b)\Big)C(1_{12}) \\
    &= \Big[C(1_{12}),\,C(2a)-C(2b)\Big]  
\end{split}
\\
\begin{split}
    C(3A)-C(3C)-C(3D)+C(3F) &= C(2g)C(1_{23}) - C(1_{23})C(2g) \\ & \qquad
    \qquad -  C(2f)C(1_{23}) + C(1_{23})C(2f) \\
    &=\Big[C(1_{23}),C(2f)-C(2g)\Big]\,.  
\end{split}
\end{align}
\end{subequations}
We now have the pieces required to independently evaluate
the left- and right-hand side of eq.~\eqref{eq:prize}:
\begin{align}
  \left.w^{(3,-2)}\right\vert_{b_n= 0}&=
  \,\,
  -\frac{1}{6}\left[w^{(1,-1)},w^{(2,-1)}
  \right]
\end{align}
for this set of diagrams.  We use $W_{(2,3,1)}$ in place of $w$ when we
only consider the contributions corresponding to this particular subset.  The
left-hand side comes directly from eq.~\eqref{eq:Wnlp}
\begin{align}
  \label{eq:W3LHS}
  \begin{split}
  W^{(3,-2)}_{(2,3,1)} &=
  -\frac16 \Big[C(1_{12}),C(2a)-C(2b)\Big] \mathcal{F}^{(2,-1)}(2a)
    \mathcal{F}^{(1,-1)}(1_{12}) \\
    & \qquad + \frac16 \Big[C(1_{23}),C(2f)-C(2g)\Big] \mathcal{F}^{(2,-1)}(2g)
    \mathcal{F}^{(1,-1)}(1_{23}).
  \end{split}
\end{align}
We now turn to the right-hand side.  At one-loop, there are two diagrams which
will contribute to this specific three-loop web:
\begin{align}
  \label{eq:w1full}
  \begin{split}
  w^{(1,-1)}\ &=\ C(1_{12})
  \mathcal{F}^{(1,-1)}(1_{12})\ +\ C(1_{23})
  \mathcal{F}^{(1,-1)}(1_{23})\,+\,\ldots\,,
  \end{split}
\end{align}
where here and in subsequent equations the ellipsis stands for those additional webs which do not eventually contribute to the three-loop web of interest. 
At two-loops, the four diagrams in figure~\ref{2six} are relevant. For each of these
the corresponding web is obtained by combining the kinematic factor with the corresponding exponentiated colour factors (see e.g.~eq.~(5.65) in \cite{Gardi:2010rn}) as follows:
\begin{align}
  \label{eq:w2exp}
  \begin{split}
  w^{(2,-1)}\ &=
  \frac12 \Big(C(2a)-C(2b)\Big)\Big(\mathcal{F}^{(2,-1)}(2a)
-\mathcal{F}^{(2,-1)}(2b)\Big) \\
&  \qquad+
  \Big(C(2g)-C(2f)\Big)\mathcal{F}^{(2,-1)}(2g)\,+\,\ldots\,,
\\
&=
  \Big(C(2a)-C(2b)\Big)\mathcal{F}^{(2,-1)}(2a)   +
  \Big(C(2g)-C(2f)\Big)\mathcal{F}^{(2,-1)}(2g)\,+\,\ldots\,,
  \end{split}
\end{align}
where in the second line we have used $\mathcal{F}^{(2,-1)}(2b)=-\mathcal{F}^{(2,-1)}(2a)$ as follows from eqs.~(\ref{eq:F2NLP_2a})
and~(\ref{eq:F2NLP_2b}). 
Now substituting these into Eq.~\eqref{eq:prize} and restricting to
contributions corresponding to the $W_{(2,3,1)}$ configuration gives
\begin{align}
  \label{eq:finalrhs}
  \begin{split}
  W^{(3,-2)}_{(2,3,1)} &=
  -\frac16 \Big[C(1_{12}),\left( C(2a)-C(2b)\right) \Big]
  \mathcal{F}^{(1,-1)}(1_{12}) \mathcal{F}^{(2,-1)}(2a) \\ 
  & \quad -  \frac16 \Big[ C(1_{23}), \left( C(2g)- C(2f) \right) \Big]
  \mathcal{F}^{(1,-1)}(1_{23}) \mathcal{F}^{(2,-1)}(2g).
  \end{split}
\end{align}
This exactly matches eq.~\eqref{eq:W3LHS}, and so for this web, the 
constraints from renormalization have been shown to be satisfied. Note, in
particular, that they have been satisfied for this web by itself, implying
no mixing with other webs. This feature is important in that it shows that
the notion of individual webs as closed sets of diagrams, first proposed
in~\cite{Gardi:2010rn}, survives after renormalization of the multi-eikonal
vertex.

In the following section, we illustrate the validity of eq.~(\ref{eq:prize})
using a second three-loop example, this time for a web that involves four eikonal lines, $W_{(1,2,2,1)}^{(3)}$.

\subsection{The four-leg three-loop web $W_{(1,2,2,1)}^{(3)}$~\label{3lwprime}}

In the previous sections, we have analysed the web of figure~\ref{3lsix} in
detail, showing how to extract the leading and subleading poles in explicit
calculations. We saw that multiple poles of web diagrams can be decomposed
into sums of products of poles of lower-order web diagrams. 
Furthermore, we explicitly verified that the leading poles, ${\cal O}(\epsilon^{-3})$, cancel through the action of the web mixing matrix, while the next-to-leading pole, ${\cal O}(\epsilon^{-2})$
is consistent with the renormalization constraint of eq.~(\ref{eq:prize}). 

In this section, we repeat this analysis for a second example, this time considering a three-loop web involving four eikonal lines,  $W_{(1,2,2,1)}^{(3)}$.
This will provide more evidence that the renormalization constraints of section~\ref{sec:renormalization} are indeed realised on a web-by-web basis. 
Given that most of the techniques and results have already been shown in the previous 
section (and in appendix~\ref{app:Fresults}), we will be much briefer here.

\begin{figure}[b]
\begin{center}
\scalebox{0.9}{\includegraphics{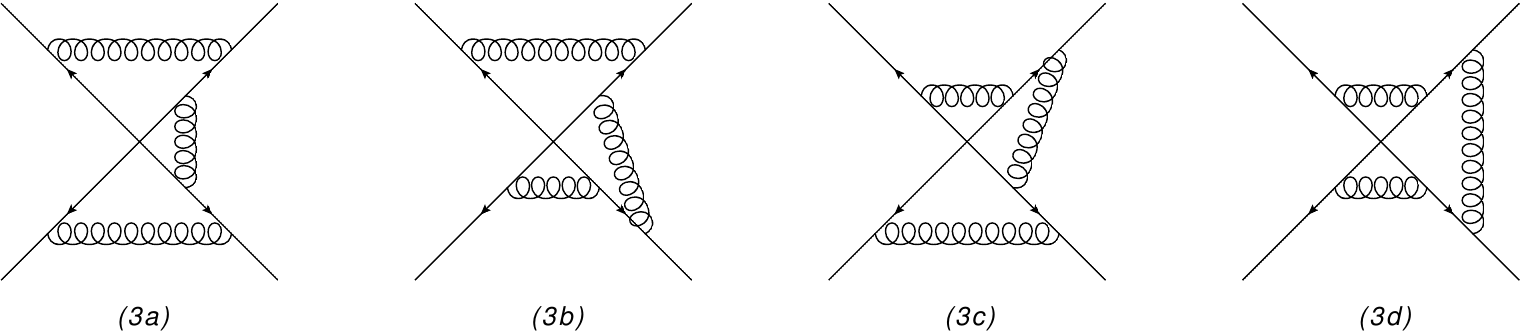}}
\caption{The four 3-loop diagrams forming the 1-2-2-1 web in which four eikonal lines are linked by three gluon exchanges, with labels as 
in~\cite{Gardi:2010rn}.}
\label{3lfour}
\end{center}
\end{figure}

The web $W_{(1,2,2,1)}^{(3)}$ is composed of four diagrams, depicted in figure~\ref{3lfour}.
This web was first  studied in~\cite{Gardi:2010rn}, where its contribution to the exponent of the eikonal scattering amplitude, including the relevant web mixing matrix, was shown to be:
\begin{equation}
W_{(1,2,2,1)}^{(3)}=\left(\begin{array}{c}{\cal F}(3a)\\{\cal F}(3b)\\{\cal F}(3c)
\\{\cal F}(3d)\end{array}\right)^T\frac{1}{6}\left(\begin{array}{rrrr}
1&-1&-1&1\\-2&2&2&-2\\-2&2&2&-2\\1&-1&-1&1
\end{array}\right)\left(\begin{array}{c}C(3a)\\C(3b)\\C(3c)\\C(3d)
\end{array}\right)\,.
\label{3lfourmat}
\end{equation}
The kinematic factors for each diagram are given in 
appendix~\ref{app:Fresults}, and the leading poles are
\begin{align}
  \label{eq:lp2}
  \begin{split}
    \mathcal{F}^{(3,-3)}&(3a)=2 \mathcal{F}^{(3,-3)}(3b) = 2
    \mathcal{F}^{(3,-3)}(3c) = \mathcal{F}^{(3,-3)}(3d) \\ 
    &= \frac13 \mathcal{F}^{(1,-1)}(1_{12}) \mathcal{F}^{(1,-1)}(1_{23})
    \mathcal{F}^{(1,-1)}(1_{34}).
  \end{split}
\end{align}
The web mixing matrix tells us that these only appear in the combination
\[
\mathcal{F}(3a)-2\mathcal{F}(3b)-2\mathcal{F}(3c)+\mathcal{F}(3d)
\] 
and therefore the leading poles automatically cancel among themselves, as they
must.  As in the previous case, this means that the contributions at
${\cal O}(\epsilon^{-2})$ from pieces common to all diagrams, e.g.~the $\Gamma(6\epsilon)$
and the $P$-functions will also cancel between themselves.  We therefore do not
consider them further here. The remaining contributions to the subleading pole are given by
\begin{subequations}
\label{eq:slp2}
\begin{align}  
  \begin{split}\label{eq:slp2_a}
    \mathcal{F}^{(3,-2)}(3a) &=\widetilde C_3
    \frac{\Gamma(6\epsilon)}{4\epsilon}\int_0^1 {\rm 
      d}x{\rm d} y {\rm d}z\ P^0(x,\gamma_{12}) P^0(y,\gamma_{23})
    P^0(z,\gamma_{34}) \ \\
    & \hspace{1.5cm} \times\left(\log(1-x) - \log(y) -\log(1-y) + \log(z) \right) 
\end{split} \\
\begin{split}\label{eq:slp2_b}
     \mathcal{F}^{(3,-2)}(3b) &=\widetilde C_3
     \frac{\Gamma(6\epsilon)}{4\epsilon}\int_0^1 {\rm 
      d}x{\rm d} y {\rm d}z\ P^0(x,\gamma_{12}) P^0(y,\gamma_{23}) P^0(z,\gamma_{34})\\
    & \hspace{1.5cm} \times \left(2\log(x)-2\log(y) +\log(1-y) - \log(1-z) \right) 
\end{split} \\
\begin{split}\label{eq:slp2_c}
     \mathcal{F}^{(3,-2)}(3c) &=\widetilde C_3
     \frac{\Gamma(6\epsilon)}{4\epsilon}\int_0^1 {\rm 
      d}x{\rm d} y {\rm d}z\ P^0(x,\gamma_{12}) P^0(y,\gamma_{23}) P^0(z,\gamma_{34})\\
    & \hspace{1.5cm} \times \left(2 \log(z)-2\log(1-y) +\log(y) - \log(1-x) \right)
\end{split} \\
\begin{split}\label{eq:slp2_d}
     \mathcal{F}^{(3,-2)}(3d) &=\widetilde C_3
     \frac{\Gamma(6\epsilon)}{2\epsilon}\int_0^1 {\rm 
      d}x{\rm d} y {\rm d}z\ P^0(x,\gamma_{12}) P^0(y,\gamma_{23}) P^0(z,\gamma_{34})\\
    & \hspace{1.5cm} \times \left(\log(1-y) - \log(z) +\log(y)-\log(1-x) \right),
  \end{split}
\end{align}
\end{subequations}
which can again be decomposed in terms of the poles of lower-order diagrams.

The relevant two-loop diagrams are ($2a$) and ($2b$) from eq.~\eqref{eq:F2NLP} and their mirror images about the horizontal.  We introduce the new labelling $2_{ijk}$ where $i,j,k$ label the ordering of the legs to which gluons are attached, considered from outside in.  So ($2a$) becomes $2_{123}$, ($2b$) is $2_{321}$, and their mirror images are respectively $2_{432}$ and $2_{234}$.  The correspondence between $x,y,z$ and $\gamma_{ij}$ can be seen
from the arguments of the $P$-functions. This notation introduces some ambiguity since by 
eqs.~(\ref{eq:F2NLP_2a}) and~(\ref{eq:F2NLP_2b}) 
\begin{equation}
\label{2_ijk_kij}
{\cal F}^{(2,-1)}(2_{ijk})=-{\cal F}^{(2,-1)}(2_{kji})
\end{equation}
and so we restrict ourselves to $2_{123}$ and $2_{432}$ here.  We find
\begin{subequations}
\label{eq:F32inF21-2}
\begin{align}  
    \mathcal{F}^{(3,-2)}(3a) &= 
+\frac13\,    \mathcal{F}^{(2,-1)}(2_{123})\mathcal{F}^{(1,-1)}(1_{34}) 
+ \frac13\, \mathcal{F}^{(2,-1)}(2_{432}) \mathcal{F}^{(1,-1)}(1_{12}) \\
    \mathcal{F}^{(3,-2)}(3b) &= 
+\frac23 \,   \mathcal{F}^{(2,-1)}(2_{123})\mathcal{F}^{(1,-1)}(1_{34}) 
- \frac13\, \mathcal{F}^{(2,-1)}(2_{432}) \mathcal{F}^{(1,-1)}(1_{12}) \\
    \mathcal{F}^{(3,-2)}(3c) &= 
 -\frac13\, \mathcal{F}^{(2,-1)}(2_{123}) \mathcal{F}^{(1,-1)}(1_{34})
+\frac23\,  \mathcal{F}^{(2,-1)}(2_{432})\mathcal{F}^{(1,-1)}(1_{12}) 
 \\
    \mathcal{F}^{(3,-2)}(3d) &= 
-\frac23\, \mathcal{F}^{(2,-1)}(2_{123}) \mathcal{F}^{(1,-1)}(1_{34}) 
-\frac23\,    \mathcal{F}^{(2,-1)}(2_{432})\mathcal{F}^{(1,-1)}(1_{12})\,.  
\end{align}
\end{subequations}
Expanding eq.~(\ref{3lfourmat}) gives
\begin{align}
  W_{(1,2,2,1)}^{(3)}=\frac16 \Big( C(3a)-C(3b)-C(3c)+C(3d) \Big) \Big(
    \mathcal{F}(3a)-2\mathcal{F}(3b) - 2\mathcal{F}(3c) + \mathcal{F}(3d)
  \Big) 
  \label{W1221all}
\end{align}
and so
\begin{align}
  \label{eq:W2}
  \begin{split}
  W_{(1,2,2,1)}^{(3,-2)}&=-\frac16 \Big( C(3a)-C(3b)-C(3c)+C(3d) \Big) \\ & \hspace{1cm} \times \left(
    \mathcal{F}^{(2,-1)}(2_{123})\mathcal{F}^{(1,-1)}(1_{34}) +
    \mathcal{F}^{(2,-1)}(2_{432})\mathcal{F}^{(1,-1)}(1_{12}) \right) \\
  &= -\frac16 \Big[ C(1_{34}),C(2_{123})-C(2_{321}) \Big]
  \mathcal{F}^{(2,-1)}(2_{123})\mathcal{F}^{(1,-1)}(1_{34}) \\& \hspace{1cm} -\frac16 \Big[
    C(1_{12}), C(2_{432})-C(2_{234}) \Big]
  \mathcal{F}^{(2,-1)}(2_{432})\mathcal{F}^{(1,-1)}(1_{12}) 
  \end{split}
\end{align}
at this subleading pole order, where the following relationships between the 
colour factors (obtained using eq.~(\ref{colrule})) have been used:
\begin{align}
  \label{eq:colours}
  \begin{split}
  C(3a)-C(3b)-C(3c)+C(3d) &= \Big[ C(1_{34}),C(2_{123})-C(2_{321}) \Big] \\ &=
  \Big[ C(1_{12}), C(2_{432})-C(2_{234}) \Big].
  \end{split}
\end{align}
Equation~(\ref{eq:W2}) is the contribution of this web to the left-hand side of equation~\eqref{eq:prize}.  We now turn to
the right-hand side,
\begin{align}
  \label{eq:prizerhs}
  -\frac{1}{6}\left[w^{(1,-1)},w^{(2,-1)}\right].
\end{align}
The relevant one-loop pieces are
\begin{align}
  \label{eq:w1full2}
  \begin{split}
  w^{(1,-1)}\ &=\ C(1_{12})
  \mathcal{F}^{(1,-1)}(1_{12})\ +\ C(1_{34})
  \mathcal{F}^{(1,-1)}(1_{34})\,+\,\ldots\,. 
  \end{split}
\end{align}
The $1_{23}$ piece does not contribute because there is no corresponding
two-loop web which it can combine with\footnote{There is of course a two-loop diagram where two gluons are exchanged between two distinct pairs of legs, $[[1],[1],[2],[2]]$, but this is not a web~\cite{Gardi:2010rn}, and thus not part of $w^{(2,-1)}$.}.
At two-loops, the diagrams are combined with the following exponentiated colour
factors (see e.g.~eq.~(5.38) in \cite{Gardi:2010rn}):
\begin{align}
  \label{eq:w2exp2}
  \begin{split}
  w^{(2,-1)}\ &=
  \frac12(C(2_{123})-C(2_{321}))\left(\mathcal{F}^{(2,-1)}(2_{123}) -
  \mathcal{F}^{(2,-1)}(2_{321})\right)  \\ & \hspace{1.5cm}+
  \frac12(C(2_{432})-C(2_{234}))\left(\mathcal{F}^{(2,-1)}(2_{432}) -
  \mathcal{F}^{(2,-1)}(2_{234})\right) \,+\,\ldots\\
  &=(C(2_{123})-C(2_{321}))\mathcal{F}^{(2,-1)}(2_{123}) +
  (C(2_{432})-C(2_{234}))\mathcal{F}^{(2,-1)}(2_{432})\,+\,\ldots\,,
  \end{split}
\end{align}
where in the second line we used eq.~(\ref{2_ijk_kij}).
Substituting eqs.~(\ref{eq:w1full2}, \ref{eq:w2exp2}) into~(\ref{eq:prizerhs}) and keeping only the parts which correspond to this web then gives 
\begin{align}
  \label{eq:Wrhsfin2}
  \begin{split}
  W_{(1,2,2,1)}^{(3,-2)}&=-\frac16 \Big[
    C(1_{12}), C(2_{432})-C(2_{234}) \Big]
  \mathcal{F}^{(2,-1)}(2_{432})\mathcal{F}^{(1,-1)}(1_{12}) \\& \hspace{1cm}
  -\frac16 \Big[ C(1_{34}),C(2_{123})-C(2_{321}) \Big]
  \mathcal{F}^{(2,-1)}(2_{123})\mathcal{F}^{(1,-1)}(1_{34})\,, 
  \end{split}
\end{align}
which agrees exactly with equation~\eqref{eq:W2}. We have thus verified in a
second example that the renormalization constraints of 
section~\ref{sec:renormalization} are satisfied on a web-by-web basis.

\section{Factorization of the leading pole of maximally reducible diagrams\label{sec:factorization}}

In the previous section we considered two specific three-loop webs and investigated how the renormalization constraints are satisfied. 
In the webs we considered, each diagram was composed exclusively of individual gluon exchanges. 
We saw that for the leading singularity of these diagrams, ${\cal O}(\epsilon^{-3})$, the kinematic dependence factorises into three one-loop kinematic factors, corresponding to sequentially shrinking the three gluon subdiagrams to the origin.
The next-to-leading singularity, ${\cal O}(\epsilon^{-2})$, can be written as a sum of products of two-loop times one-loop subdiagrams.
These factorization properties have proven crucial in satisfying the renormalization constraints, namely the cancellation of the leading singularity and the non-trivial relation~(\ref{eq:prize}) between three-loop webs and one- and two-loop webs.

In this section we take a step towards an all-order generalization of these results. We consider multi-eikonal diagrams at any order ${\cal O}(\alpha_s^n)$, but restrict ourselves to ones that are composed \emph{exclusively of individual gluon exchanges.} For this class of diagrams we will show that a similar factorization property holds for the leading singularity ${\cal O}(\epsilon^{-n})$:
in a given web, for all diagrams that have an ${\cal O}(\epsilon^{-n})$ pole, this singularity factorizes into $n$ one-loop integrals. Moreover, these leading singularities are all the same up to a combinatorial \emph{symmetry factor} $s(D)$, a non-negative integer number which determines how many different ways there are in a given diagram $D$ to sequentially shrink the individual subdiagrams to the origin. 
The quantity $s(D)$ is defined to be zero for those diagrams which do not have a leading singularity. A more general definition and some examples for the symmetry factor will be given in section~\ref{sec:conjecture} (and in appendix~\ref{app:sD}) where we explain its role in the context of the web mixing matrix.

Our first observation is that a web of diagrams which are made of individual gluon exchanges satisfies the requirement:
\begin{equation}
\sum_{D}{\cal F}(D)=\prod_{k=1}^n{\cal F}^{(1)}(\gamma_{i_kj_k}) \,,
\label{sumFD}
\end{equation}
where the sum is over all the diagrams in the web.
This relation, which holds to any order in epsilon, may be derived as follows. Each diagram has a kinematic part, after
rescaling the distance variables associated with each gluon emission 
according to $s_1\rightarrow\sqrt{-\beta_1^2}$ etc., of the
form
\begin{align}
{\cal F}(D)&=\left(\frac{g_s^2(\mu^2)^{\epsilon}{\cal N}}{2}\right)^n(\gamma_{i_1j_1}
\ldots\gamma_{i_nj_n})
\int_0^\infty ds_1\ldots\int_0^\infty ds_n\int_0^\infty dt_1
\ldots\int_0^\infty dt_n\notag\\
&\times e^{-\sum_{j=1}^nm(s_j+t_j)}
\Big[\left(s_1^2+t_1^2-s_1\,t_1\gamma_{i_1j_1}\right)\ldots\left(s_n^2+t_n^2-s_n\,t_n
\gamma_{i_nj_n}\right)\Big]^{\epsilon-1}
\notag\\&\times
\left[\prod\Theta\right](\{s_i\},\{t_i\}).
\label{FD1}
\end{align}
Here we have assumed that the $k^{\rm th}$ gluon is exchanged between parton lines
$i_k$ and $j_k$, and is labelled by distance parameters $(s_k,t_k)$. 
The last line denotes the fact
that there is a product of Heaviside functions, whose role is to order the
gluons proceeding outwards from the hard interaction vertex. This is the only
difference between the diagrams $\{D\}$, and summing over all $D$ gives the 
complete sum of Heaviside functions
\begin{equation}
\sum_D\left[\prod\Theta\right](\{s_i\},\{t_i\})=1\,,
\label{sumtheta}
\end{equation}
such that one has
\begin{align}
\label{sum_lead_sing}
\sum_D{\cal F}(D)&=\prod_{k=1}^n\frac{g_s^2}{2} (\mu^2)^{\epsilon} {\cal N}\gamma_{i_kj_k}\int_0^\infty
ds_k\int_0^\infty dt_k\,e^{-m(s_k+t_k)}\Big[s_k^2+t_k^2-s_k\,t_k\gamma_{i_kj_k}
\Big]^{\epsilon-1}\,.
\end{align}
This is recognisable as the product of one-loop graphs occurring on the 
right-hand side of eq.~(\ref{sumFD}). This result will be useful in what follows.

Let us now consider a particular maximally reducible diagram $D$, namely one having an ${\cal O}(\epsilon^{-n})$ pole, thus a non-zero symmetry factor $s(D)$ as defined above. 
Contributions to the leading singularity of this diagram come only from the $s(D)$ different integration regions for which the $n$ single-gluon subdiagrams are ordered, and can thus be sequentially shrunk to the multi-eikonal vertex (the origin, where the hard interaction takes place).

\begin{figure}
\begin{center}
\scalebox{0.8}{\includegraphics{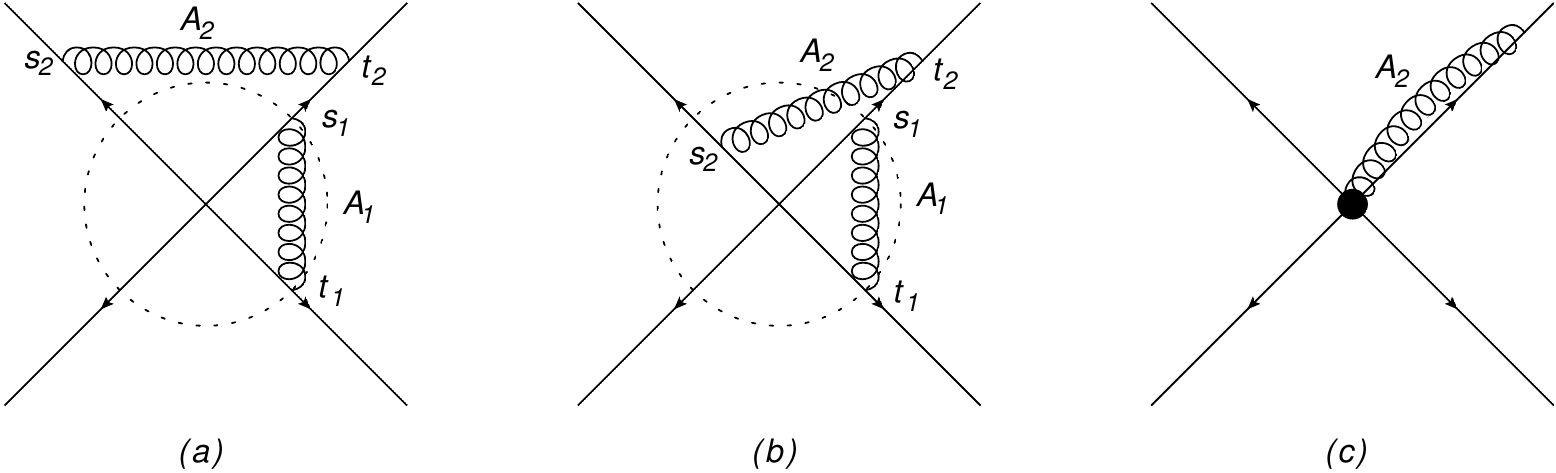}}
\caption{The web diagram (2a) of figure~\ref{2six}, where we depict: 
(a) the integration region $s_2>s_1$; (b) the integration region $s_2<s_1$;
(c) the effect of shrinking gluon $A_1$ to the origin when $s_2<s_1$.}
\label{2afig}
\end{center}
\end{figure}

To illustrate this point, consider the web diagram (2a) shown in 
figure~\ref{2six}. We redraw this in figure~\ref{2afig}, labelling the
gluon emissions by $A_i$, such that the corresponding distance parameters
for the two ends of the gluon are $(s_i,t_i)$. The Heaviside functions in 
this case impose $s_1<t_2$, and there are then two distinct integration 
regions for $s_2$, namely $s_2>s_1$ and $s_2<s_1$. These conditions operate on distance variables on different parton lines. However, it is meaningful to compare distances on different lines given that we have rescaled distance variables according to $s_1\rightarrow s_1/\sqrt{-\beta_1^2}$ etc.
The two regions are shown schematically in figure~\ref{2afig}(a)
and (b). Although these diagrams are topologically equivalent, they have 
different properties as the gluon $A_1$ is shrunk to the origin. In the first
case ($s_2>s_1$), $A_1$ may be shrunk to the origin independently of $A_2$, which generates a leading singularity (in this case an $\epsilon^{-2}$ pole).
In the second case, the requirement $s_2<s_1$ means that as $A_1$ is shrunk to the origin, the end of $A_2$ labelled by $s_2$ is ``trapped'' at the hard interaction vertex, such that the gluon $A_2$ becomes collinear to $\beta_2$, as shown schematically in figure~\ref{2afig}(c). 
This would generate an additional (collinear) singularity were the parton lines massless. However, we consider massive parton lines in this paper, so that the contribution from figure~\ref{2afig}(c) has no further singular behaviour, thus has only a single pole, corresponding to the shrinking of $A_1$.

Let us now denote by $A_1<A_2$ the integration region in which 
\begin{equation}
\label{eq:order}
s_1<s_2,\,t_2\,,\qquad{\rm and}\qquad t_1<s_2,\,t_2 \,,
\end{equation}
corresponding to the ordered region of figure~\ref{2afig}(a). From the above discussion, we see that the region $A_1<A_2$ generates the leading pole of the web diagram, but that not all subleading poles will be captured, as for the latter, one would also need to include configurations such as that shown in figure~\ref{2afig}(b). Armed with this notation, we can now proceed to analyse the leading pole of a general diagram consisting of single gluon emissions.

The above argument is straightforwardly generalised to diagrams containing any number $n$ of individual gluon exchanges, which contribute at leading pole, ${\cal O}(\epsilon^{-n})$. That is, the leading pole only receives contributions from the integration region in which all subdiagrams are ordered in the manner of figure~\ref{2afig}(a). 
Denoting the gluon subdiagrams by $A_k$ with associated distance parameters $s_k$ and $t_k$ along the Wilson lines, a given ordering can be written as $A_{\pi_1}<A_{\pi_2}<\ldots <A_{\pi_n}$ where 
$\pi=(\pi_1,\pi_2,\ldots\pi_n)$ is a permutation of $(1,2,\ldots,n)$ and where
the ``$<$" symbol has the same meaning as described in eq.~(\ref{eq:order}).

We may then write the contribution to diagram $D$ from the ordering $\pi$ as
\begin{align}
&{\cal F}(\left.D\right\vert \pi)\equiv \left(\frac{g_s^2(\mu^2)^\epsilon{\cal N}}{2}\right)^n
\left[\prod_{k=1}^n\gamma_{i_kj_k}\right]
\iint\limits_{A_{\pi_1}<A_{\pi_2}<\ldots<A_{\pi_n}}d\{s_k\}\,d\{t_k\}\notag\\
&\times e^{-\sum_{j=1}^nm(s_j+t_j)}
\Big[\left(s_1^2+t_1^2-s_1\,t_1\gamma_{i_1j_1}\right)\ldots\left(s_n^2+t_n^2-s_n\,t_n
\gamma_{i_nj_n}\right)\Big]^{\epsilon-1}.
\label{FD3}
\end{align}
This definition is such that 
\begin{equation}
\label{summing_over_pi}
\sum_{\pi\in D} {\cal F}(D\vert \pi)={\cal F}(D)\,
\left(1+{\cal O}(\epsilon)\right)\,,
\end{equation}
where the sum is over all orderings $\pi$ which are consistent with diagram $D$. 
In this way one recovers the full leading pole of $D$ by combining ordered regions as defined above.
Note, however, that the sum does not exactly reproduce the Heaviside functions in eq.~(\ref{FD1}) due to the fact that it neglects contributions from integration 
regions such as that shown in figure~\ref{2afig}(b). These regions will only contribute at subleading orders in $\epsilon$, hence the ${\cal O}(\epsilon)$ corrections in eq.~(\ref{summing_over_pi}).

We now argue that each such unique ordering gives \emph{the same} leading pole. First we make 
$n$ paired variable transformations of the form
\begin{equation}
\left(\begin{array}{c}s_k\\t_k\end{array}\right)=\lambda_k
\left(\begin{array}{c}x_k\\1-x_k\end{array}\right),
\label{nvars}
\end{equation}
such that eq.~(\ref{FD3}) becomes
\begin{align}
{\cal F}(\left.D\right\vert \pi)&=\left(\frac{g_s^2
{\cal N}}{2}\frac{\mu^{2\epsilon}}{m^{2\epsilon}}\right)^n\prod_{k=1}^n\gamma_{i_kj_k}
\int_0^1dx_k\,P(x_k,\gamma_{i_kj_k})
\int_0^{\lambda_k^{\rm max}}d\lambda_k\ \lambda_k^{2\epsilon-1}e^{-\lambda_k},
\label{FD4}
\end{align}
where we have scaled $\lambda_k\rightarrow\lambda_k/m$,
and $\lambda_k^{\rm max}$ is determined from the Heaviside functions, 
depending upon the parameters $\lambda_{k+1}$, $x_{k+1}$ and $x_k$ occurring in (\ref{nvars}). 

The next step is to consider the product of $\lambda_k$ integrals on the 
right-hand side. These may be written as
\begin{align}
\label{lambdaprod}
\prod_{k=1}^n \int_0^{\lambda_k^{\rm max}}d\lambda_k\ \lambda_k^{2\epsilon-1}e^{-\lambda_k}
&=\int_0^\infty d\lambda_n\int_0^{\lambda_n\alpha_n}d\lambda_{n-1}
\ldots\int_0^{\lambda_2\alpha_2}d\lambda_1\,\left(\lambda_1\lambda_2\ldots
\lambda_n\right)^{2\epsilon-1}e^{-\sum_{k=1}^n\lambda_k}.
\end{align}
Here we have used the strict ordering ${A_{\pi_1}<A_{\pi_2}<\ldots<A_{\pi_n}}$, where from dimensional considerations each upper limit of integration has been expressed as 
\begin{equation}
\lambda_k^{\rm max}=\lambda_{k+1}\,\alpha_{k+1}\,,
\label{lkmaxdef}
\end{equation}
where $\alpha_{k+1}$ is a function of $x_{k+1}$ and $x_k$, and for later convenience we define $\alpha_1=1$. 
 
Equation (\ref{lambdaprod}) can be computed by making the following change of variables, 
by analogy with (\ref{trans2}):
\begin{equation}
\sigma_k\,=\,\sum_{i=1}^k\,\lambda_k\,,\qquad\quad
z_{k-1}\equiv \frac{\sigma_{k-1}}{\sigma_k}
\end{equation} 
where, for convenience, $z_0\equiv 0$. 
It then follows that the original variables $\lambda_k$ can be expressed as
\begin{equation}
\lambda_k\,=\,\sigma_k\,(1-z_{k-1})\,=\,z_kz_{k+1}\ldots z_{n-1}\,\sigma_n\,(1-z_{k-1})\,.
\end{equation} 
The Jacobian is
\begin{equation}
\frac{\partial(\lambda_1,\lambda_2,\ldots,\lambda_n)}{\partial(z_1,z_2,\ldots,z_{n-1},\sigma_n)}
\,=\,\sigma_2\sigma_3\ldots\sigma_n\,=
\,z_2 z_3^2 z_4^3\ldots z_{n-1}^{n-2}\,\sigma_n^{n-1}
\end{equation}
and one also has
\[
\lambda_1\lambda_2\ldots\lambda_n=z_1z_2^2z_3^3\ldots z_{n-1}^{n-1}\,(1-z_1)(1-z_2)\,\ldots(1-z_{n-1})\,\sigma_n^{n}
\]
so that eq.~(\ref{lambdaprod}) becomes:
\begin{align}
\label{lambdaprod1}
\begin{split}
\prod_{k=1}^n\int_0^{\lambda_k^{\rm max}}d\lambda_k\ \lambda_k^{2\epsilon-1}e^{-\lambda_k}
&=\int_0^\infty d\sigma_n \, \sigma_n^{2n\epsilon-1}\, e^{-\sigma_n}
\\&\times 
\prod_{k=1}^{n-1} \int_0^{1}dz_k\,z_k^{2k\epsilon-1}\,(1-z_k)^{2\epsilon-1}
\,\Theta\left(\frac{z_{k}}{1-z_{k}}<\frac{\alpha_{k+1}}{1-z_{k-1}}\right)\,.
\end{split}
\end{align}
The $\sigma_n$ integral represents the overall UV divergence corresponding to simultaneously shrinking all gluons to the origin. The remaining product of $z_k$ integrals can be sequentially evaluated starting from $z_{n-1}$ and working backwards to $z_1$. Each such integral depends upon a single Heaviside function, and may be carried out using the transformation $\rho_k=z_k/(1-z_k)$. Similarly to (\ref{rint1}) each integration results in a hypergeometric function, which becomes trivial upon expansion in $\epsilon$. One ultimately finds 
 \begin{align}
\label{lambdaprod2}
\begin{split}
\prod_{k=1}^n\int_0^{\lambda_k^{\rm max}}d\lambda_k\ \lambda_k^{2\epsilon-1}e^{-\lambda_k}
&=\,\Gamma(2n\epsilon)\,\frac{\alpha_2^{2\epsilon}}{2\epsilon}\,\frac{\alpha_3^{4\epsilon}}{4\epsilon} \ldots
\frac{\alpha_{n-1}^{2(n-2)\epsilon}}{2(n-2)\epsilon}\,
\frac{\alpha_{n}^{2(n-1)\epsilon}}{2(n-1)\epsilon}\,\left(1+{\cal O}(\epsilon^2)\right)\,.
\end{split}
\end{align}
Substituting this back into (\ref{FD4}) yields
\begin{align}
{\cal F}(\left.D\right\vert \pi)&=\left(\frac{g_s^2
{\cal N}}{2}\frac{\mu^{2\epsilon}}{m^{2\epsilon}}\right)^n
\,\frac{1+{\cal O}(\epsilon^2)}{(2\epsilon)^n\,n!}
\prod_{k=1}^n\gamma_{i_kj_k}
\int_0^1dx_k\,P(x_k,\gamma_{i_kj_k})
\,\, \left(\alpha_k(x_k,x_{k-1})\right)^{2(k-1)\epsilon}\,
\,,
\label{FD5}
\end{align}
where we explicitly wrote the arguments of $\alpha_k=\alpha_k(x_k,x_{k-1})$.
At leading pole one may ignore subleading contributions from the factors $\alpha_{k}^{2(k-1)\epsilon}$, and by comparing with (\ref{eq:Fijoneloop}) one observes that the leading singularity arising from this ordering is 
\begin{align}
{\cal F}(\left.D\right\vert \pi)=\frac{1+{\cal O}(\epsilon)}{n!}\,\prod_k{\cal F}^{(1)}(\gamma_{i_kj_k})\,.
\label{FD6}
\end{align}
Given that there are $s(D)$ such orderings, eq.~(\ref{summing_over_pi}) yields the following leading singularity of $D$:
\begin{equation}
{\cal F}^{(n,-n)}(D)=\frac{s(D)}{n!}\prod_{k=1}^n{\cal F}^{(1,-1)}(\gamma_{i_kj_k})\,.
\label{FD7}
\end{equation}
Note that this result is consistent with the expression obtained above for the sum over all diagrams $D$ in the set, eq.~(\ref{sumFD}), given that the total number of orderings of the $n$ subdiagrams is 
\begin{equation}
\sum_D s(D)=n!\,\,.
\label{sumsD}
\end{equation}

Equation~(\ref{FD7}) is the main result of this section. It shows that the leading pole
of a maximally reducible diagram composed of single-gluon exchanges 
can be explicitly related to the product of the $n$ constituent one-loop diagrams. 
The leading poles of different diagrams in the web are equal up to the factor $s(D)$ which counts the number of orders of sequentially shrinking the different subdiagrams to the origin. 
In the next section we will see that this property, in conjunction with the structure of the web mixing matrix, explains the cancellation of the leading poles in these webs, as required by renormalization.

Before doing so, it is worth noting that there is an interesting constraint on the structure of subleading 
poles, stemming from eq.~(\ref{sumFD}). This relation implies that the sum 
over all kinematic parts of a given web gives a complete factorised product
of connected subdiagrams. Consequently, whereas individual diagrams have subleading poles which depend upon the ordering of their connected subdiagrams, the sum does not. 
This in turn implies that subleading poles associated with non-connected subdiagrams 
such as those on the r.h.s. of eqs.~(\ref{eq:master3NLP}) must vanish upon summing over all diagrams in the web. One may indeed check that the coefficients of
the products $\mathcal{F}^{(2,-1)}(2A)\mathcal{F}^{(1,-1)}(\gamma_{12})$
and $\mathcal{F}^{(2,-1)}(2G)\mathcal{F}^{(1,-1)}(\gamma_{23})$ sum separately
to zero after substituting the results into eq.~(\ref{sumFD}). Likewise, the
coefficients of $\mathcal{F}^{(2,-1)}(2_{123})\mathcal{F}^{(1,-1)}(1_{34})$ 
and $\mathcal{F}^{(2,-1)}(2_{432}) \mathcal{F}^{(1,-1)}(1_{12})$
sum to zero in eqs.~(\ref{eq:F32inF21-2}). 
Note that this cancellation is not specific to the particular class of diagrams considered here.
For webs containing $n_c$ connected subdiagrams that may be non-trivial (containing three gluon vertices etc.) one may use the appropriate generalization of eq.~(\ref{sumFD}), which is
\begin{equation}
\sum_{D}{\cal F}(D)=\prod_{k=1}^{n_c} {\cal F}(A_k) \,,
\label{sumFD_gen}
\end{equation}
where $A_k$ is a connected subdiagram with kinematic part ${\cal F}(A_k)$.

\section{A new conjecture for web mixing matrices\label{sec:conjecture}}

The main aim of this paper has been to investigate the renormalization of
multiparton webs using the renormalizability of Wilson-line correlators. We have
seen how to renormalize webs, and learned that a tight singularity structure is
realised on a web by web basis.  The explicit examples considered in section~\ref{sec:3_loop_examples} confirm our expectation that the leading pole of any given web cancels as a consequence of the properties of web mixing matrices in conjunction with the factorization properties of the kinematic factors.
We further saw in section~\ref{sec:factorization} that the factorization of the leading poles generalises to all orders for a particular class of diagrams made of individual gluon exchanges. 
Since we also expect the cancellation of leading singularities to persist to all orders -- see eq.~(\ref{conjecture}) -- it is clear that the cancellation mechanism we observed at three loops must be general.
Building upon this insight, in this section we formulate a new conjecture for web mixing matrices which explains these cancellations.
This conjecture, which we formulate below as a weighted column sum-rule, appears to be a general combinatorial property of the web mixing matrices.

First, we need to precisely define the notion of maximal reducibility. In
a given web, not all diagrams have leading poles. One example is the
web of figure~\ref{3lsix}, where only three of the six diagrams are
${\cal O}(\epsilon^{-3})$. A \emph{maximally reducible} diagram has 
a number of irreducible subdiagrams equal to the number of connected 
subdiagrams. It is easily checked that this condition is verified by 
diagrams ($3D$)--($3F$) in figure~\ref{3lsix}, but not by the first three
diagrams: ($3A$) and ($3C$) have two irreducible subdiagrams and ($3B$) has only
one. All diagrams, though, have three connected pieces, which in this case are
all individual gluons.
Another three-loop example, one with a three-gluon vertex, is that of figure~\ref{appendixfig2}. 
These diagrams have just two connected pieces.
The leftmost and central diagrams are maximally reducible, as the two subdiagrams may be sequentially shrunk to the origin. This is not the case for the rightmost diagram.

Having established the terminology, we may state our conjecture as follows:\\

\noindent\emph{Consider a single web $W$. For a given diagram $D$ in $W$, 
let $s(D)$ be the number of distinct ways in which the irreducible 
subdiagrams of $D$ may be shrunk to the origin, where $s(D)=0$ if $D$ is
not maximally reducible. 
Then the web mixing matrix satisfies the weighted column sum rule
\begin{equation}
\sum_{D}s(D)R_{DD'}=0\qquad\forall D'\,,
\label{columnrule}
\end{equation}
where the sum is over all the diagrams in $W$.
}\\

As an illustration of this result, consider the web of figure~\ref{3lsix}.
This web has three maximally reducible diagrams: ($3D$)--($3F$). Each of these has three
irreducible subdiagrams, each of which is a single gluon exchange. Furthermore,
there is only one sequence in each diagram by which one may shrink all the 
gluons to the origin: one must start with the innermost gluon, and work 
sequentially outwards. The set of values $s(D)$ for the six diagrams may then
be written as the row vector
\begin{equation}
s^{(2,3,1)}_D=\left(\begin{array}{rrrrrr}
0 & 0 & 0 & 1 & 1 & 1\end{array}\right).
\label{sD1}
\end{equation}
Equation~(\ref{columnrule}) then amounts to the statement
\begin{equation}
\left(\begin{array}{rrrrrr}
0 & 0 & 0 & 1 & 1 & 1\end{array}\right)
\left(\begin{array}{rrrrrr}3&0&-3&-2&-2&4\\-3&6&-3&1&-2&1\\-3&0&3&4&-2&-2
\\0&0&0&1&-2&1\\0&0&0&-2&4&-2\\0&0&0&1&-2&1\end{array}\right)
=\left(\begin{array}{rrrrrr} 0 & 0 & 0 & 0 & 0 & 0\end{array}\right),
\label{crule1}
\end{equation}
where we have taken the web mixing matrix from eq.~(\ref{1-6mat}). A second,
and perhaps less trivial, example is provided by the web of 
figure~\ref{3lfour}, all of whose diagrams are maximally reducible. However,
whereas diagrams ($3b$) and ($3c$) have only one ordering by which one may shrink
the gluons to the origin, in diagrams ($3a$) and ($3d$) this ordering is not unique.
This is because the upper and lower gluons act on completely different parton
lines, so that the corresponding subdiagrams commute with each other. In both
cases one may choose either to shrink the upper gluon followed by the lower 
one, or vice versa. Thus, the vector of multiplicity factors $s(D)$ is given
in this case by
\begin{equation}
s_D^{(1,2,2,1)}=\left(\begin{array}{rrrr} 2 & 1 & 1 & 2\end{array}\right),
\label{sD2}
\end{equation}
and eq.~(\ref{columnrule}) for this web is 
\begin{equation}
\left(\begin{array}{rrrr} 2 & 1 & 1 & 2\end{array}\right)
\left(\begin{array}{rrrr}
1&-1&-1&1\\-2&2&2&-2\\-2&2&2&-2\\1&-1&-1&1
\end{array}\right)=
\left(\begin{array}{rrrr} 0 & 0 & 0 & 0\end{array}\right).
\label{crule2}
\end{equation}
Further examples of these symmetry factors may be found in 
appendix~\ref{app:sD}. 

As justification for this conjecture, let us first consider the special class
of diagrams considered in section~\ref{sec:factorization},  consisting 
entirely of individual gluon exchanges. There we were able to derive a result
for the leading pole of any maximally-reducible diagram. We saw in eq.~(\ref{FD7}) that the leading pole for any such diagram $D$ reduces to the product of the $n$ corresponding one-loop subdiagrams times a symmetry factor $s(D)$. Thus, the leading poles of the maximally-reducible diagrams in the set are all equal up to this symmetry factor. 

One may then use the fact, discussed in section~\ref{sec:renormalization}, that in order to satisfy the renormalization constraints in the conformal limit, eq.~(\ref{wnn_no_beta}), leading poles, ${\cal O}(\epsilon^{-n})$, must cancel.
Assuming that there is no cancellation between different webs (which looks rather unlikely given the different kinematic dependence)  eq.~(\ref{wnn_no_beta}) translates into a constraint on individual webs,
and given the independence of colour factors of different diagrams in a web,  this can only be realised if the mixing matrix admits eq.~(\ref{conjecture}), 
\begin{equation}
\label{conjecture_}
\sum_{D}\,{\cal F}^{(n,-n)}(D) \,R_{(n_1,\ldots,n_L)}^{(n)}(D,D')\,=\,0\,,\qquad \quad \forall D'\,.
\end{equation}
Substituting the leading singularities according to~(\ref{FD7}) into (\ref{conjecture_}) then immediately implies that 
\begin{equation}
\label{W02}
\sum_{D}\,s(D) \,R_{(n_1,\ldots,n_L)}^{(n)}(D,D')\,=\,0\,,\qquad \quad \forall D'\,.
\end{equation}
This is the weighted column sum rule of eq.~(\ref{columnrule}) which, for
this class of diagrams, is implied by renormalization.
The above argument does not extend, however, to diagrams with non-trivial connected subdiagrams involving three or four gluon vertices or fermion loops, in which case the number of connected pieces in the diagrams, $n_c$, is smaller than the order $n$.
There are two reasons for this. 
Firstly, the maximal pole obtained by shrinking individual subdiagrams to the origin would be in this case ${\cal O}(\epsilon^{-n_c})$; therefore this singularity does not relate to the constraint (\ref{wnn_no_beta}) corresponding to ${\cal O}(\epsilon^{-n})$ in $w$, but rather to renormalization constraints of lower orders, as displayed in (\ref{w_nk_relations}), which necessarily involve commutator terms. 
Secondly, the factorization of the ${\cal O}(\epsilon^{-n_c})$ pole obtained upon shrinking all subdiagrams to the origin for such a diagram takes a more complicated form, which does not depend solely on the symmetry factor $s(D)$.

Nevertheless, it is interesting to examine whether the weighted column sum rule
of eq.~(\ref{columnrule}) still applies, even in the case of webs containing
gluon interactions off the eikonal lines. Some examples are shown in 
appendix~\ref{app:sD}, and we find, somewhat remarkably, that the sum rule 
holds in all cases! Given that we presently lack an explanation for this from 
renormalization properties alone, it would be interesting to study the 
implications of this result in more detail. The weighted column sum rule is
analagous to other properties of web mixing matrices (e.g. idempotence and
zero sum rows) whose physical implications are better understood, and
which have been demonstrated to be general using combinatoric 
methods~\cite{Gardi:2010rn,Gardi:2011wa}. Two clear goals for future study are
thus to attempt to understand the physics embodied by eq.~(\ref{columnrule}), 
and also to seek a combinatoric proof, which may involve methods similar to 
those utilised in~\cite{Gardi:2011wa}. 

Note that even the justification of eq.~(\ref{columnrule}) for the special
case of diagrams with individual gluon emissions falls short of a rigorous
proof. The reason for this, as stated above, is that we have assumed that 
no cancellations occur between different webs in writing eq.~(\ref{conjecture}). 
This complication would be circumvented upon producing a combinatoric proof
which relies purely on the properties of individual webs, as has been carried
out for the zero sum row property in~\cite{Gardi:2011wa}. 

In this section, we have used the insight gained in previous sections from explicit calculation of multiparton webs in order to arrive at a general conjecture regarding the columns of web mixing matrices. This complements the zero sum row property,
and hints at the existence of further combinatoric structures underlying web
mixing matrices. This concludes, however, the analysis which we will undertake
in this paper. In the following section, we discuss our results and conclude.

\section{Conclusions\label{sec:conclusions}}

In this paper, we have considered the renormalization properties of multiparton
webs, taking a further step in developing the diagrammatic approach~\cite{Gardi:2010rn,Mitov:2010rp,Gardi:2011wa} to soft-gluon exponentiation in multiparton scattering.
Webs constitute an efficient way of calculating the exponents of eikonal 
amplitudes directly. They are therefore an important tool in the study of multiparton amplitudes, and in particular their infrared singularity structure.

Building upon~\cite{Mitov:2010rp,Gardi:2010rn}, we combined in this paper two complementary approaches to soft-gluon exponentiation, one which stems from the multiplicative renormalizability of Wilson-line correlators and one which utilises a diagrammatic interpretation. 

According to the diagrammatic approach the fundamental objects composing the exponent at any given order in perturbation theory are webs. In the multiparton case, these are not individual diagrams, but rather closed sets of diagrams that are related to each other by permuting the gluon attachments to the Wilson lines. The colour and kinematic factors of these diagrams are entangled through mixing matrices.  

The problem of determining soft singularities in amplitudes is equivalent to the problem of computing the renormalization of a corresponding eikonal amplitude. Multiplicative renormalizability implies that all soft singularities are encodable in a finite anomalous dimension. 
Owing to recent progress the soft anomalous dimension in multiparton scattering is known to two-loop order. Determining its higher-loop corrections and understanding its all-order properties is important from both a general field-theory perspective, and a pragmatic collider-physics one. The present paper paves the way for such calculations. There are several different aspects to this: 
\begin{itemize}
\item{} Firstly, eikonal calculations in dimensional regularization suffer from an inherent problem of giving rise to scaleless integrals involving both infrared and ultraviolet singularities. 
Here we provide a prescription to disentangle these singularities at any loop order: by using non-lightlike eikonal lines we avoid collinear poles, and by introducing an exponential regulator along the Wilson lines (\ref{FRmod}) we eliminate soft ones. This facilitates a direct computation of the anomalous dimension associated with the multi-eikonal vertex. 
\item{} We have shown that multiplicative renormalizability, and the subsequent finite anomalous dimension, imply a highly constrained singularity structure for the exponent of the eikonal amplitude. Specifically, \emph{all multiple poles at any given order are fully determined by lower-loop webs}. 
Higher-order poles appear due to two distinct reasons: running coupling corrections and, in the multi-eikonal case, the non-commuting nature of the colour generators. 
Both these effects are fully understood and are incorporated in the renormalization constraints, which are summarised in eqs.~(\ref{w_nk_relations}) through four loops. These relations offer powerful checks on any multiloop eikonal computation.
\item{} Furthermore, we have demonstrated that these renormalization constraints are satisfied on a web by web basis.
Thus, the combinations formed through the operation of the web mixing matrices have a singularity structure that is determined by renormalization, where all multiple poles can be computed using lower-order webs. Only the single pole terms at each order contain new information, which in turn allows one to determine the corresponding contribution to the soft anomalous dimension coefficient $\Gamma^{(n)}$ according to eqs.~(\ref{Gamma_n}).
\item{} A major difficultly in multiloop computations in dimensional regularization is the increasingly high power in $\epsilon$ to which the expansion needs to be made. Being able to characterize, a priori, the singularity structure of webs, and consequently isolate the single-pole contributions to the anomalous dimension facilitates efficient higher-loop computations by both analytical and numerical methods.
\end{itemize}

We investigated in this paper the way in which the renormalization constraints of  eqs.~(\ref{w_nk_relations}) are satisfied focusing on the class of diagrams composed of individual gluon exchanges. This class of diagrams has been convenient in that all divergences in any diagram are related to the renormalization of the multi-eikonal vertex, rather than to the renormalization of the strong coupling. As a consequence, this analysis is directly valid in any gauge theory. 

Our main findings can be summarised as follows: the renormalization constraints 
are realised on a web-by-web basis through the operation of the web mixing matrix, in conjunction with the fact that multiple poles in each diagram reduce to sums of products of lower-order diagrams. Each term in this sum corresponds to a particular decomposition of the original diagram into subdiagrams.
At leading pole, any diagram reduces to a product of all connected subdiagrams (single gluon exchanges in the case considered here) while at subleading poles several terms occur, corresponding to different decompositions of the diagram into its subdiagrams, where the latter may be reducible. 

At subleading pole, and starting from three-loop order, the renormalization of the multi-eikonal amplitude exponent becomes more complicated owing to the non-commuting nature of the colour generators. As a consequence the renormalization constraints include commutators of lower-order webs.  
The first non-trivial relation of this kind, eq.~(\ref{eq:prize}), expresses the next-to-leading pole of three-loop contributions to the exponent in terms of a commutator of two-loop and one-loop webs. In section~\ref{sec:3_loop_examples} we verified by explicit calculations that this relation is satisfied on a web-by web basis. A key ingredient is the fact that 
the next-to-leading poles can indeed be written as a sum over decompositions of any three-loop diagrams into two-loop times one-loop subdiagrams. The coefficients in front of each product are clearly combinatoric in nature, and are still somewhat mysterious. A full analysis of subdivergences would benefit from a more detailed investigation of these numbers.  

The factorization property of the leading singularity has been established in section~\ref{sec:factorization} for a general $n$-loop multi-eikonal diagram composed of individual gluon exchanges. We found that in a given web, the leading singularities of all maximally reducible diagrams are the same up to an overall symmetry factor, $s(D)$, which counts the number of ways of shrinking individual subdiagrams to the multi-eikonal vertex.
We have seen that the cancellation of the leading singularities in this class of webs can be understood through a property of the corresponding mixing matrix: the zero column sum rule of eq.~(\ref{columnrule}).

Furthermore,  the sum rule (\ref{columnrule}) itself appears to be completely general, working also for web diagrams which contain non-trivial connected subdiagrams. It is not yet clear what the physics of this result is in terms of renormalization properties. Further insight may be gained by corroborating the sum rule via a combinatoric proof, as has been achieved for the zero sum row property in~\cite{Gardi:2011wa}. In any case, the weighted column sum rule may act as a useful springboard for further investigation of the combinatorics underlying webs. 

In summary, soft-gluon exponentiation can both be derived from the renormalizability of Wilson-line correlators \emph{and} be constructed diagrammatically via webs. Putting these two pictures together allowed us to determine the singularity structure of webs and gain further insight to the structure of subdivergences. This understanding and the methods we have developed pave the way for higher-loop computation of the soft anomalous dimension and for a more complete understanding of infrared singularities in scattering amplitudes.

\acknowledgments

We are grateful to Eric Laenen and Gerben Stavenga for correspondence during early stages of this project, and to Gregory Korchemsky, Lorenzo Magnea and George Sterman for useful discussions. CDW is supported by the STFC Postdoctoral Fellowship ``Collider Physics at the 
LHC'', and thanks the School of Physics and Astronomy at the University of 
Edinburgh for hospitality on a number of occasions.


\appendix

\section{Results for kinematic factors}
\label{app:Fresults}
In this appendix, we collect results for the kinematic factors of various
web diagrams up to three loops, which are used throughout the paper in order
to verify the renormalization constraint of eq.~(\ref{eq:prize}). All results
have been obtained using the method outlined in section~\ref{sec:F3D}. 

Firstly, one may consider the one-loop web of figure~\ref{1loopfig}, whose 
kinematic part is given by (\ref{eq:Fijoneloop}), 
\begin{align}
  \label{eq:F12F23}
  \begin{split}
    \mathcal{F}^{(1)}(1_{ij})\ &=\ \left(\frac{\mu^2}{m^2} \right)^\epsilon\
    \frac{g_s^2}2\ \frac{\Gamma(1-\epsilon)\Gamma(2\epsilon)}{4\pi^{2-\epsilon}}\
    \gamma_{ij} \int_0^1 {\rm d}x\ P(x,\gamma_{ij})\,.\\
  \end{split}
\end{align}

Next up are various two-loop webs, shown in figure~\ref{2six}, where we use
labels introduced in~\cite{Gardi:2010rn}.
Their kinematic parts for the leading and first sub-leading poles are given by
\begin{subequations}
\label{eq:twoloop}
\begin{align}  
    \mathcal{F}^{(2)}(2a)\ &=\ \mathcal{C}_2\
    \frac{\gamma_{12} \gamma_{23}}{2\epsilon}\int_0^1 {\rm d}x {\rm d}w\
    \left(\frac{1-x}w\right)^{2\epsilon} P(x,\gamma_{12})P(w,\gamma_{23})
\left(1+{\cal O}(\epsilon^2)\right)\\
    \mathcal{F}^{(2)}(2b)\ &=\ \mathcal{C}_2\
    \frac{\gamma_{12} \gamma_{23}}{2\epsilon}\int_0^1 {\rm d}x {\rm d}w\
    \left(\frac{w}{1-x}\right)^{2\epsilon} P(x,\gamma_{12})P(w,\gamma_{23})
\left(1+{\cal O}(\epsilon^2)\right)\\
    \mathcal{F}^{(2)}(2f)\ &=\ \mathcal{C}_2\
    \frac{\gamma_{12}^2}{2\epsilon}\int_0^1 {\rm d}x {\rm d}y\
    \ \left( {\rm
        min}\left(\frac{x}{y},\frac{1-x}{1-y}\right)\right)^{2\epsilon}\
    P(x,\gamma_{12})P(y,\gamma_{12})\left(1+{\cal O}(\epsilon^2)\right)\\
    \mathcal{F}^{(2)}(2g)\ &=\ \mathcal{C}_2\ \frac{\gamma_{12}^2}{2\epsilon}\int_0^1 {\rm d}x {\rm d}y\
    \ \left( \left(\frac{x}{y}\right)^{2\epsilon} - \left(
        \frac{1-x}{1-y}\right)^{2\epsilon} \right) \Theta(x-y)
      P(x,\gamma_{12})P(y,\gamma_{12})\notag\\
&\times\left(1+{\cal O}(\epsilon^2)\right),
\end{align}
\end{subequations}
where we defined the common factor
\begin{align}
  \label{eq:C2def}
  \mathcal{C}_2=\left(\frac{\mu^2}{m^2} \right)^{2\epsilon}
    \frac{g_s^4}4\ \frac{\Gamma(1-\epsilon)^2\Gamma(4\epsilon)}{(4\pi^{2-\epsilon})^2}.
\end{align}
These expressions, and those that follow in eqs.~\eqref{eq:3loops} and
\eqref{eq:threeloopsecondweb}, are valid up to 
$\mathcal{O}(\epsilon^2)$ corrections, which is all we require in this study.  We note in
passing that there is zero contribution at ${\cal O}(\epsilon^{-2})$ for diagram (2g) as
we expect from the fact that this diagram is irreducible.

In section~\ref{3lw} we consider the three-loop web shown in
figure~\ref{3lsix}. The kinematic parts of each diagram for leading and first
subleading pole are found to be
\begin{subequations}
\label{eq:3loops}
\begin{align}
\allowdisplaybreaks
  \begin{split}
    \mathcal{F}^{(3)}(3A)\ &=\ \mathcal{C}_3\
    \frac{\gamma_{12}^2\gamma_{23}}{4\epsilon^2} \int_0^1{\rm d}x{\rm d}y{\rm
      d}w\ \left( \frac{1-x}{w} \right)^{2\epsilon} \left(
      \left(\frac{x}{y}\right)^{2\epsilon} - \left( 
        \frac{1-x}{1-y}\right)^{2\epsilon} \right) \\ & \qquad \qquad \times \Theta(x-y)
      \ P(x,\gamma_{12})P(y,\gamma_{12}) P(w,\gamma_{23})
\left(1+{\cal O}(\epsilon^2)\right) 
  \end{split}
\\
\notag
\\
\begin{split}
    \mathcal{F}^{(3)}(3B)\ &=\ \mathcal{C}_3\
    \frac{\gamma_{12}^2\gamma_{23}}{8\epsilon^2} \int_0^1{\rm d}x{\rm d}y{\rm
      d}w\ \left[   \left(\frac{1-y}w \right)^{2\epsilon} \left( \left(
        \frac{x}{y}\right)^{4\epsilon} - \left(  
        \frac{1-x}{1-y}\right)^{4\epsilon} \right) \right. \\
    & \hspace{5cm} \left. -2\left( \frac{1-x}{w} \right)^{2\epsilon} \left(
      \left(\frac{x}{y}\right)^{2\epsilon} - \left( 
        \frac{1-x}{1-y}\right)^{2\epsilon} \right) \right] \\ & \qquad \qquad
  \times \Theta(x-y)\ 
      P(x,\gamma_{12})P(y,\gamma_{12}) P(w,\gamma_{23}) 
\left(1+{\cal O}(\epsilon^2)\right)
\end{split}
\\
\notag
\\
\begin{split}
      \mathcal{F}^{(3)}(3C)\ &=\ \mathcal{C}_3\
      \frac{\gamma_{12}^2\gamma_{23}}{8\epsilon^2} \int_0^1{\rm d}x{\rm d}y{\rm
        d}w\ \left( \frac{1-y}{w} \right)^{-4\epsilon} \left(
        \left(\frac{1-x}{1-y}\right)^{-2\epsilon} - \left( 
          \frac{x}{y}\right)^{-2\epsilon} \right) \\ & \qquad \qquad \times \Theta(x-y)
      \ P(x,\gamma_{12})P(y,\gamma_{12}) P(w,\gamma_{23})  
\left(1+{\cal O}(\epsilon^2)\right)
\end{split}
\\
\notag
\\
\begin{split}
      \mathcal{F}^{(3)}(3D)\ &=\ \mathcal{C}_3\
      \frac{\gamma_{12}^2\gamma_{23}}{8\epsilon^2} \int_0^1{\rm d}x{\rm d}y{\rm
        d}w\ \left( \frac{1-y}{w} \right)^{2\epsilon} \left( {\rm
        min}\left(\frac{x}{y},\frac{1-x}{1-y}\right) \right)^{4\epsilon}\\ &
    \qquad \qquad \times P(x,\gamma_{12})P(y,\gamma_{12})
    P(w,\gamma_{23}) 
\left(1+{\cal O}(\epsilon^2)\right)
\end{split}
\\
\notag
\\
\begin{split}
    \mathcal{F}^{(3)}(3E)\ &=\ \mathcal{C}_3\
    \frac{\gamma_{12}^2\gamma_{23}}{8\epsilon^2} \int_0^1{\rm d}x{\rm d}y{\rm
      d}w\ \left[ 2\left( \frac{1-x}{w} \right)^{2\epsilon} \left( {\rm
          min}\left(\frac{x}{y},\frac{1-x}{1-y}\right) \right)^{2\epsilon}
    \right. \\ & \hspace{5cm} \left. -\left( \frac{1-y}{w} \right)^{2\epsilon}
      \left( {\rm 
          min}\left(\frac{x}{y},\frac{1-x}{1-y}\right) \right)^{4\epsilon} \right]\\ &
    \qquad \qquad \times P(x,\gamma_{12})P(y,\gamma_{12}) P(w,\gamma_{23}) 
\left(1+{\cal O}(\epsilon^2)\right)
\end{split}
\\
\notag
\\
\begin{split}
    \mathcal{F}^{(3)}(3F)\ &=\ \mathcal{C}_3\
    \frac{\gamma_{12}^2\gamma_{23}}{8\epsilon^2} \int_0^1{\rm d}x{\rm d}y{\rm
      d}w\ \left( \frac{w}{1-x} \right)^{4\epsilon} \left( {\rm
        min}\left(\frac{x}{y},\frac{1-x}{1-y}\right) \right)^{2\epsilon}\\ &
    \qquad \qquad \times P(x,\gamma_{12})P(y,\gamma_{12}) P(w,\gamma_{23})
\left(1+{\cal O}(\epsilon^2)\right)
\end{split}
\end{align}
\end{subequations}
where the common factor $\mathcal{C}_3$ is as in eq.~\eqref{eq:C3def}:
\begin{align}
  \label{eq:C3defapp}
  \mathcal{C}_3=\left(\frac{\mu^2}{m^2} \right)^{3\epsilon}
    \frac{g_s^6}8\ \frac{\Gamma(1-\epsilon)^3\Gamma(6\epsilon)}{(4\pi^{2-\epsilon})^3}\,.
\end{align}

In section~\ref{3lwprime} we consider the three-loop web of 
figure~\ref{3lfour}. The kinematic parts in this instance at leading and first
subleading pole are found to be
\begin{subequations}
  \label{eq:threeloopsecondweb}
\begin{align}
\allowdisplaybreaks
  \begin{split}
    \mathcal{F}(3a) &=\widetilde C_3 \frac{\Gamma(6\epsilon)}{8\epsilon^2}\int_0^1 {\rm
      d}x{\rm d} y {\rm d}z\ P(x,\gamma_{12}) P(y,\gamma_{23})
    P(z,\gamma_{34}) \\
    & \hspace{1.5cm} \times \left(3\left(\frac{1-x}y \right)^{2\epsilon} -\left(
      \frac{1-y}{z} \right)^{2\epsilon}\left(\frac{1-x}y
    \right)^{4\epsilon}\right)\left(1+{\cal O}(\epsilon^2)\right) 
  \end{split}
\\
  \begin{split}
     \mathcal{F}(3b) &=\widetilde C_3 \frac{\Gamma(6\epsilon)}{8\epsilon^2}\int_0^1 {\rm
      d}x{\rm d} y {\rm d}z\ P(x,\gamma_{12}) P(y,\gamma_{23}) P(z,\gamma_{34})\
    \left(\frac{x}y \right)^{4\epsilon} \left(
      \frac{1-y}{1-z} \right)^{2\epsilon} \left(1+{\cal O}(\epsilon^2)\right)
\end{split}
\\
  \begin{split}
     \mathcal{F}(3c) &=\widetilde C_3 \frac{\Gamma(6\epsilon)}{8\epsilon^2}\int_0^1 {\rm
      d}x{\rm d} y {\rm d}z\ P(x,\gamma_{12}) P(y,\gamma_{23}) P(z,\gamma_{34})\
    \left(\frac{z}{1-y} \right)^{4\epsilon} \left(
      \frac{y}{1-x} \right)^{2\epsilon} \\
&\times\left(1+{\cal O}(\epsilon^2)\right)
\end{split}
\\
  \begin{split}
     \mathcal{F}(3d) &=\widetilde C_3 \frac{\Gamma(6\epsilon)}{4\epsilon^2}\int_0^1 {\rm
      d}x{\rm d} y {\rm d}z\ P(x,\gamma_{12}) P(y,\gamma_{23}) P(z,\gamma_{34})\
    \left(\frac{1-y}z \right)^{2\epsilon} \left(
      \frac{y}{1-x} \right)^{2\epsilon}\\
&\times\left(1+{\cal O}(\epsilon^2)\right)\,,
  \end{split}
\end{align}
\end{subequations}
where 
\begin{align}
  \label{eq:Ctilde3}
  \widetilde C_3 = \left(\frac{\mu^2}{m^2} \right)^{3\epsilon}
    \frac{g_s^6}8\
    \frac{\Gamma(1-\epsilon)^3}{(4\pi^{2-\epsilon})^3}\,\gamma_{12}\gamma_{23}\gamma_{34}\,
\end{align}
differs from eq.~(\ref{eq:C3def}) only in the $\gamma_{ij}$ factors.

\section{Further examples of symmetry factors\label{app:sD}}

In section~\ref{sec:conjecture} we formulate a weighted column sum rule for web mixing matrices, eq.~(\ref{columnrule}). This 
involves the symmetry factors $s(D)$, which for a given diagram $D$ quantify 
the number of ways in which individual connected subdiagrams may be shrunk to 
the origin. In this appendix, we give some further examples of these symmetry
factors for various webs together with the corresponding mixing matrices. We shall see that in each case the weighted column sum rule of eq.~\eqref{columnrule} is admitted.

\begin{figure}
\begin{center}
\scalebox{0.8}{\includegraphics{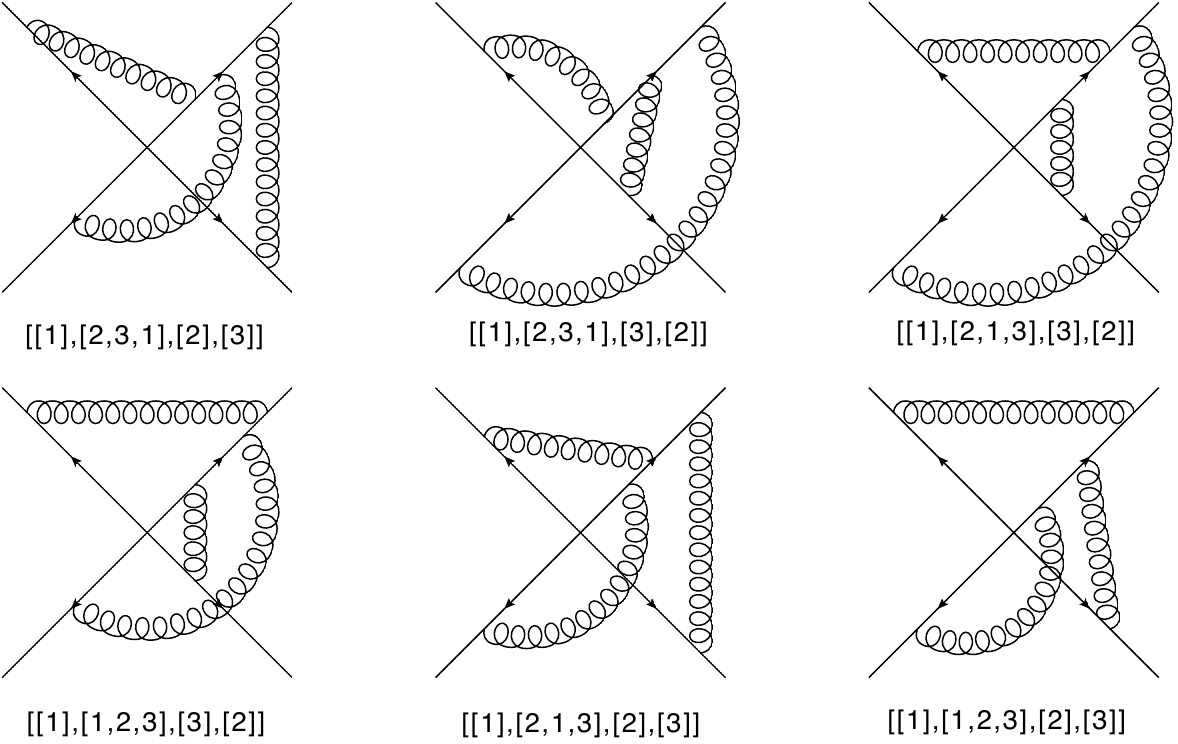}}
\caption{Web whose mixing matrix is given by eq.~(\ref{1311mat}).}
\label{1311fig}
\end{center}
\end{figure}

Firstly, we consider the web shown in figure~\ref{1311fig}, whose diagrams are
labelled as in~\cite{Gardi:2010rn}. The web mixing matrix is
\begin{align}
\frac{1}{6}
\left[ \begin {array}{rrrrrr} 2&-1&-1&2&-1&-1\\ \noalign{\medskip}-1&
2&-1&-1&-1&2\\ \noalign{\medskip}-1&-1&2&-1&2&-1\\ \noalign{\medskip}2
&-1&-1&2&-1&-1\\ \noalign{\medskip}-1&-1&2&-1&2&-1
\\ \noalign{\medskip}-1&2&-1&-1&-1&2\end {array} \right] 
 \left[ \begin {array}{c} [[1],[2,3,1],[2],[3]]\\ \noalign{\medskip}[[
1],[2,3,1],[3],[2]]\\ \noalign{\medskip}[[1],[2,1,3],[3],[2]]
\\ \noalign{\medskip}[[1],[1,2,3],[3],[2]]\\ \noalign{\medskip}[[1],[2
,1,3],[2],[3]]\\ \noalign{\medskip}[[1],[1,2,3],[2],[3]]\end {array}
 \right], \label{1311mat}
\end{align}
where we include the vector of labels on the right-hand side so as to 
define the ordering of the diagrams in the matrix. The symmetry factors $\{s(D)\}$
for this web are
\begin{displaymath} 
s^{(1,3,1,1)}_D=\left[ \begin {array}{cccccc} 1&1&1&1&1&1\end {array} \right]\,.
\end{displaymath}
It is straightforward to see that with this $s(D)$, the mixing matrix of~(\ref{1311mat}) satisfies eq.~\eqref{columnrule}.

\begin{figure}
\begin{center}
\scalebox{0.8}{\includegraphics{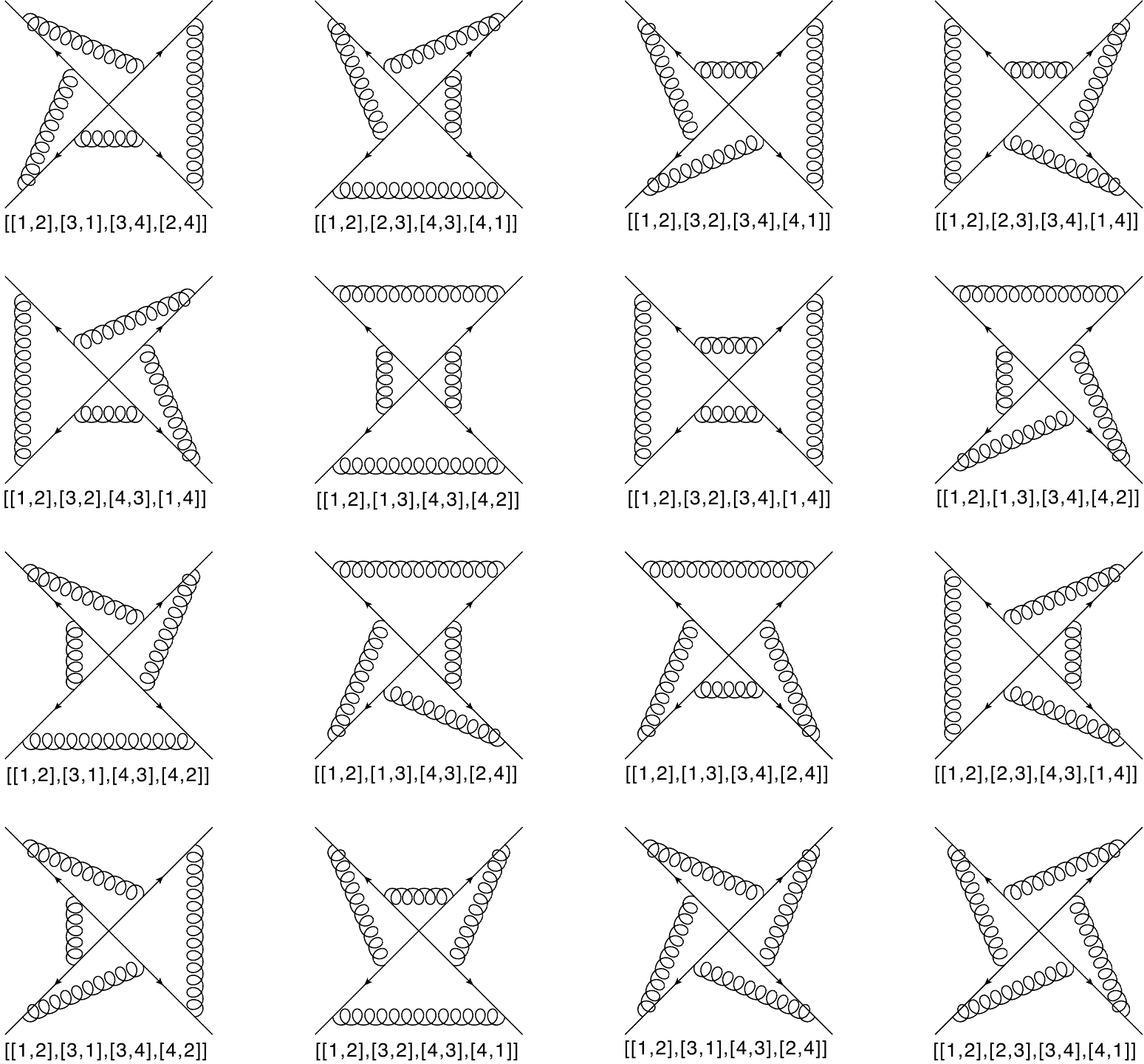}}
\caption{Web whose mixing matrix is given by eq.~(\ref{2222mat}).}
\label{2222fig}
\end{center}
\end{figure}

Next, we consider the four loop web of figure~\ref{2222fig}, whose web mixing
matrix is given by
{\footnotesize
\begin{align}
\frac{1}{24}
\left[ \begin {array}{rrrrrrrrrrrrrrrr} 24&0&-18&-6&-6&-18&-18&-6&-18
&-6&12&12&12&12&12&12\\ \noalign{\medskip}0&24&-6&-18&-18&-6&-6&-18&-6
&-18&12&12&12&12&12&12\\ \noalign{\medskip}0&0&6&-6&2&-2&-2&2&-2&2&4&4
&-4&-4&-4&4\\ \noalign{\medskip}0&0&-6&6&-2&2&2&-2&2&-2&-4&-4&4&4&4&-4
\\ \noalign{\medskip}0&0&2&-2&6&2&-6&-2&2&-2&-4&4&-4&4&-4&4
\\ \noalign{\medskip}0&0&-2&2&2&6&-2&-6&-2&2&-4&4&-4&4&4&-4
\\ \noalign{\medskip}0&0&-2&2&-6&-2&6&2&-2&2&4&-4&4&-4&4&-4
\\ \noalign{\medskip}0&0&2&-2&-2&-6&2&6&2&-2&4&-4&4&-4&-4&4
\\ \noalign{\medskip}0&0&-2&2&2&-2&-2&2&6&-6&-4&-4&4&4&-4&4
\\ \noalign{\medskip}0&0&2&-2&-2&2&2&-2&-6&6&4&4&-4&-4&4&-4
\\ \noalign{\medskip}0&0&2&-2&-2&-2&2&2&-2&2&4&0&0&-4&0&0
\\ \noalign{\medskip}0&0&2&-2&2&2&-2&-2&-2&2&0&4&-4&0&0&0
\\ \noalign{\medskip}0&0&-2&2&-2&-2&2&2&2&-2&0&-4&4&0&0&0
\\ \noalign{\medskip}0&0&-2&2&2&2&-2&-2&2&-2&-4&0&0&4&0&0
\\ \noalign{\medskip}0&0&-2&2&-2&2&2&-2&-2&2&0&0&0&0&4&-4
\\ \noalign{\medskip}0&0&2&-2&2&-2&-2&2&2&-2&0&0&0&0&-4&4\end {array}
 \right] 
 \left[ \begin {array}{c} [[1,2],[3,1],[4,3],[2,4]]
\\ \noalign{\medskip}[[1,2],[2,3],[3,4],[4,1]]\\ \noalign{\medskip}[[1
,2],[3,1],[4,3],[4,2]]\\ \noalign{\medskip}[[1,2],[2,3],[3,4],[1,4]]
\\ \noalign{\medskip}[[1,2],[1,3],[3,4],[4,2]]\\ \noalign{\medskip}[[1
,2],[3,1],[3,4],[2,4]]\\ \noalign{\medskip}[[1,2],[3,2],[4,3],[1,4]]
\\ \noalign{\medskip}[[1,2],[2,3],[4,3],[4,1]]\\ \noalign{\medskip}[[1
,2],[1,3],[4,3],[2,4]]\\ \noalign{\medskip}[[1,2],[3,2],[3,4],[4,1]]
\\ \noalign{\medskip}[[1,2],[1,3],[3,4],[2,4]]\\ \noalign{\medskip}[[1
,2],[2,3],[4,3],[1,4]]\\ \noalign{\medskip}[[1,2],[3,1],[3,4],[4,2]]
\\ \noalign{\medskip}[[1,2],[3,2],[4,3],[4,1]]\\ \noalign{\medskip}[[1
,2],[1,3],[4,3],[4,2]]\\ \noalign{\medskip}[[1,2],[3,2],[3,4],[1,4]]
\end {array} \right] \label{2222mat}\,.
\end{align}
}
The symmetry factors in this case are more complicated, and are given by
\begin{displaymath}
s_D^{(2,2,2,2)}=
 \left[ \begin {array}{cccccccccccccccc} 0&0&1&1&1&1&1&1&1&1&2&2&2&2&4
&4\end {array} \right] \,.
\end{displaymath}
Again, one can easily verify that with this $s(D)$, the mixing matrix of~(\ref{2222mat}) satisfies eq.~\eqref{columnrule}.
Note that two diagrams are not maximally reducible, as already pointed out 
in~\cite{Gardi:2010rn}. Furthermore, there are 3 topologies of maximally
reducible diagram equivalent under reflections and rotations, which give rise
to 3 possible values for $s(D)$, each with a nontrivial multiplicity.

\begin{figure}
\begin{center}
\scalebox{0.8}{\includegraphics{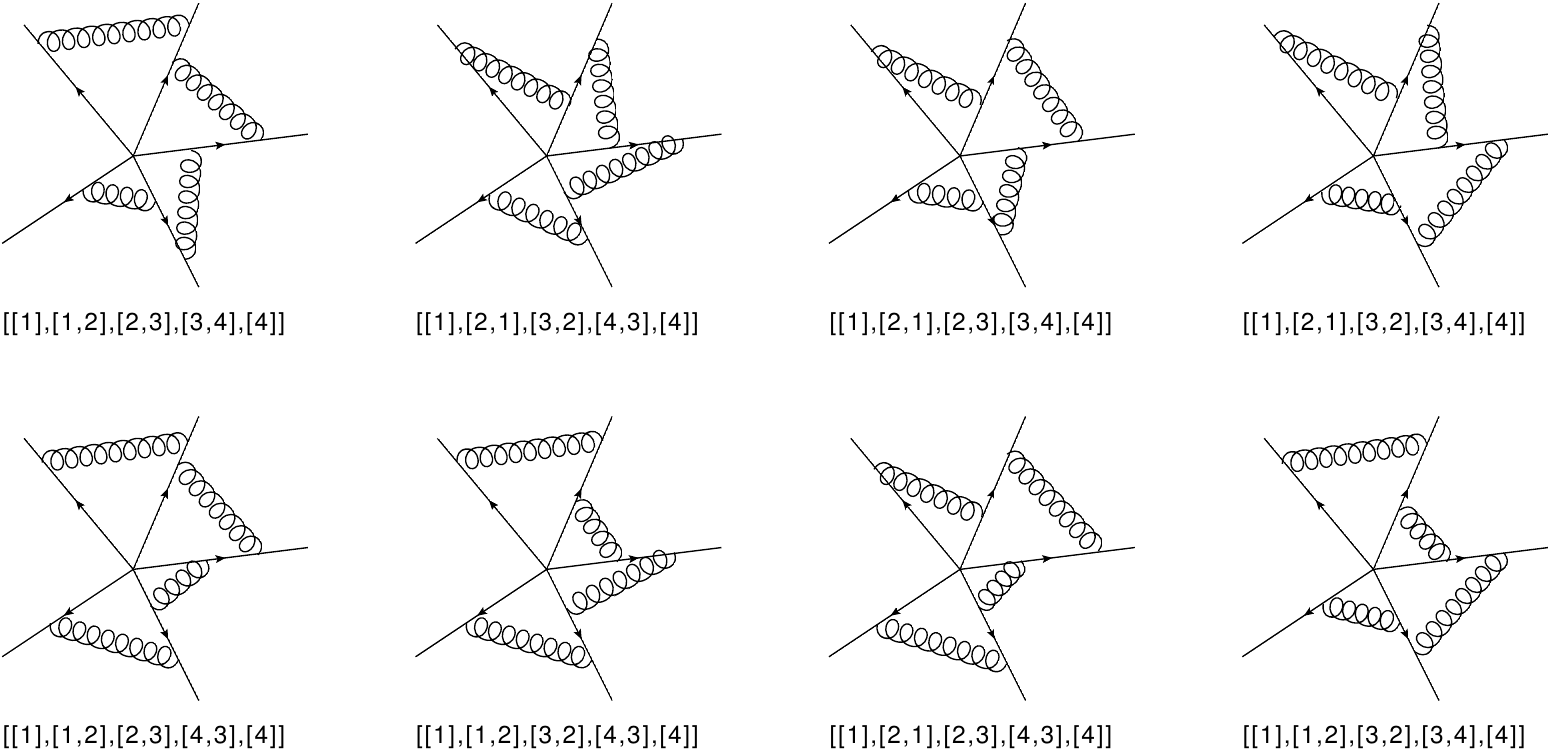}}
\caption{Web whose mixing matrix is given by eq.~(\ref{12221mat}).}
\label{12221fig}
\end{center}
\end{figure}

Next we consider a second four loop example, namely the web of 
figure~\ref{12221fig}.
In this case the mixing matrix is
\begin{align}
\frac{1}{6}\left[ \begin {array}{rrrrrrrr} 6&-6&-6&6&-6&6&6&-6
\\ \noalign{\medskip}-6&6&6&-6&6&-6&-6&6\\ \noalign{\medskip}-2&2&2&-2
&2&-2&-2&2\\ \noalign{\medskip}2&-2&-2&2&-2&2&2&-2
\\ \noalign{\medskip}-2&2&2&-2&2&-2&-2&2\\ \noalign{\medskip}2&-2&-2&2
&-2&2&2&-2\\ \noalign{\medskip}2&-2&-2&2&-2&2&2&-2
\\ \noalign{\medskip}-2&2&2&-2&2&-2&-2&2\end {array} \right] 
 \left[ \begin {array}{c} [[1],[1,2],[2,3],[3,4],[4]]
\\ \noalign{\medskip}[[1],[2,1],[3,2],[4,3],[4]]\\ \noalign{\medskip}[
[1],[2,1],[2,3],[3,4],[4]]\\ \noalign{\medskip}[[1],[2,1],[3,2],[3,4],
[4]]\\ \noalign{\medskip}[[1],[1,2],[2,3],[4,3],[4]]
\\ \noalign{\medskip}[[1],[1,2],[3,2],[4,3],[4]]\\ \noalign{\medskip}[
[1],[2,1],[2,3],[4,3],[4]]\\ \noalign{\medskip}[[1],[1,2],[3,2],[3,4],
[4]]\end {array} \right] \label{12221mat},
\end{align}
and the corresponding symmetry factors are
\begin{displaymath}
s_D^{(1,2,2,2,1)}=
 \left[ \begin {array}{cccccccc} 1&1&3&3&3&3&5&5\end {array} \right]. 
\end{displaymath}
There are again three different possible values of $s(D)$. Interestingly
in this case, these values are all odd numbers. Also here the sum rule of eq.~(\ref{columnrule}) is satisfied.

\begin{figure}[htb]
\begin{center}
\scalebox{1.0}{\includegraphics{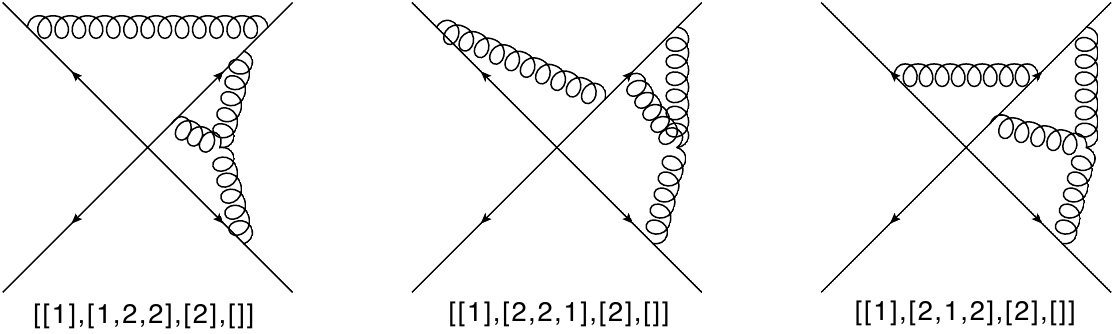}}
\caption{Diagrams contributing to eq.~(\ref{appex2}).}
\label{appendixfig2}
\end{center}
\end{figure}

Let us now consider a few examples of multiparton webs that contain one or more three-gluon vertices. A simple one is the three-loop web $W^{(3)}_{(1,3,1)}$ shown in figure~\ref{appendixfig2}.
The mixing matrix is given by
\begin{align}
\frac{1}{6}\left[ \begin {array}{rrr} 3&-3&0\\ \noalign{\medskip}-3&3&0
\\ \noalign{\medskip}-3&-3&\,\,6\end {array} \right] 
 \left[ \begin {array}{c} [[1],[1,2,2],[2],[]]\\ \noalign{\medskip}[[1
],[2,2,1],[2],[]]\\ \noalign{\medskip}[[1],[2,1,2],[2],[]]\end {array}
 \right]
\label{appex2}
\end{align}
and the symmetry factors by 
\begin{displaymath}
s_D^{(1,3,1)}=
 \left[ \begin {array}{cccccccc} 1&1&0\end {array} \right],
\end{displaymath}
admitting the sum rule of eq.~(\ref{columnrule}). 

\begin{figure}[htb]
\begin{center}
\scalebox{1.12}{\includegraphics{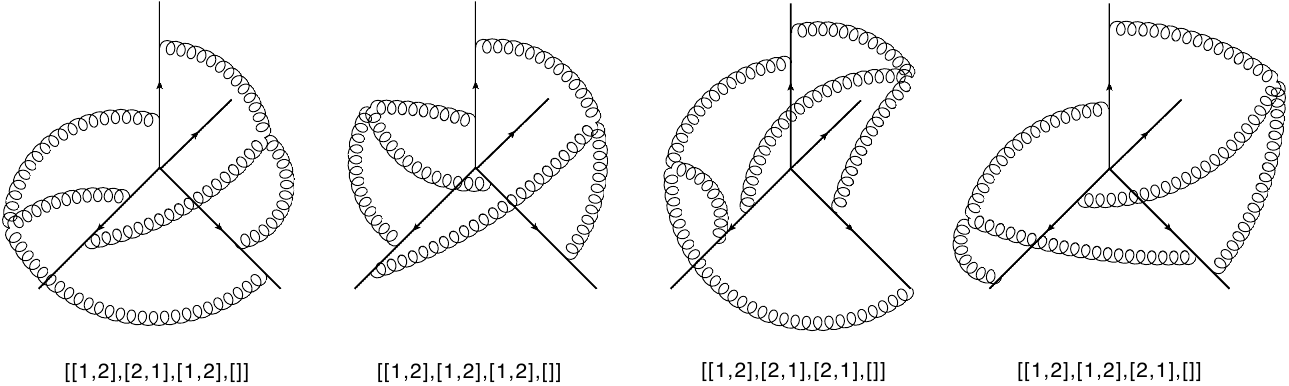}}
\caption{Diagrams contributing to eq.~(\ref{appex7}).}
\label{appendixfig7}
\end{center}
\end{figure}

Next consider the four-loop webs involving two three-gluon vertices, $W^{(4)}_{(2,2,2)}$, shown in figure~\ref{appendixfig7}.
The mixing matrix in this case is given by:
\begin{equation}
 \frac{1}{24}\,\left[ \begin {array}{rrrr} 24&-24&0&0\\ \noalign{\medskip}0&0&0&0
\\ \noalign{\medskip}0&-24&24&0\\ \noalign{\medskip}0&-24&0&24
\end {array} \right] 
 \left[ \begin {array}{c} [[1,2],[2,1],[1,2],[]]\\ \noalign{\medskip}[[1,
2],[1,2],[1,2],[]]\\ \noalign{\medskip}[[1,2],[2,1],[2,1],[]]
\\ \noalign{\medskip}[[1,2],[1,2],[2,1],[]]\end {array}
 \right]
\label{appex7}
\end{equation}
and the symmetry factor is 
\begin{displaymath}
s_D^{(1,3,1)}=
 \left[ \begin {array}{cccccccc} 0&1&0&0\end {array} \right]\,.
\end{displaymath}
Because three out of the four diagrams have no subdivergences, leaving just one maximally reducible diagram, [[1,2],[1,2],[1,2],[]],  the sum rule of eq.~(\ref{columnrule}) reduces to the statement 
that each element in the second row of (\ref{columnrule}) must vanish. This is indeed the case.

\begin{figure}[htb]
\begin{center}
\scalebox{0.8}{\includegraphics{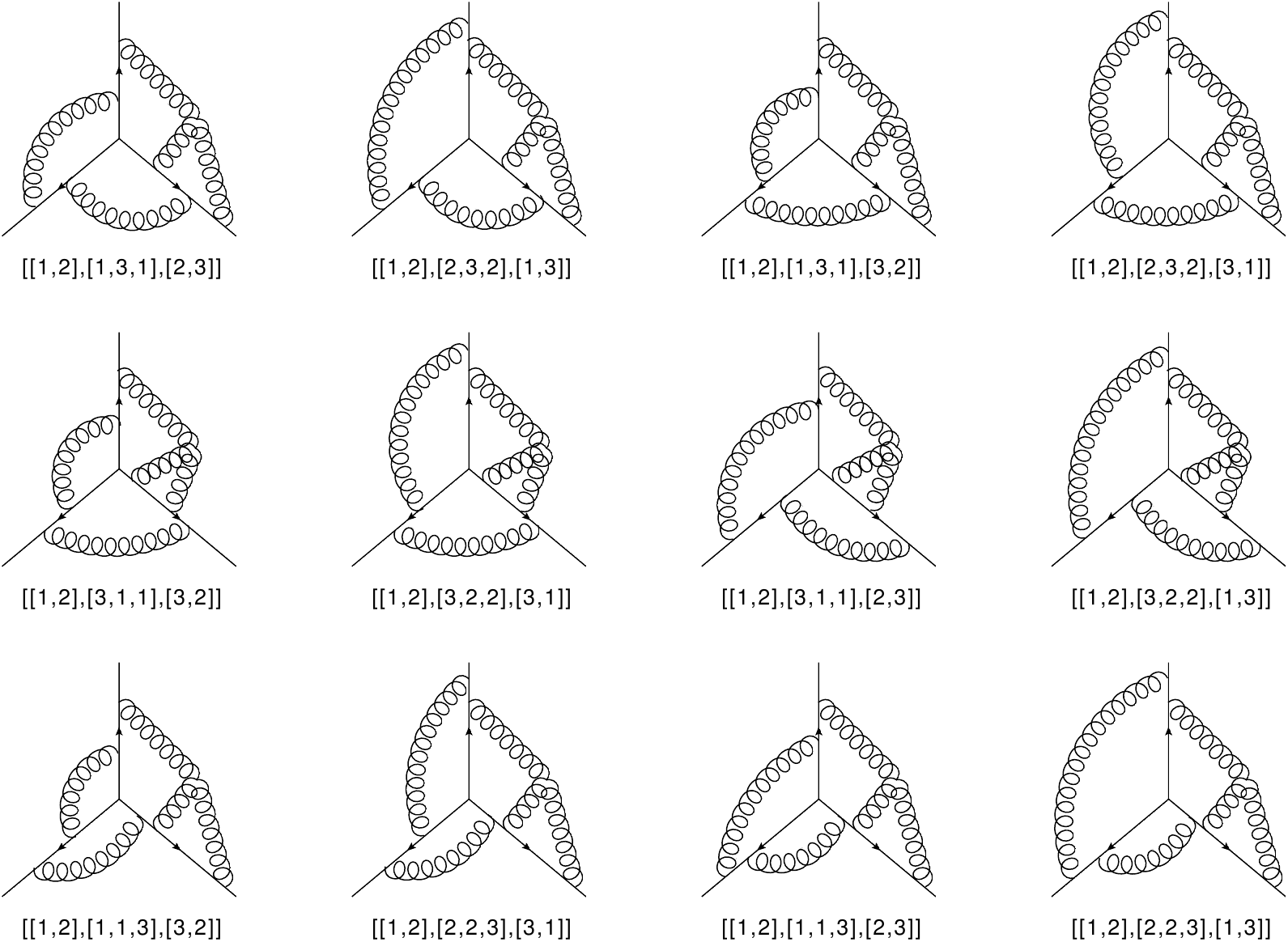}}
\caption{Diagrams contributing to eq.~(\ref{3l3gv}).}
\label{3l3gvfig}
\end{center}
\end{figure}

\begin{figure}[htb]
\begin{center}
\scalebox{0.7}{\includegraphics{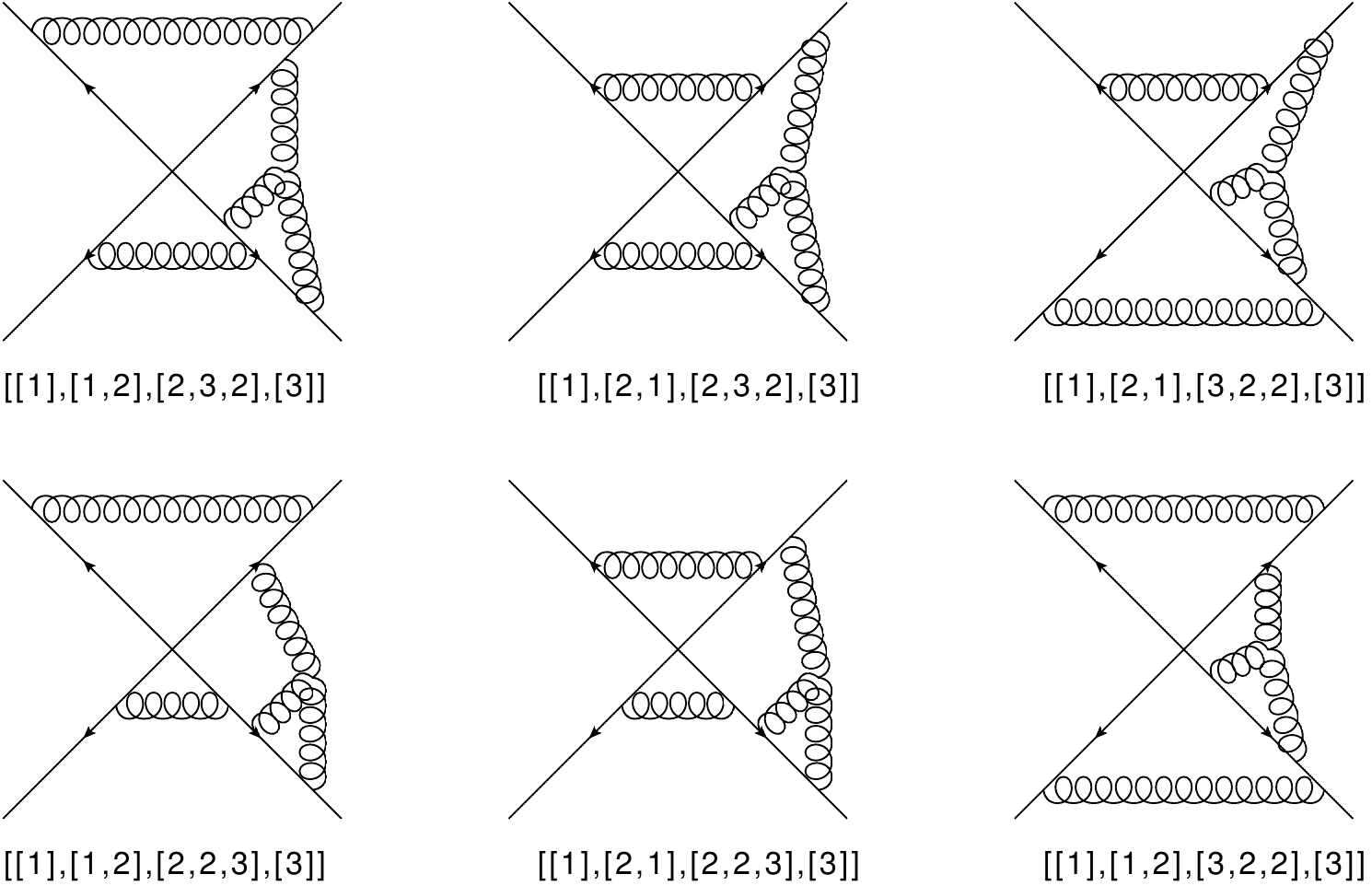}}
\caption{Diagrams contributing to eq.~(\ref{3lsix_3gv}).}
\label{3lsix_3gvfig}
\end{center}
\end{figure}

Next, consider the four loop example of figure~\ref{3l3gvfig} with mixing matrix 
\begin{align}
\label{3l3gv}
\frac{1}{24}
\left[ \begin {array}{rrrrrrrrrrrr} 24&0&0&0&-12&-12&-16&-4&8&-4&8&8
\\ \noalign{\medskip}0&24&0&0&-12&-12&8&8&-4&8&-16&-4
\\ \noalign{\medskip}0&0&24&0&0&0&-16&-16&8&-16&8&8
\\ \noalign{\medskip}0&0&0&24&0&0&8&8&-16&8&-16&-16
\\ \noalign{\medskip}0&0&0&0&12&-12&-4&8&-4&-4&-4&8
\\ \noalign{\medskip}0&0&0&0&-12&12&-4&-4&8&8&-4&-4
\\ \noalign{\medskip}0&0&0&0&0&0&8&-4&-4&-4&8&-4\\ \noalign{\medskip}0
&0&0&0&0&0&-4&8&8&-4&-4&-4\\ \noalign{\medskip}0&0&0&0&0&0&-4&8&8&-4&-
4&-4\\ \noalign{\medskip}0&0&0&0&0&0&-4&-4&-4&8&-4&8
\\ \noalign{\medskip}0&0&0&0&0&0&8&-4&-4&-4&8&-4\\ \noalign{\medskip}0
&0&0&0&0&0&-4&-4&-4&8&-4&8\end {array} \right] 
 \left[ \begin {array}{c} [[1,2],[1,3,1],[2,3]]\\ \noalign{\medskip}[[
1,2],[2,3,2],[3,1]]\\ \noalign{\medskip}[[1,2],[3,1,1],[2,3]]
\\ \noalign{\medskip}[[1,2],[2,2,3],[3,1]]\\ \noalign{\medskip}[[1,2],
[2,3,2],[1,3]]\\ \noalign{\medskip}[[1,2],[1,3,1],[3,2]]
\\ \noalign{\medskip}[[1,2],[1,1,3],[2,3]]\\ \noalign{\medskip}[[1,2],
[3,1,1],[3,2]]\\ \noalign{\medskip}[[1,2],[2,2,3],[1,3]]
\\ \noalign{\medskip}[[1,2],[3,2,2],[1,3]]\\ \noalign{\medskip}[[1,2],
[3,2,2],[3,1]]\\ \noalign{\medskip}[[1,2],[1,1,3],[3,2]]\end {array}
 \right] 
\end{align}
and symmetry factors 
\begin{equation}
\label{sD232}
s_D^{(2,3,2)}=
 \left[ \begin {array}{cccccccccccc} 0&0&0&0&0&0&1&1&1&1&1&1
\end {array} \right],
\end{equation}
such that the column sum rule is satisfied. All of the examples involving
three gluon vertices that we have so far considered involve $s(D)$ values of
1 or 0 only. As a final non-trivial example, it is thus instructive to consider
the web of figure~\ref{3lsix_3gvfig}.
Here the mixing matrix is
\begin{align}
\label{3lsix_3gv}
\frac{1}{24}
\left[ \begin {array}{rrrrrr} 12&-12&8&-4&4&-8\\ \noalign{\medskip}-
12&12&-4&8&-8&4\\ \noalign{\medskip}0&0&8&8&-8&-8\\ \noalign{\medskip}0
&0&8&8&-8&-8\\ \noalign{\medskip}0&0&-4&-4&4&4\\ \noalign{\medskip}0&0
&-4&-4&4&4\end {array} \right] 
 \left[ \begin {array}{c} [[1],[1,2],[2,3,2],[3]]\\ \noalign{\medskip}
[[1],[2,1],[2,3,2],[3]]\\ \noalign{\medskip}[[1],[2,1],[3,2,2],[3]]
\\ \noalign{\medskip}[[1],[1,2],[2,2,3],[3]]\\ \noalign{\medskip}[[1],
[2,1],[2,2,3],[3]]\\ \noalign{\medskip}[[1],[1,2],[3,2,2],[3]]
\end {array} \right] 
\end{align}
and the symmetry factors are
\begin{equation}
\label{sD1231}
s_D^{(1,2,3,1)}= \left[ \begin {array}{rrrrrr} 0&0&1&1&2&2\end {array} \right]. 
\end{equation}
Even in this case, the weighted column sum rule is satisfied.

Beyond presenting evidence for the sum rule of eq.~(\ref{columnrule}), we believe that 
these examples will be useful for further study of the symmetry factors and the role the sum rule plays in constraining the corresponding web mixing matrices. 
It is clear that some interesting combinatoric patterns underly the possible values of these objects. It would be both interesting and useful to know what these are.

\bibliographystyle{JHEP}
\bibliography{refs2}

\end{document}